\documentstyle[12pt]{article}

\def\({\c c}
\def\ii {\'\i}

\def\nl {\par \noindent }

\begin{document}

 \sloppy

\hoffset = -1truecm
\voffset = -2.5truecm  


\title{{\begin{flushright}
\nl {\normalsize SLAC-PUB-8219}
\nl {\normalsize August 1999}
\end{flushright}}
\vspace{0.4cm}
\large\bf  PERSPECTIVES OF LIGHT-FRONT QUANTIZED FIELD THEORY:   
{ \it  SOME NEW RESULTS}\thanks{Research partially supported by the Department 
of Energy under contract DE-AC03-76SF00515.} \thanks{Invited article for:  $\,$
{\sl Saga of Field Theory: From Points to Strings}, Ed., A.N. Mitra,  
 {\it Indian National Science Academy-INSA}, New Delhi, India.}}

\vspace{1.5cm}

\author{
{\large \bf
Prem P. Srivastava\thanks{E-mail: (1) prem@lafexsu1.lafex.cbpf.br; 
(2) prem@uerj.br; (3) prem@slac.stanford.edu. \quad 
On leave of absence from {\it Instituto de 
F\ii sica, UERJ- Universidade do Estado de Rio de Janeiro}, RJ, Brasil. } }\\
\it Stanford Linear Accelerator Center, Stanford University, Stanford,\\
{\it CA 94309, USA.} }

\date{}

\maketitle

\vspace{0.7cm}

\begin{abstract}
A  review  of some  basic topics in the light-front (LF) quantization of 
relativistic field theory is made. It is argued that 
the LF quantization is {\it equally appropriate} as the 
conventional one and that they lead,  
assuming the microcausality principle, to the same 
physical content. This is confirmed  in the studies on the LF  
 of   the  
spontaneous symmetry breaking (SSB), of the degenerate vacua in
  Schwinger model (SM) and  
  Chiral SM (CSM), of the chiral boson theory, and of the 
  QCD in covariant gauges among others.
The discussion on the LF is more economical 
 and  more transparent than that found in  the 
 conventional  equal-time quantized  theory.  
   The removal of the constraints on the LF phase space by 
following the Dirac method, in fact, 
  results in  a substantially 
reduced number of independent dynamical variables. Consequently, 
the  descriptions  of the 
physical Hilbert space and the vacuum structure, for example, 
become more tractable. 
 In the context of the Dyson-Wick perturbation theory the 
relevant popagators in the {\it front form} theory are causal. 
 The Wick rotation can then  be performed to employ the 
 Euclidean space integrals in momentum space.  The lack of
manifest covariance becomes   tractable, 
 and  still more  so  if we 
employ, as discussed in the text,  
the Fourier transform of the fermionic field
based on a special construction  of the  LF spinor.  
The fact that the hyperplanes $x^{\pm}=0$ 
constitute characteristic surfaces of the hyperbolic partial 
differential equation is found irrelevant 
in the quantized theory; it seems sufficient to quantize the theory 
on one of the 
charateristic hyperplanes.

\end{abstract}
\newpage

\tableofcontents                                                       

\vspace{0.3cm}

\nl {\bf Acknowledgements} \hfill  {\bf 36 }                               \\

\nl {\bf Appendix A:\qquad} {\bf  Poincar\'e Generators on the LF }\hfill {\bf
36}            \\

\nl {\bf Appendix B:\qquad} {\bf  LF Spin Operator. Hadrons in LF Fock
basis}\hfill {\bf 37}   \\

\nl {\bf Appendix C:\qquad} {\bf  BRS-BFT Quantization on the LF of the CSM
}\hfill {\bf 39}   \\ 

\nl {\bf References} \hfill  {\bf 46}

\newpage

\section{Introduction }\label{intro}
\vspace{1cm}

Half a centuary ago,  Dirac \cite{dir}  discussed the unification, 
in a relativistic theory,  of 
the principles of the quantization and the special relativity theory which were
by then firmly established. 
The Light-Front (LF) 
quantization which studies the  relativistic 
quantum dynamics of  physical system on the hyperplanes  
: $x^{0}+x^{3}\equiv {\sqrt{2}}x^{+}=const.$,  called 
the {\it front form} theory, was also proposed and some of 
its advantages pointed out.  
The {\it instant form} or the 
conventional equal-time theory 
on the contrary uses the $x^{0}=const.$ hyperplanes.  
The LF coordinates $x^{\mu}: (x^{+},x^{-},x^{\perp} 
)$,  where $x^{\pm}=(x^{0}{\pm} x^{3}) 
/{\sqrt 2}=x_{\mp}$ and   $ x^{\perp} = 
(x^{1}, x^{2})=(-x_{1},-x_{2})$,   are convenient to use in the {\it front form}
theory.  
They are  {\it not related by a finite Lorentz transformation} 
to the coordinates $(x^{0}\equiv t,x^{1},x^{2},x^{3})$ 
usually employed in the {\it instant form } theory and as such the 
descriptions of the same physical content in a dynamical theory 
on the LF, which studies the evolution of the system in $x^{+}$ in place of
$x^{0}$, 
 may come out to be different from that given in the conventional 
treatment. This was   found to be the case, for example, 
  in the description of the 
spontaneous symmetry breaking (SSB) mechanism (Sec. 3) some time ago 
and in the recent study (Sec. 6)  of some soluble 
two-dimensional gauge theory models,
where it was also demonstrated 
that LF quantization is very economical in displaying the
relevant degrees of freedom, leading  directly to a physical
Hilbert space.  The LF quantized field theory may perhaps  also be of some 
relevance in the understanding of the unification of the 
principles of the quantization with that of the general 
covariance \cite{kurs}. 

We recall that the  field theory at infinite momentum   was  employed   in the  
context of the  current algebra sum rules \cite{furlan}. 
The Feynman rules adapted for infinite momentum frame (IMF), which were  used by 
 Weinberg \cite{wei},  showed  substantial simplifications 
  in the context of the old fashioned perturbation theory computations. 
In the deep inelastic region with the IMF limit 
Bjorken \cite {bj1}
 predicted the scaling of the deep inelastic structure functions. The 
 parton model \cite{fey} of Feynman was also formulated in the IMF. 
At the same time the connection between the use of the LF variables and the 
IMF limit was being  noticed by several authors \cite{susskind}, 
which prompted  gradually the interest in the study of the 
{\it front form} theory as proposed by Dirac.   
 
More recently,  
  the interest in LF quantization  has been revived \cite{bro, ken, pre}   
 due to the difficulties encountered in the computation, in the conventional 
framework,  of the nonperturbative effects in the context of  QCD 
and  the problem of the relativistic bound states of  
fermions \cite{bro, ken} 
in the presence of the complicated vacuum. 
Studies show \cite{ken, bro, perry} that the application of 
Light-front Tamm-Dancoff method  may be feasible here.
 The technique of the 
regularization on the lattice has been 
quite successful for some problems but it cannot handle, for example, the  
 bound states of light ( chiral) fermions and  has not  been able yet 
to demonstrate, for example,  the confinenment of 
quarks. 
The problem of reconciling 
the standard constituent quark model and the QCD to 
describe the hadrons is also not satisfactorily resolved. In the former 
we employ few valence quarks while in the latter the QCD vacuum state itself 
contains, in the conventional theory, 
an infinite sea  of constituent quarks and gluons ( partons) 
with the density 
of low momentum constituents getting very large in view of the infrared 
slavery. 
  The {\it front form} dynamics may serve as a 
complementary tool to study such probelms,  since  we 
may possibly arrange to have a simple vacuum in it while transfering the 
complexity of the problem  to the 
LF Hamiltonian. In the case of the scalar field theory, for example, 
the corresponding LF  Hamiltonian is, in fact, 
found \cite{pref}  to be {\it nonlocal} due to the presence  of 
{\it constraints} on the {\it LF phase space}. 

The LF quantization of QCD in its Hamiltonian form provides an
alternative approach to lattice gauge theory for the computation of
nonperturbative quantities, such as \cite{bro, ken}  the
spectrum and the LF Fock state wavefunctions of relativistic bound
states.  LF variables have found natural applications in several contexts, 
for example, in 
the quantization of (super-) string theory and M-theory \cite{susskind1}.  
They have  also been
 employed in the nonabelian
bosonization \cite{wit} of the field theory of $N$ free Majorana fermions.  
The vacuum structures \cite{pre1,pre2} in the  LF quantized Schwinger model (SM) and the 
Chiral SM (CSM) have been recently studied  in a transparent 
fashion. The LF quantized QCD in covariant gauges has also 
 been studied \cite{pre3} in the context of the Dyson-Wick perturbation theory,  
where it is shown that  the  lack of manifest covariance 
in the  calculations  becomes  more tractable 
 thanks to a useful form of the LF spinor introduced (Sec. 4). The relevant 
 propagators are shown to be causal  and the Wick rotation can 
 be performed \cite{pre4} 
  to go over to the 
 Euclidean space integrals allowing for  the dimensional regularization 
 to be  used. 
 The {\it front form} theory has also found applications in the nonperturbative
 sector of QCD in the context of the Bethe-Salpeter dynamics. The Covariant 
 Instaneity ansatz (CIA) \cite{mit1} introduced earlier, which invokes the 
 Markov-Yukawa Transversality Principle, has been extended now to the covariant 
 null plane (CNPA) \cite{mit2,carbonel}.  


\subsection{Light-Front Quantized  Theory }

We will make the {\it convention} to regard
$x^{+}\equiv \tau$ as the 
LF-time coordinate while $x^{-}$ as the {\sl longitudinal 
spatial} coordinate. We note that   
  $\left[x^{+}, i\partial^{-} \right]
= \left[x^{-},{i}\partial^{+} \right]= -i $ where $\partial^{\pm}=
\partial_{\mp}=(\partial^{0}\pm \partial^{3})/\sqrt {2}$ etc. 
so that the coordinates $x^{+}$ and $x^{-}$ appear in a symmetric fashion.  
 In terms of 
the null vector $n^{\mu}=(1,0,0,1)/{\sqrt 2}$ and 
its dual ${\tilde n }=(1,0,0,-1)/{\sqrt 2}$, with  
$n\cdot n= {\tilde n }\cdot{\tilde n }=0$, ${\tilde n }\cdot n =1$, 
they  may be written  also as $x^{+}=n\cdot x$ and $x^{-}= {\tilde n}\cdot x$  
(See also Sec. 5).   
The temporal  evolution in $x^{0}$ or 
$x^{+}$ of the system is 
generated by the Hamiltonians which are different 
in the two {\it forms} of the theory.

Consider \cite{pre2, pre} the invariant distance between two spacetime points 
: $ (x-y)^{2}=(x^{0}-y^{0})^{2}-(\vec x-\vec y)^2= 2 
(x^{+}-y^{+}) (x^{-}-y^{-}) - (x^{\perp}-y^{\perp})^{2}$. 
On an equal $x^{0}=y^{0}=const. $ hyperplane the points have  
spacelike separation  except for if  they 
are {\it coincident} when it becomes lightlike one.  
On the LF with $x^{+}=y^{+}=const.$ 
the distance becomes  {\it independent of}  $(x^{-}-y^{-})$ and 
the seperation is again spacelike; it becomes lightlike one 
when  $x^{\perp}=y^{\perp}$ but with the difference that 
now the points need {\it not}  
necessarily be coincident along the longitudinal direction. 
The LF field theory hence {\it need not necessarily be local 
in} $x^{-}$, even if the corresponding 
{\it instant form} theory is formulated as a local one. 
For example, the commutator 
$[A(x^{+},x^{-},{x^{\perp}}),B(0,0,0^{\perp})]_{x^{+}=0}$ 
of two scalar observables would vanish on the grounds of
microcausality principle in relativistic field theory for 
$ x^{\perp}\ne 
0$ when  $x^{2}\vert_{x^{+}=0}$ is spacelike. 
Its value  would  hence be proportional to  $\,\delta^{2}(x^{\perp})\, $ 
and a finite number of its derivatives,  
implying locality only in $x^{\perp}$ but not necessarily so 
in $x^{-}$. Similar arguments in 
the {\it instant form} theory lead to the locality 
in all the three spatial coordinates. 
In view of the microcausality both  of the commutators 
 $[A(x),B(0)]_{x^{+}=0}$ and 
$[A(x),B(0)]_{x^{0}=0}$ are nonvanishing    
only on the light-cone, $x^{2}=0$. The possibility of nonlocality in the 
longitudinal direction in the {\it front form} treatment  seems to allow  us 
  to display  in some cases the structures parallel to those found 
  in  string theory (Sec. 4.6).  

We note  that in the LF 
quantization we time order with 
respect to   $\tau$ (which is a monotonic parameter as well) 
rather than  $t$.  
The  microcausality principle, 
however, ensures that the retarded commutators  
$[A(x),B(0)]\theta(x^{0})$ and  $[A(x),B(0)]\theta(x^{+})$,  
which appear \cite{ryd} in the S-matrix elements of relativistic 
field theory, do not  lead to disagreements in the two formulations. 
In the regions 
$x^{0}>0, x^{+}<0$ and $x^{0}<0, x^{+}>0$, where the commutators  
seem different  the $x^{2}$  is spacelike and both of them vanish. 
Hence, admitting \cite{weif}
 the  microcausality 
principle to hold,  the LF hyperplane  
seems  {\it equally valid and appropriate} as the conventional  one 
of the {\it instant form} theory for the canonical quantization. This is
demonstrated to be so, for example, in the context of SSB, SM, CSM, and QCD 
in covariant gauges discussed in this article.  

We note that the hyper planes $x^{\pm}=0$ define the characteristic 
surfaces of hyperbolic partial differential equation. It is known from their 
mathematical theory \cite{sne} that a solution exists if we specify the (Cauchy)
initial data on both of the hyperplanes. From  the actual 
computations in the  {\it front form} theory we come to 
conclusion \cite{ pre2} that  
(barring some massless theories) it is sufficient in  the 
canonical quantization of the {\it front form}  theory to 
select  one of the hyperplanes. The
information on the commutators on the other characteristic hyperplane 
seems to be already contained \cite{pre1} in the quantized theory.

A distinguishing feature of the {\it front form} theory is that it gives
rise to a constrained dynamical system \cite{dir1}. After the elimination of the
phase space constraints in the Hamiltonian formulation it leads to an
appreciable reduction in the number of independent field operators which would 
describe the Hilbert space of the theory.  
The vacuum structure, for example, 
then becomes more tractable and the 
computation of physical quantities simpler. This is, for example, 
 verified \cite{pre1, pre2, pre3} in the studies of the LF quantized SM, 
 CSM, and QCD in covariant gauges reviewed in Secs. (6, 7).

\subsection{LF Poincar\'e  and   IMF Generators. LF Spin Operator}

The structure of the {\it LF phase space} 
 is different from that of the one in the  conventional theory.  
A different description of 
 the same physical content, 
 compared to that  found in the conventional treatment, may emerge 
in the {\it front form} theory.
  For example, the SSB gets a different description \cite{ fno3, pre} and   
  the broken continuous symmetry is now inferred 
from the study of the residual unbroken symmetry of the  LF Hamiltonian 
operator while the symmetry of the LF vacuum remains intact. However, the 
expression which counts the number of Goldstone bosons 
present in the {\it front form} theory,  comes out to be the same as 
that found in the the discussion of 
equal-time quantized theory. A new proof of the
Coleman's theorem \cite{col} on the absence of the Goldstone bosons 
in two dimensional 
theory also emerges \cite{fno3, pre}.     
The LF vacuum 
is generally  found to be simpler  and in many cases 
the interacting theory vacuum 
is seen to coincide with the perturbation theory 
one \cite{fno1}.

Another  important advantge pointed out by Dirac of the {\it front form} 
theory is that here {\it seven} out of the ten   
Poincar\'e generators are {\it kinematical}, e.g., they leave 
the plane $x^{+}=0 $ invariant \cite{dir}. 
In the standard notation, {\it viz.},  
$K_{i}=-M^{0i}, J_{i}=-(1/2)\epsilon_{ijk}
M^{kl}, i,j,k=1,2,3$, they are : \quad    
$P^{+},P^{1},P^{2},\, 
M^{12}=-J_{3},\, M^{+-}= M^{03}= -K_{3},\,
M^{1+}=(K_{1}+J_{2})/\sqrt{2}$, and 
$ M^{+2}=(K_{2}-J_{1})/\sqrt{2}$. 
In the conventional theory on the other hand  
 only six such ones,  ${\vec P}$ and  
$M^{ij}=-M^{ij} $,  leave the hyperplane $x^{0}=0$ 
invariant. Expressed otherwise,  the generator 
$K_{3}$ is dynamical one in the {\it instant form} theory but it turns 
out to be   kinematical on the LF in the sense that there it 
generates \cite{pre1}  simply the scale 
transformations of the LF components of 
 $P^{\mu}$ and  $M^{\mu\nu}$, and $x^{\mu}$, with  $\mu,\nu=+,-,1,2$. 
 
There is also an interesting correspondence of the 
LF components of the Poincar\'e generators  
with the generators in the IMF.  
 Consider the inertial frame ${\cal S}'$ moving along the 3-axis with velocity 
$v/c =\tanh \eta$ relative to the inertial frame ${\cal S}$.   
From $({M'}^{\mu\nu}, {P}'^{\mu})$
$= exp(-i\eta K_{3})\;({M}^{\mu\nu}, {P}^{\mu})\;exp(i\eta K_{3})$ 
 we derive (Appendix A)
\begin{eqnarray}
{J'}_{1}&=&J_{1}\cosh \eta +K_{2} \sinh \eta, \quad 
{J'}_{2}=J_{2}\cosh \eta -K_{1} \sinh \eta, \quad {J'}_{3}=J_{3}
\nonumber \\
{K'}_{1}&=&K_{1}\cosh \eta -J_{2} \sinh \eta, \quad
{K'}_{2}=K_{2}\cosh \eta +J_{1} \sinh \eta, \quad {K'}_{3}=K_{3}
\nonumber \\
{(P_{0}+P_{3})}'&=&{e^{\eta}}(P_{0}+P_{3})\qquad
{(P_{0}-P_{3})}'={e^{-\eta}}(P_{0}-P_{3})\qquad {P'}_{1}=P_{1}\quad 
{P'}_{2}=P_{2} 
\end{eqnarray}
When $v/c \to 1 (-1)$ or $\eta\to \infty (-\infty)$ the Lorentz transformation 
becomes singular. However, we may define the {\it renormalized } 
generators, $\;{J'}_{a}/\cosh \eta$, $\,{K'}_{a}/\cosh \eta\,$, and 
$\; {e^{\mp\eta}}(P_{0}\pm P_{3})'$  which have well defined limits. 
The generators thus obtained coincide in the limit 
with the LF components of the  Poincar\'e generators. 
We note also that to  particle at rest in ${\cal S}$ corresponds  
 the  four-momentum ${p'}^{\mu}$ in the inertial frame ${\cal S'}$:  
$\;{p'}_{\mu}/(m_{0}\cosh \eta ),\,$  which tends to a null vector.

It is also worth remarking that the $\,+\,$ component 
of the Pauli-Lubanski pseudo-vector $W^{\mu}$ is special in that it 
contains only the LF kinematical generators. 
We may define   the  {\it LF Spin operator} by
 $\;{\cal J}_{3}=-W^{+}/P^{+}$. In the masssive case   
  the other two components of 
$\vec {\cal J}$, generating together  an $SU(2)$ algebra,  are shown to be 
 ${\cal J}_{a}=-({\cal J}_{3}P^{a}
+W^{a})/\sqrt{P^{\mu}P_{\mu}} \,$,  $ a=1,2$, which, however,  do  
carry in them  also the  LF {\it dynamical generators} 
$P^{-}, M^{1-}, M^{2-} $.  The case of both the massive and massless 
 fermions is discussed in detail in Sec. 4; the general case is considered 
 in Appendix B.

\section{LF quantized scalar theory}
\setcounter{equation}{0}
\renewcommand{\theequation}{2.\arabic{equation}}

\subsection{Covariant Phase Space Factor on the LF}
 
Some interesting insight on the {\it front form} quantized field theory 
may already be gained by considering 
the Lorentz invariant phase space-{\it LIPS} or {\it Covariant 
phase space} \cite{premsud} factor, which is found  relevant  
 in the analysis of the physical processes,  introduced first 
   in the context of the covariant version 
 of the statistical model of Fermi \cite{fermi}.  
On the LF the dispersion relation associated with the free massive particle  
is $2p^{+}p^{-}=(p^{\perp}p^{\perp}+m^{2})>0$.
 It has  no  square root,  like 
in the conventional one. 
The {\it signs}, for example,  
of $p^{+}$ and $p^{-}$ are {\it correlated}  since 
 $\;p^{+}p^{-}>0\,$ \cite{fno2}.      
 The {\it LISP} factor in the LF coordinates is thus  
defined as:   
 $\int d^{4}p \;\theta(\pm p^{+})\theta(\pm p^{-}) \delta(p^{2}-m^{2})
 =\int d^{2}p^{\perp}dp^{+}
\int dp^{-}\;\theta(\pm p^{+})\theta(\pm 
p^{-})\,\delta (2p^{+}p^{-}-\left[m^{2}
+{p^{\perp}}^{2}\right]\,)=\int {d^{2}p^{\perp}dp^{+} \theta(p^{+})/(2p^{+})}$,  
compared   to  the conventional  one:  
$\int d^{4}p \,\theta(\pm p^{0})\delta(p^{2}-m^{2})= 
\int {d^{3}{\vec p}/(2E_{p})}$ with  $E_{p}=+\sqrt{{\vec p}^2+m^2}>0$. 
A distinguishing feature in the case of the LF is thus  the appearence of 
$\;\theta(p^{+})/(2 p^{+})\;$ in the phase space factor. Such considerations 
are relevant, for example, in writing the Fourier transform of the fields and  
 the discussion of chiral boson theory (Sec. 3.4).

\subsection{LF Commutator }

    Consider, for example, 
a  real massive free scalar field $\phi(\tau, x^{-},x^{\perp})$, 
satisfying $({\Box}+ m^{2}) \phi=0$ where $\Box = (2\partial_{+}
\partial_{-}-\partial_{\perp}\partial_{\perp})$.   
For $p^{+}>0$, and consequently $p^{-}>0$, the complete set of plane wave 
solutions of the equation of motion are $ e^{+ip\cdot x}\,$  and 
$ e^{-ip\cdot x}\,$ where  $p\cdot x=p^{-}x^{+}+p^{+}x^{-}-p^{\perp}x^{\perp}$ 
and $\tau\equiv x^{+}$ indicates the LF-time coordinate. 
The Fourier transform  of $\phi$ on the LF may clearly   be  written as, 
  \begin{equation} 
 {\phi}(x)={\frac {1}{ {\sqrt {(2\pi)^{3}}}}}
              \int d^{2}p^{\perp}
	     {{ dp^{+}}\over {\sqrt {2p^+}}}
	      \theta(p^{+})\left[a(p^{+}, p^{\perp}) e^{-ip\cdot x}+
          {a^{\dag}}(p^{+}, p^{\perp}) e^{ip\cdot x}\right]
\end{equation}	 
where  we have 
isolated $\sqrt {2p^{+}}$  only for latter  convenience 
and $p^{\perp}$ as well as $p^{+}$ are to be integrated 
  from $-\infty$ to $\infty$,   which is
very convenient when we deal with generalized functions like $\theta(p^{+})$. 
  In the quantized theory  
 $a(p)$  and $a^{\dag}(p)$ denote the  creation and annihilation operators 
 of the quantum excitations associated with the quantized 
 field operator $\phi$.  
 They  are assumed to satisfy  
 the canonical commtation relations, with the nonvanishing one given by 
$\,\left [a(k), 
 a^{\dag}(p)\right]= \delta(k^{+}-p^{+})\delta^{2}(k^{\perp}-p^{\perp})\equiv
 \delta^{3}(k-p)$.  The Fock space is constructed employing these 
 operators. 
 
  The equal-LF-time  commutator  
of the field operator can be computed by employing  
its Fourier transform expression 
\begin{eqnarray}
 \left[\phi(x),\phi(y)\right]_{\tau}
 &=& 
 \frac{1}{(2\pi)^{3}} \int d^{2}p^{\perp}d^{2}k^{\perp}\,
 \frac {dp^{+}dk^{+}\theta(p^{+})\theta(k^{+})}{\sqrt{2p^{+}2k^{+}}}\,
 \nonumber \\[1ex]
&& \times
\left[e^{-i(p.x-k.y)}-	e^{i(p.x-k.y)}\right]_{x^{+}=y^{+}=\tau}\, \delta^{3}(p-k)
 \nonumber \\[1ex] 
&=&
 \frac{\delta^{2}(x^{\perp}-y^{\perp})}{(2\pi)} \int 
 \frac {dp^{+}}{2p^{+}}\,
\left[\theta(p^{+})+\theta(-p^{+})\right]
e^{-ip^{+}(x^{-}-y^{-})}
 \nonumber \\[1ex] 	
&=& -\frac {i}{4\pi} \epsilon(x^{-}-y^{-})
\delta^{2}(x^{\perp}-y^{\perp}). 
\end{eqnarray}
Here we have used the free particle dispersion relations for $k^{\mu}$ and 
$p^{\mu}$, made use of the delta function in the integrand,    
 set $\left[\theta(p^{+})+\theta(-p^{+})\right]= 1$ (or rather 
the Cauchy principal value-CPV) in the sense of the 
distribution theory, and used the integral representation of the  {\it sign}
function 
 $\,\epsilon(x)=1 $ or $=- 1$  according as $x>0$ or $x<0$.   
The equal-$\tau$ commutator obtained here,  often termed {\it  the LF
commutator},  
 is  {\it nonlocal}   
along the longitudinal direction $x^{-}$, which  as we argued before is 
not  unexpected in the {\it front form} theory.    
It  vanishes for the spacelike distances, 
and is nonvanishing  only on the light-cone for  $x^{-}\neq y^{-}$, when 
 we {\it assume } $\epsilon(0)=0 $.   

\subsection{Length Dimensions $\,L_{\perp}$ 
and $\,L_{\|}$ }

It is natural and suggested also, for example,  from the expression  of the 
LF commutator 
to introduce \cite{ken} {\it two distinct  
units of length dimensions}, $\,L_{\perp}$ 
and $\,L_{\|}$ in the {\it front  form} theory. 
 Indicating the dimension of a quantity by $[..]$ we write: 
 $\,[x^{\perp}]=L_{\perp}$, $\,[ x^{-}]=L_{\|}$, $ [\partial_{-}]=1/L_{\|}$. 
     Requiring that $p^{\perp}\cdot
x^{\perp}$ be dimensionless we find $[p^{\perp}]= [m]=1/L_{\perp}$, if we
 recall   the dispersion relation. 
Making  similar arguments we find  $[p^{+}] = 1/L_{\|}$, 
$[p^{-}] = L_{\|}/(L_{\perp})^{2}$, $[x^{+}] = (L_{\perp})^{2}/L_{\|}$ 
while   
$[H^{lf}]\equiv [P^{-}]=L_{\|}/(L_{\perp})^{2}$ for the LF Hamiltonian and  
 $[{\cal H}^{lf}]=1/(L_{\perp})^{4}$ for the Hamiltonian density. Similar 
considerations apply to the other composite operators like 
current densities and we remark that 
$\theta(x)$ and $\epsilon(x)$ do not carry any dimensions. The dimensional 
analysis is useful in finding \cite{ken}  the possible counter terms required in 
the renormalization of the theory. 
From the LF commutator (2.2) we conclude  that $[\phi]=
1/L_{\perp}$,  which is also found to be the case for the gauge field  but for 
the fermionic field we have  $[\psi_{+}]=1/(L_{\perp}{\sqrt L_{\|}})$, where 
$\psi_{+}$ indicates the dynamical component of the fermion field. 

\subsection{LF Hamiltonian. Dirac Procedure }

 The free scalar theory  is
 described by the Lagrangian density ${\cal L}= \partial_{+}\phi
\partial_{-}\phi-(1/2) \partial_{\perp}\phi\partial_{\perp}\phi-m^{2}
\phi^{2}/2$. 
 It is first order in $\partial_{+}\phi$ and 
the canonical momenta defined as  
$\,\pi=\partial {\cal L}/\partial(\partial_
{+}\phi)= \partial_{-}\phi\,$  describes   a  constraint 
on the  phase space dynamics of the {\it front form} scalar theory. 
We have here  a constrained dynamical system \cite{dir1}. 
The canonical Hamiltonian density is found to be 
$\,{\cal H}_{c}= m^{2}\phi^{2}/2$.  There is 
a systematic procedure\footnote{See also Secs. 5,6, and Appendix C.} 
- called the Dirac method \cite{dir1}- which allows us to 
construct the self-consistent Hamiltonian formulation, required to 
canonically quantize the  theory with phase space constraints. 
The primary constraint above is written as
\begin{equation} 
 \chi \equiv (\pi-\partial_{-}\phi) \approx 0
 \end{equation}
  where $\,\approx\,$ stands for 
weak equality, meaning that it should  not be employed inside the Poisson brackets, but only 
 after they  have been computed.    

We  define next an extended Hamiltonian density 
 ${\cal H}_{e}= {\cal H}_{c} + u \chi $  
 where  $u$ is a Lagrange multiplier 
 field. Hamiltons equations of motion employ  $H_{e}
 \equiv \int d^{2}x^{\perp}dx^{-} {\cal H}_{e}$ 
 and  we require the persistency condition on the constraint, e.g.,  
 $\{\chi(\tau,x^{-},x^{\perp}), H_{e}(\tau)\}\approx 0$. In the simple
  case under study 
  we are  led to a differential  equation which would determine the 
 multiplier field $u$. In the gauge theory considered below 
  new secondary constriants may arise.  We  now include them also 
  in the extended Hamiltonian  and repeat the procedure,  untill no more 
  constraints show up. For the computational purposes we may initially 
 start with    the 
 standard Poisson brackets at equal-LF-time $\tau$,  with  the 
  nonvanishing  one  defined by\footnote{In the context of the 
  canonical quantization we mostly  
   deal with equal-$\tau$ brackets and commutators. We  will  fequently  
 suppress $\tau$ from writing and write   
occasionally   $ x$ to indicate  the set  $(x^{-},x^{\perp})$. } 
 $\{\pi(\tau,x), \phi(\tau,y)\}=- \delta^{3}(x-y)\equiv -\delta^{2}(x^{\perp}-y^{\perp}) 
 \delta(x^{-}-y^{-})$.

  The nature of the set of constraints found  in the theory 
is then analyzed. A  constraint is first class if it has  vanishing 
Poisson brackets with itself,  with the  the other constraints,   and 
 with  the Hamiltonian; otherwise it is 
 a second class one.  Corresponding to a first class constraint we 
 may be required   
to add in  the theory some appropriate and
accessible gauge-fixing external constraints \cite{dir1}.   
In the present case their is one  local constraint $\,\chi\approx 0$. From 
the constraint matrix     
\begin{equation}
\{\chi(\tau,x^{-},x^{\perp}),
 \chi(\tau,y^{-},y^{\perp})\} =-2 \delta^{2}(x^{\perp}-y^{\perp}) 
 \partial_{-}\delta(x^{-}-y^{-}). 
\end{equation}
we conclude  that it  is second class by itself,  since the right hand side
is nonvanishing. There is, in fact,   also a first class
constraint in the theory in the form of the zero-momentum-mode of 
$\chi$; we will comment on it in Sec. (2.6).

 We go over now from the Poisson to the modified Poisson brackets, 
 called Dirac brackets,   which have the property that 
 we are allowed to set $\chi=0$ as a strong equality 
  relation, valid 
 even inside the Dirac brackets. The Hamilton's equations 
  employ \cite {dir1} now  the modified brackets. We construct first  the
   {\it inverse} 
   of the constraint matrix:  $\{\chi(\tau,x^{-},x^{\perp})$, 
 $\chi(\tau,y^{-},y^{\perp})\}^{-1} =\,-\delta^{2}(x^{\perp}-y^{\perp})
 \epsilon(x^{-}-y^{-})/4$.  The  modified 
 bracket is then defined by 
 \begin{equation}
 \{f(x),g(y)\}_{D}=\{f(x),g(y)\}- \int \int d^{3}u d^{3}v
 \{f(x),\chi(u)\}\{\chi(u),\chi(v)\}^{-1}\{\chi(v),g(y)\}
\end{equation}
 In  view of its very constructuion the Dirac bracket 
of any dynamical variable with $\chi$ is seen to vanish identically.

  It is clear  that in place of 
 $H_{e}$ we may then employ the {\it reduced Hamiltonian} obtained by setting 
 $\chi\equiv (\pi-\partial_{-}\phi)=0$  in it, which would also 
  remove 
 the Lagrange multiplier field, while $\pi $  becomes  now  
 a dependent variable, e.g., removed from the theory. For the independent 
 field $\phi$ which survives in  the {\it front form} scalar theory 
 here considered we find  
 \begin{equation}
 \{\phi(\tau,x),\phi(\tau, y)\}_{D}=- 
  \frac{1}{4}\epsilon(x^{-}-y^{-})\delta^{2}(x^{\perp}-y^{\perp})
 \end{equation}
The Hamilton's equation: $\,\dot \phi(\tau,x)=
\{\phi(\tau,x), H_{c}\}_{D}$,  where an overdot indicates the derivation 
with respect to $\tau$,   does recover also   the Lagrange equation.   
  The theory is canonically  quantized by the 
correspondence $\,i\{f,g\}_{D}\to  [f,g]\,$, 
 the commutator of the corresponding 
quantized operators. The Hamiltons equations correspond to the 
 equations of motion of  field operators, e.g., 
 $idf/d{\tau}=[f,H]$,  in the Heisenberg picture. 
The  commutator of the scalar field operators on the LF is thus given by 
\begin{equation}
\left[\phi(\tau,x),\phi(\tau,y)\right]= \frac{-i}{4}
 \epsilon(x^{-}-y^{-})\delta^{2}(x^{\perp}-y^{\perp})
 \end{equation}
which is the same as found above  by the simple arguments based on 
the  Fourier 
expansion of the field in the {\it front form} theory. Employing 
this commutator we recover in the present case 
the Lagrange equation of motion for the field operator as well.    

\subsection{Scalar Field Propagator in momentum space}

The Fourier expansion (2.1) may also be regarded as furnishing the momentum 
space realization of the commutator (2.7) and the 
 propagator in  momentum space is easily derived. 
 The propagator in configuration
space is defined by 
\begin{equation}
 \left\langle 0 \vert \,T(\phi(x)\phi(0))\, \vert 0\right\rangle= 
\theta(\tau)\,\left\langle 0 \vert(\phi(x)\phi(0))\,\vert 0\right\rangle
+\theta(-\tau)
\left\langle 0 \vert\,(\phi(0)\phi(x))\,\vert 0\right\rangle. 
\end{equation}  
It follows that 
\begin{eqnarray}
\left\langle 0\vert\,T(\phi(x)\phi(0))\,  \vert 0\right\rangle 
&=&
{1\over {(2\pi)^{3}}}\int d^{3}p 
\frac {\theta(p^{+})}{{2p^{+}}}\left[ \theta(\tau) e^{-ip.x}
+ \theta(-\tau) e^{ip.x}\right]
\nonumber  \\
&=&
\frac{i}{(2\pi)^{4}}\int  
{d^{3}p }d{\lambda}\, e^{-i(\lambda \tau+p^{+}x^{-}-p^{\perp}x^{\perp})}
 \frac {\left[\theta(p^{+})+\theta(-p^{+})\right]}{(m^2+p^{\perp}p^{\perp}-2p^{+}\lambda -i\epsilon)}
                       \nonumber  \\ 
&=&
\frac{i}{(2\pi)^{4}}\int d^{4}p \frac{e^{-ip.x}}{(p^{2}-m^2+i \epsilon)}
\end{eqnarray}
Here we   have used the integral representations\footnote{
$\theta(\tau)e^{-ip^{-}\tau}= {1}/{(2i\pi )}\int d \lambda {e^{-i\lambda \tau}}
/{(p^{-}-\lambda-i\epsilon)}$}
 of $\theta({\pm \tau})$ and  
 performed the well known standard manipulations.  The 
  factor $\,\left[\theta(p^{+})+
 \theta(-p^{+})\right]\,$ in the integrand has been set 
  to unity   and 
    the dummy integration variable  
  $\lambda$ has been renamed as   $p^{-}$ for convenience in the last line.  
 The  $d^{4}p$ stands for $ d^{2}p^{\perp}dp^{+}dp^{-}$,  
 with the understanding, as is clear from the derivation above,  
 that the integration over 
 the  $p^{-}$ has to be performed first. 
 The range of integration is 
 from $-\infty$ to $\infty$ for all of these variables. 
 
 The momentum space 
 representations of the energy-momentum tensor are also found easily and we
 check that $\,N(p)=a^{\dag}(p)a(p)\,$ has the usual interpretation of a number 
 operator. In fact, 
 \begin{eqnarray}
 {H^{lf}}_{c}&\equiv &P^{-}= \int d^{2}x^{\perp}d x^{-}
 :\left[\frac {m^{2}}{2} \phi^{2}+ \frac{1}{2}\partial_{\perp}\phi 
 \partial_{\perp}\phi \right]: \nonumber \\
 &=&\frac{1}{2} \int d^{2}p^{\perp}\,
  {dp^{+}\theta(p^{+})}\!  : \left[ a^{\dag}(p)a(p)+ a(p)a^{\dag}(p)\right]: 
   \frac{m^{2}+p^{\perp}p^{\perp}}{2p^{+}}\nonumber \\
   &=& \! \int d^{2}p^{\perp}\,
  {dp^{+}\theta(p^{+})}  \left [ a^{\dag}(p)a(p) \right ] \; p^{-}
 \nonumber \\  
 P^{+} &= & \int d^{2}x^{\perp}d x^{-}
 : {(\partial_{-}\phi)}^{2} : \nonumber \\
&=&\int d^{2}p^{\perp}\,
  {dp^{+}\theta(p^{+})}   \; \left[ a^{\dag}(p)a(p) \right]  \; p^{+} 
\end{eqnarray}

\subsection{First class constraint.  
  Symmetry in $x^{+}$ and $x^{-}$ }

 It is worth making an {\it important remark}.  There is, in fact, 
present \cite{girotti} in the scalar
theory discussed above {\it still another constraint which is first class}.  
 We easily show that the zero-longitudinal-momentum mode 
$\,\sqrt{2\pi}\tilde{\chi}(\tau, k^{+}=0)=\int dx^{-} \chi\,$, 
represents a first class constraint in the theory.  For example, considering 
for simplicity the two dimensional theory,  (2.4) reads in 
the momentum space as  
\begin{equation}
\{\tilde{\chi}(\tau, k^{+}),\tilde{\chi}(\tau, p^{+})\}= 
-2ik^{+} \delta(k^{+}+p^{+}). 
\end{equation}
It clearly indicates the presence of the  
first class constraint $\tilde{\chi}(\tau, k^{+}=0)\approx 0$ in the theory. 
Such a constraint or symmetry  requires us to introduce in the theory an external
(gauge-fixing ) 
 constraint \cite{dir1}, such that the pair becomes a second class set. 
   We will take  advantage of this gauge freedom  
 in order to decompose 
 the scalar field into the  {\it bosonic condensate} variable   and the 
 quantum fluctuation field. 
 When combined with the {\it standard} Dirac procedure it allows
 us to build \cite{fno3, pre} a description  of the  SSB mechanism on the LF.  
  
We also note  that 
the {\it front form} formulation of {\it relativistic} theory 
is inherently symmetrical with respect to 
$x^{+}$ and $x^{-}$ 
and it is a matter of {\it convention} that we take the plus component as
the LF-time   while the other as a spatial coordinate. 
The  theory quantized at $x^{+}=const.$  hyperplanes seems  already 
to incorporate \cite{pre1} in it the information on the 
equal-$x^{-}$ commutation relations.  
For example, 
we easily derive from (2.1) the following equal$-x^{-}$ commutator 
\begin{eqnarray}
\lefteqn{
\left[\phi(x^{+},x^{-}, x^{\perp}),\phi(y^{+},x^{-},y^{\perp} ) 
 \right]= } \nonumber \\
 & & 
 \frac{1}{(2\pi)^{3}} \int\! d^{2}p^{\perp}\,
 \frac {dp^{+}\theta(p^{+})}{2p^{+}}\,
\left[e^{-ip^{-}(x^{+}-y^{+})+ip^{\perp}(x^{\perp}-y^{\perp})}-	
e^{ip^{-}(x^{+}-y^{+})-ip^{\perp}(x^{\perp}-y^{\perp})}\right]\!. 
\end{eqnarray} 
In view of the free particle dispersion relation we may replace the measure 
$ {dp^{+}\theta(p^{+})}/{2p^{+}}$ by $ {dp^{-}\theta(p^{-})}/{2p^{-}}$. 
 The equal-$x^{-}$ commutator is then given
by $\, ({-i}/{(4\pi)}) \epsilon(x^{+}-y^{+})
\delta^{2}(x^{\perp}-y^{\perp})$. 
In two dimensional space-time it is customary to define 
 the right  and the left movers  by 
$\phi(0,x^{-})\equiv \phi^{R}(x^{-})$, 
and $\phi(x^{+},0)\equiv \phi^{L}(x^{+})$. We find   
$\left [\phi^{R}(x^{-}),\phi^{R}(y^{-})\right]=(-i/4)\epsilon(x^{-}-y^{-})$
while $\left[\phi^{L}(x^{+}),\phi^{L}(y^{+})\right]=
(-i/4)\epsilon(x^{+}-y^{+})$. The symmetry under discussion seems responsible 
for appreciable simplifications in the {\it front form} quantized 
 theory. 


\section{SSB Mechanism, Topological Kink Solution, and  Chiral Boson theory 
on the LF}

\setcounter{equation}{0}
\renewcommand{\theequation}{3.\arabic{equation}}
\subsection{SSB in two dimensional scalar theory }

 The conventional {\it instant form} description of the tree level 
 SSB is based on the space and time independent solutions 
 of the Lagrange equation, 
 $\phi_{class}\equiv \omega$,  such that they also  
 minimize  the Hamiltonian functional;  
invoking  the (external) physical considerations. 
 We do not apparently  have much physical intuition on the LF 
to avail of such arguments. The constrained dynamical system on the LF seems, 
however, to already contain in it  the corresponding relevant constraints. 
    For simplicity we consider first  the 
two dimensional theory\footnote{Here $\tau=(x^{0}+x^{1})/{\sqrt 2}$,  
$x\equiv x^{-}=(x^{0}-x^{1})/{\sqrt 2}$. 
 An overdot indicates 
the LF-time derivative while a prime indicates derivative with respect to 
$x^{-}$; 
the generalization to $3+1$ dimensions is discussed in Sec. (3.2).} with 
$\; {\cal L}=(\partial_{+}\phi)(\partial_{-}\phi)-V(\phi)\;$   

This is probably  the simplest example of a constrained dynamical system 
in the context of field theory. It is reasonable to expect that the 
well tested  Dirac procedure, when applied to it,  {\it must } 
result  in a satisfactory  description of SSB on the LF.

   The Lagrange equation, $\,2\dot{\phi^\prime}= - V'(\phi) $,   
   is of first order in LF-time $\tau$. 
The left hand side remains unaltered under $\phi\to \phi+ c(\tau)$   
and  $\phi=const.$ are clearly possible solutions\footnote{The self-dual 
{\it kink} solution which depends on $x^{-}$ as well is 
discussed in Sec. (3.3).}. Integrating over the 
space variable and assuming appropriate   boundary conditions we 
are led to the following constraint \cite{fno3, pre} on the potential 
 \begin{equation}
\int  dx^{-}\;\frac{\delta V(\phi)}{\delta \phi}=0. 
\end{equation}
We show now that this constraint is also present   on the phase space 
and in the quantized theory. The description of SSB then follows from the 
discussion on the structure of the Hilbert space.    

In order to take care of the  first class constraint $\,\int dx^{-} 
\chi \approx 0\;$ 
mentioned  in Sec. (2.6) we make the following {\it separation} 
of the dynamical (collective) bosonic {\it condensate } 
variable $\omega(\tau)$ from  
the (quantum) fluctuation variable $\varphi(\tau,x)$
\begin{equation} 
\;\phi(\tau,x)=\omega(\tau)+\varphi(\tau,x).
\end{equation}
Here  we also  set  $\int dx^{-} \varphi(\tau,x)=0$ 
 so that the fluctuation field carries no zero-longitudinal-momentum mode 
in it.
 The {\sl  separation thus corresponds in a sense 
 to an external  gauge-fixing constraint } which we must impose \cite{dir1} 
 in the theory. 
  It was  introduced \cite{fno3} 
 originally on physical considerations and $\omega$ was termed   as the  
 dynamical {\it bosonic condensate} variable. 
 
 We  apply now   the  standard Dirac procedure   to construct  
LF Hamiltonian formulation.  The  canonically quantized 
theory  results \cite{pre, fno3} in the following commutators 
\begin{eqnarray}
\left[\varphi(x,\tau),\varphi(y,\tau)\right]&=&-{i\over 4}\epsilon(x-y),\\
\nonumber \\
\left[\omega(\tau),\varphi(x,\tau)\right] &=&0.  
\end{eqnarray}
and for $V(\phi)=
(\lambda/4)(\phi^{2}-m^{2}/\lambda)^{2}$, with a 
negative sign for the mass term and    $\lambda \geq 0$, $m\neq 0$,  
the  LF Hamiltonian is given by  
\begin{equation}
{H^{lf}\equiv 
 P^{-}}= \int  dx \,\Bigl [\omega(\lambda\omega^2-m^2)\varphi+
{1\over 2}(3\lambda\omega^2-m^2)\varphi^2+
\lambda\omega\varphi^3+{\lambda\over 4}\varphi^4 
\Bigr].  
\end{equation}
 We recover also 
the  {\it constraint equation} (3.1) now  as a second class 
constraint   on the phase space: 
\begin{equation}
{\omega(\lambda\omega^2-m^2)}+
\quad{1\over R}
 \int_{-R/2}^{R/2} dx \Bigl[ \,(3\lambda\omega^2-m^2)\varphi + 
 \lambda (3\omega\varphi^2+\varphi^3 ) \,
\Bigr]=0  
\end{equation}
where $R\to\infty$ and  the Cauchy principle value of $\int_{-\infty}^
{\infty} dx \,f(x)$ is 
defined by $lim_{R\to \infty}\int_{-R/2}^{R/2} dx \,f(x)$.

 The
commutation relations indicate that the operator $\omega$ is a c-number or a
background field.  
Eliminating $\omega$ would lead  to  {\it LF 
 Hamiltonian which is nonlocal} \cite{fno3, pre} along the 
 longitudinal coordinate $x^{-}$ even though     
   the  scalar  theory  written  
in the conventional coordinates is  local.

At the tree or classical level, $\varphi$ are bounded ordinary functions 
in $x^{-}$ and  when $R\to \infty$ only the first term survives  in the 
constraint equation leading to  
${\omega(\lambda\omega^2-m^2)}=0$.  
  This result is the same  
   as that obtained  in the conventional theory.  
There, however,  it is essentially added to the theory, 
on  physical considerations,  which require the energy functional to 
attain its minimum (extremum) value. 
The {\it stability property}, say,  of a particular constant solution 
may be inferred as usual from the  analysis of the  
classical partial differential equation of motion. For example,  
$\omega=0$ is shown to be an unstable solution on the LF 
for the potential  considered above,  
while the other two 
solutions with $\omega\neq 0$ give rise to the  stable 
phases.  A similar analysis, it is clear,  of the corresponding 
{\it partial} differential  equations 
 in the conventional coordinates  can also 
be made; the Fourier transform theory is convenient to use.  
Also the new ingredient in the form of the 
  constraint equation on the LF  does  have  
its counterpart in the conventional {\it instant form} framework as 
is shown  in \cite{pre4}.  
It is 
{\it remarkable} that the {\it front form} theory seems to contain inside   
it all 
the necessary ingredients in order to  describe the SSB, 
 when we follow the Dirac 
procedure to handle the constrained LF dynamics of the scalar field.

 We could have employed   the DLCQ \cite{pauli}, 
including the   condensate term  also in  it. The existence of the {\it
continuum } limit of DLCQ theory adding to it also the dynamical condensate 
variable was demonstrated \cite{fno3, pre5}, contradicting 
the then prevalent notion on the contrary\footnote{See,   
T. Maskawa and K. Yamawaki, Prog. Theo. Phys. 56 (1976) 270; 
K. Nakanishi and K. Yamawaki, Nucl. Phys. B122 (1977) 15.    
  A history of the so called zero mode problem is traced \cite{pre} in 
hep-th/9312064.}.  
 The demonstration assures \cite{parisi} us of the 
self-consistency  of the {\it front form} theory itself. 
 In the theory described in 
finite volume, the commutator of $\omega$ with    $\varphi$ is found 
nonvanishing  and as such it is an  {\it operator};  only when 
$R \to \infty$ does it becomes a classical background field.

  It is worth  stressing  
that in our discussion   the condensate variable is introduced as a dynamical 
variable.  The Dirac {\it procedure must  decide} whether it comes out to be 
  c- or  q-number. In the discussions of the  bosonized 
SM and CSM  models the operator   $\omega$ 
is not a background field, like in the scalar theory.    It turns out to be 
an  operator  and  plays an  important role in 
describing the structure of the Hilbert space 
and the degenerate vacua in these gauge theories (Sec. 6).

The field commutator obtained above can be realized 
in momentum space through  
 the   Fourier transform of $\varphi$: \quad  
$\varphi(x,\tau)= {(1/{\sqrt{2\pi}})}\int dk\; 
{\theta(k)}\;
[a(k,\tau)$ $e^{-ikx}+{a^{\dag}}(k,\tau)e^{ikx}]/(\sqrt {2k})$, 
where $k^{+}\equiv k $ and the operators $a(k,\tau)$ and ${a^{\dag}}(k,\tau)$ 
satisfy the canonical equal-$\tau$ commutation relations, with the nonvanishing
one given by  
$[a(k,\tau),{a^{\dag}}(k^\prime,\tau)]=\delta(k-k^\prime)$.  

The (perturbative) vacuum state is  defined 
by  $\,a(k,\tau){\vert vac\rangle}=0\,$, 
$k> 0$. The tree level  
description of the {\it SSB} is  given as
follows. The values of $\omega=
\,{\langle\vert \phi\vert\rangle}_{vac}\;$ obtained from  
$V'(\omega)=0 $ characterize 
the different   vacua in the theory. 
Distinct  Fock spaces corresponding to different values of $\omega$ 
are built as usual by applying the creation operators on the corresponding
vacuum state.  The $\omega=0$ corresponds to  {\it symmetric phase} 
since the Hamiltonian operator is then symmetric under 
$\varphi\to -\varphi$. For $\omega\ne 0$ this symmetry is 
violated and the  system is said to be 
in a {\it broken or asymmetric phase}.

 The constraint equation (3.6) also  shows that the value of  
$\omega$ would be altered from its tree level value 
in view of the quantum corrections,  arising from the other terms. The 
renormalization of the two 
dimensional scalar theory was discussed  \cite{pre4}   
 to one-loop order by  employing the 
Dyson-Wick  expansion based on LF-time ordering. It was 
 found   that it is   convenient to 
derive \cite{pre4} the renormalized constraint equation instead   of 
solving    the constraint equation first, which would require the 
difficult job of dealing with     nonlocal and nonlinear Hamiltonian. 
     
     In the  supernormalizable theory here  
the two renormalized equations, viz, the mass renormalization  condition and 
the renormalized constraint equation, 
 allow us to 
to study \cite{pre4} the phase 
transition in the two dimensional scalar theory, 
which was  conjectured long time ago by Simon and Griffiths \cite{simon}.

\subsection{Spontaneously broken continuous symmetry}

The {\it extension  to 
$3+1$ dimensions} and to the global {\it continuous symmetry } 
is straightforward \cite{pre}.  
Consider  real scalar fields 
$\phi_{a} (a=1,2,..N)\,$ which form  an isovector of global 
internal symmetry group 
 $O(N)$. We now  write
 $\phi_{a}(x,x^{\perp},\tau)
=\omega_{a}+\varphi_{a}(x,x^{\perp},\tau)$ and 
the Lagrangian density is  ${\cal L}=[{\dot\varphi_{a}}{\varphi'_{a}}-
{(1/ 2)}(\partial_{\perp}\varphi_{a})(\partial_{\perp}\varphi_{a})-V(\phi)]$.   
 The Taylor series 
expansion of the constraint equations $\beta_{a}=0$ gives a set of coupled
equations $R\,V'_a(\omega)+
\,V''_{ab}(\omega)\int dx \varphi_{b}+\,
V'''_{abc}(\omega)\int dx \varphi_b\varphi_c/2+...=0$. Its discussion at 
the tree level leads to the conventional theory results. 
The  LF symmetry generators are found to be 
$G_{\alpha}(\tau)=
-i\int d^{2} {x^{\perp}} dx \varphi'_{c}(t_{\alpha})_{cd}\varphi_{d}$ 
$=\,\int d^{2}{k^{\perp}}\,dk \, \theta(k)\, 
{a_{c}(k,{k^{\perp}})
^{\dag}} (t_{\alpha})_{cd} a_{d}(k,{k^{\perp}})$ where 
$\alpha,\beta=1,2,..,N(N-1)/2\, $,  are the group indices, 
$t_{\alpha}$ are hermitian and antisymmetric generators of $O(N)$, and 
${a_{c}(k,{k^{\perp}})^{\dag}}$ ($ a_{c}(k,{k^{\perp}})$) is creation (
destruction)  operator,   contained in the momentum space expansion 
of $\varphi_{c}$. These  are to be contrasted with the generators 
in the equal-time theory,  
$ Q_{\alpha}(x^{0})=\int d^{3}x \, J^{0}
=-i\int d^{3}x (\partial_{0}\varphi_{a})(t_{\alpha})_{ab}\varphi_{b} 
-i(t_{\alpha}\omega)_{a}\int d^{3}x 
({{d\varphi_{a}}/ dx_{0}})$.  
All  the symmetry generators thus  
annihilate the LF vacuum  and the SSB is  
now seen  in the 
broken symmetry of the quantized theory Hamiltonian. The expression which  
 counts the number of Goldstone bosons in the {\it front form} theory 
 is found to be identical to that  in the conventional theory. 
In contrast, the first term 
on the right hand side of $Q_{\alpha}(x^{0})$, which is similar to the 
one on the LF,  does  annihilate  
the conventional theory vacuum  but 
the second term  gives now  non-vanishing contributions  
for some of the (broken) generators. The  symmetry of the
conventional theory vacuum 
is thereby broken  while the quantum Hamiltonian remains invariant.  
The {\it physical content}  of SSB in the {\it instant form}
and the {\it front form}, however, is the same 
though achieved  by different descriptions. Alternative proof  
 on the LF,  
in two dimensions, can be given  
of the Coleman's theorem related to the absence of 
Goldstone bosons; we are unable \cite{pre} to implement 
the second class constraints over the phase space. The tree level Higgs 
mechanism may also be discussed straightforwardly \cite{pre}. 
We remark that the simplicity of the LF vacuum is in a sense 
compensated by the involved nonlocal Hamiltonian. The latter, however, 
may be treatable using advance computational techniques. 
 Also in connection with renormalization it may not 
 be necessary \cite{pre4};
we may instead obtain the renormalized 
constraint equations.

\subsection {Kink solution and Topological quantum number}

The  classical Lagrange equation of the two dimensional  self-interacting
theory, $\,2\partial_{-}\partial_{+}\phi= -V'(\phi)\,$, with the $V(\phi)$ 
given above, 
 is known to have finite energy topological soliton solutions \cite{leite}
  called 
 {\it kink } solutions. The theory has an internal symmetry, 
 $\phi \to -\phi$.  They can  
be recovered in the {\it front form} theory as well. 
The  {\it kink} corresponds to   the  {\it self-dual}  
solution satisfying $ \partial_{-}\phi=  -\partial_{+}\phi $ and given by  
\begin{equation}
\phi_{kink}= 
{\pm }\frac{m}{\sqrt \lambda} \tanh \left [ \frac{m}{2} (x^{+}-x^{-})
\right]
\end{equation}
where the upper (lower)  sign corresponds  to the kink (anti-kink) solutions. 
The kink on the LF carries both the LF energy and longitudinal momentum such
that $P^{+}=P^{-}$ and\footnote{ $P^{-}=\int d x^{-} V(\phi)$ and 
$P^{+}= \int d x^{-} (\partial_{-}\phi)^{2}$. } 
its mass is determined to be $\,M= \sqrt{2P^{+}P^{-}}=
{\sqrt 8}m^{3}/(3\lambda)$. The kink interpolates between the two vacuua of the
theory: $\phi_{kink}(0,x^{-}=\infty)=-m/{\sqrt \lambda} $ and 
$\phi_{kink}(0,x^{-}=-\infty)= m/{\sqrt \lambda} $. 
The topological charge may be defined by $Q= \int d x^{-} j^{+}$ 
where $j^{\mu}=- ({\sqrt \lambda}/(2m)) \epsilon^{\mu\nu} \partial_{\nu}
\phi$, with $\epsilon^{+-}=-\epsilon^{-+}=\epsilon_{01}=1$,  
is the  conserved topological current density. The toplogical charges of kink,
anti-kink and vacuum solutions are $1$, $-1$, and $0$ respectively. The $Q$ 
is absolutely conserved prohibiting the decay of the kink into vaccum. 
Similar  (topological) quantum numbers on the LF  arise also, 
for example,  in the context 
of the structure of the 
   degenerate vacua in the canonical quantiztion of SM and CSM models discussed 
below.

\subsection {Chiral Boson theory on the LF}

The chiral boson (or self-dual scalar) field in $1+1$ dimensions plays 
an important role, for example,  in the formulation 
of  string theories \cite{schw},  
in the description \cite{stone} 
of boundary excitations  of the quantum Hall state, and in a number of two-
dimensional statistical systems which are related to the Coulomb-gas model.

We recall that the free massive  theory with  
$\;{ \cal L}= \partial^{\mu}\phi \partial_{\mu}\phi /2-m^{2}\phi^{2}/2\;$
has the LF Hamiltonian $\;m^{2}\phi^{2}/2\;$. The dispersion relation 
 $2p^{+}p^{-}=m^{2}>0$ governs  the {\it correlation} between the signs of 
  $p^{+}$ and $p^{-}$. In the massless theory,     
at the      classical level,  a chiral boson 
solution, $\partial_{0}\phi=\partial_{1}\phi $ (and an  anti-chiral one, 
$\partial_{0}\phi=-\partial_{1}\phi $) is obtained.  
Several  quantized theory models \cite {siegel, 
flor, sriv} of chiral boson have been proposed. 
 The {\it front form} theory of chiral boson looks  more appropriate 
 and transparent \cite{prep} when compared to the conventional one.

The Floreanini and Jackiw (FJ) model \cite{flor}  is  based on the 
 following {\it manifestly  non-covariant Lagrangian}
\begin{eqnarray}
{ \cal L}&= & (\partial_{0}\phi-\partial_{1}\phi)\partial_{1}\phi 
\nonumber \\
&= &
\frac{1}{2}{\eta}^{\mu\nu} \partial_{\mu}\phi \partial_{\nu}\phi -
\frac{1}{2} (\partial_{0}\phi-\partial_{1}\phi)^{2}.  
\end{eqnarray}
where $\phi$ is a real scalar field and 
$\eta^{00}=- \eta^{11}=1, \; \eta^{01}=\eta^{10}=0$. 

  In the {\it instant form }  
frame work it leads \cite{flor, costa} to  the following equal-time commutator 
\begin{equation}
\left[\phi(x^{0},x^{1}), \phi(x^{0}, y^{1})\right]= \frac{-i}{4}
\epsilon (x^{1}-y^{1}).  
\end{equation}
The commutator  is nonvanishing, is nonlocal, 
and violates the microcausality principle, 
contrary to what we encounter  usually  
in the   conventional theory \cite {weif}.     
 These {\it objections   disappear} when  we consider the theory 
quantized in  the LF coordinates. 

We will consider a {\it modified} FJ chiral boson  model with the following  
Lagrangian density written in the LF coordinates 
\begin{eqnarray}
{ \cal L}&= & (\partial_{+}\phi-\frac{1}{\alpha}
\partial_{-}\phi)\;\partial_{-}\phi 
\nonumber \\
&= &
\frac{1}{2} {\eta}^{\mu\nu} \partial_{\mu}\phi \partial_{\nu}\phi - 
 \frac {1}{\alpha}(\partial_{-}\phi)^{2},  
\end{eqnarray}
where $\eta^{+-}=\eta^{-+}=1, \eta^{++}=\eta^{--}=0$ and 
$\alpha$ is a fixed parameter. For 
$\alpha=1 $ it coincides with  (3.8) in the conventional 
coordinates.  
 
The LF quantization  of the scalar theory 
with a potential term included in it has been discussed in Sec. (2.4). From 
(3.10) we derive   
\begin{eqnarray}
{ H}^{lf}&=& \int d x^{-} \; \frac{1}{\alpha} (\partial_{-}\phi)^{2}\nonumber \\
\left[\phi(\tau,x^{-}),\phi(\tau,y^{-})\right]&= &\frac{-i}{4}
 \epsilon(x^{-}-y^{-})
 \end{eqnarray}
The  LF commutator (3.11), which is nonlocal in $x^{-}$ and  
  nonvanishing only on the light-cone, does  not 
   conflict with the microcausality  (Sec. 1.1 and \cite{weif}) unlike 
   (3.9).  
  
The Heisenberg equation of motion for the field operator is 
\begin{equation}
\partial_{+}\phi=\frac{1}{i}\left[\phi, H^{lf}\right]=
 \frac{1}{\alpha}\partial_{-}\phi  
\end{equation}
and   the Lagrange equation 
\begin{equation}
  \partial_{-}\left[\partial_{+}\phi - \frac{1}{\alpha}\partial_{-}\phi
 \right]=0.
 \end{equation}
is recovered.

The commutator (3.11) can be realized in momentum space through the following 
Fourier transform ($x^{+}\equiv \tau$ )
\begin{equation} 
\phi(x^{+},x^{-})=\frac{1}{\sqrt{2\pi}}\!
\int dk^{+}\;\!\frac{\theta(k^{+})}{\sqrt{2k^{+}}}
\left[a(x^{+}, k^{+}) e^{-ik^{+}x^{-}}+ 
{a^{\dag}}(x^{+}, k^{+}) e^{ik^{+}x^{-}}\right],  
\end{equation}
if  the operators $a$ and $a^{\dag}$ are  assumed to satisfy the equal-$\tau$ 
canonical commutation relations, with the nonvanishing one given by 
 $\;\left[a(x^{+}, k^{+}), 
{a^{\dag}}(x^{+}, p^{+})\right]= \delta(k^{+}-p^{+})$.
On using the equation of motion (3.12) we derive easily   
\begin{equation}
a(x^{+}, k^{+})= e^{-ik^{-}x^{+}} a(k^{+}), \qquad 
a^{\dag}(x^{+}, k^{+})= e^{ik^{-}x^{+}} a^{\dag}(k^{+}).
\end{equation} 
where we  set 
\begin{equation}
k^{-}=\frac{1}{\alpha} k^{+}, \qquad \mbox{\rm implying}
\qquad 2k^{+}k^{-}=\frac{2}{\alpha}(k^{+})^{2}.  
\end{equation}
The dispersion relation for the free  FJ chiral boson is different from 
that  for a free scalar particle with (finite $k^{+}$ but)  vanishing mass,  
except for when $|\alpha| \to \infty$.     

The Fourier transform now assumes  the form 
\begin{equation} 
\phi(x^{+},x^{-})=\frac{1}{\sqrt{2\pi}}\!\int dk^{+}\;
\!\frac{\theta(k^{+})}{\sqrt{2k^{+}}}
\left[a( k^{+}) e^{-ik \cdot x}+ 
{a^{\dag}}( k^{+}) e^{ik \cdot x}\right]. 
\end{equation}
where $k\cdot x\equiv k^{-}x^{+}+ k^{+}x^{-}
= k^{+} (x^{-}+x^{+}/\alpha)$ and 
the nonvanishing commutator satisfies  $\;\left[a( k^{+}), 
{a^{\dag}}( p^{+})\right]= \delta(k^{+}-p^{+})$. 

The components of the classical canonical energy-momentum tensor $T^{\mu \nu}$ 
following from the noncovariant Lagrangian density (3.10) are 
found to be 
\begin{eqnarray}
T^{+-}=-T^{-+}= \frac{1}{\alpha} T^{++}&=& 
\frac{1}{\alpha} (\partial_{-}\phi)^{2},\nonumber\\
T^{--}&= &(\partial_{+}\phi)^{2}-\frac{2}{\alpha}(\partial_{+}\phi)
(\partial_{-}\phi). 
\end{eqnarray}
The    on shell conservation equations  
\begin{equation}
 \partial_{\mu}T^{\mu \pm}= 2 (\partial_{\mp}\phi)\,
 \partial_{-}\left[\partial_{+}\phi - \frac{1}{\alpha}\partial_{-}\phi
 \right]=0
\end{equation} 
   may be easily
checked. They allow us to  define, if the surface integrals can be dropped, 
 the conserved translation generators $P^{\pm}$  
\begin{equation}
P^{+}= \int d x^{-} :T^{++}:\;= \int d x^{-} :(\partial_{-}\phi)^{2}:\;=
\int dk^{+} \theta(k^{+}) \; N(k^{+})\; (k^{+}) 
\end{equation}
and 
\begin{equation}
P^{-}\equiv H^{lf}= \int d x^{-} :T^{+-}:\;= \frac{1}{\alpha}\; P^{+}
\end{equation}
where $N(k^{+})={a^{\dag}}( k^{+})a( k^{+})$ is the number operator and 
$:\;\; :$ indicates the normal ordering.

From (3.18) and in  virtue of $ \;(T^{+-}+T^{-+})=0\,$   
we may derive the following  relation 
\begin{equation}
\partial_{+}\left[x^{-} T^{++}+ x^{+} T^{+-}\right]+
\partial_{-}\left[x^{-} T^{-+}+ x^{+} T^{--}\right]= 0. 
\end{equation}
which is valid on shell.  
We may hence define another  conserved generator
\begin{equation}
M=  x^{+} P^{-} + \int dx^{-}\; x^{-} T^{++}
\end{equation}

The generators $M, P^{+}, P^{-}$ form a closed algebra: \quad  
  $\left[ M, P^{+}\right]
=-iP^{+}$,   $\; \left[ M, P^{-}\right]=-iP^{-}$, and 
$\;  \left[ P^{+}, P^{-}\right]= 0$. The operator  $M$ thus 
generates the scale (boost) transformations on $P^{\pm}$ 
by the same amount which leaves $P^{+}/P^{-}$ invariant.   
The mass operator $\,2P^{+}P^{-}$, however, gets scaled and 
is not invariant  under $M$.  
 The usual (kinematical)  
Lorentz boost generator $\, M^{+-}\equiv - x^{+} P^{-}+ \int dx^{-}\;
x^{-} T^{++}\,$ has similar properties. It is, however, as   
seen from (3.19), 
 is {\it not conserved}  in the manifestly {\it noncovariant model}  
  under consideration.  
The Lagrange equation is shown to be form invariant under the 
infinitesimal transformation \cite{flor, west} 
$\; \phi\to \phi+ \epsilon ( x^{-}+ x^{+}/\alpha )\partial_{-}\phi\;$   
generated by $M$. 

 In the limit  when $|\alpha| \to \infty$  we 
find $\phi \to {\phi_{R}}(x^{-})$ while $H^{lf}\to 0$,  which corresponds to 
the LF Hamiltonian of  free  massless scalar theory, Sec. (2.6). 
The field $\phi_{R}$ 
satisfies: $\; \left[{\phi_{R}}(x^{-}),{\phi_{R}}(y^{-})\right]= 
-i \epsilon(x^{-}-y^{-})/4$. The limiting case is thus seen  
to describe  a right (moving) chiral boson theory with  the Lagrangian density 
as given in (3.10).   

An alternative form of the Lagrangian density may also be employed in our
context.    
We recall that  in the  quantization of  gauge theory
    it is  found useful (Sec. 5) to  introduce  an auxiliary 
 field $B(x)$ of canonical mass dimension two (in $3+1$ dimensions) 
 and add $(B\partial_{\mu}A^{\mu}
 +\alpha B^{2}  )$ as the gauge-fixing term to the Lagrangian density. 
In the two dimensional theory under consideration 
 it is also  possible to follow 
this procedure,     
since the corresponding   $B(x)$  field here 
carries the canonical mass dimension one.  The 
discussion parallel to the one given above may thus   be  based also on  
  the following \cite{barcel, kim} Lagrangian density
\begin{equation}
 {\cal L}=\frac{1}{2} {\eta}^{\mu\nu} \partial_{\mu}\phi \partial_{\nu}\phi 
 +{\sqrt 2} B(x)(\partial_{-}\phi) + \frac{\alpha}{2} B(x)^{2}. 
\end{equation}
The elimination of the auxiliary field using its equation of motion leads to 
(3.10) and the conclusions reached are the same.

 We  make only brief comments on  other models. 
   Siegel's \cite{siegel} theory  which employs
\begin{equation}
{ \cal L}= \frac{1}{2}{\eta}^{\mu\nu} \partial_{\mu}\phi \partial_{\nu}\phi +
B(x) (\partial_{0}\phi-\partial_{1}\phi)^{2} 
\end{equation}
is afflicted by anomaly which is to be eliminated by the addition of 
a Wess-Zumino term. The resulting theory does not describe \cite{imb}
pure  chiral bosons since they are coupled to the gravity. 
 In this model the 
auxiliary field carries vanishing canonical dimension and, for 
example, a $B^{2}$ term cannot be added without introducing the 
dimensionful parameters.

The  model  based on the idea of implementing the 
chiral constraint through a linear constraint \cite{sriv, barcel},  
\begin{equation}
{\cal L}=\frac{1}{2} {\eta}^{\mu\nu} \partial_{\mu}\phi \partial_{\nu}\phi   
 +B_{\mu}({\eta}^{\mu\nu}-\epsilon^{\mu\nu})
\partial_{\nu}\phi,
\end{equation}
where $B_{\mu}$ is Lagrange multiplier field,    
does not seem to exhibit physical excitations \cite{rivel}. We note that the 
 field  $B_{\mu}$ carries dimension one and that this is the usual procedure 
in the classical theory which, however,  seems to  break down 
at the quantum level.

To summarize, the simple  procedure  of separating first the {\it condensate 
variable}, which in fact corresponds to a gauge-fixing condition  
needed on the phase space in the context of Dirac procedure, 
before applying the standard procedure itself,  is found to be successful 
 in describing \cite{fno3, pre} the SSB, the phase transition 
in two dimensional scalar theory, 
the SSB of continuous symmetry in $3+1$ dimensional theory, in furnishing 
a new proof of the Coleman's theorem, and 
in the description  of (the tree level) Higgs mechanism. 
It is also  found  successful in 
showing \cite{pre1, pre2} transparently and economically the vacuum structures 
in the SM and CSM models 
as will be reviewed in Sec. 5. The  self-duality constraint 
in the interacting theory leads to the well known {\it kink } solution in the 
{\it front form} theory as well.    
The     chiral boson theory discussion becomes   transparent and the 
 LF commutator 
 does not conflicts with the microcausality. A model  
   of the   chiral boson theory seems to emerge in the context of 
the {\it modified} FJ theory.  
   
We  will next review  the essentials of  
 the LF quantization of the Dirac and Maxwell   fields.

\section{ LF quantized  Dirac field}
\setcounter{equation}{0}
\renewcommand{\theequation}{4.\arabic{equation}}

\subsection{Anticommutators  }

On the LF there is a natural decomposition of the spinor space. 
The LF components \cite{notation} $\,\gamma^{\pm}$, where 
$\,\gamma^{\pm}=(\gamma^{0}\pm \gamma^{3})/{\sqrt {2}}\,$ have the properties  
$\,({\gamma^{+}})^{2}=({\gamma^{-}})^{2}=0$, $\,\gamma^{0}\gamma^{+}=
\gamma^{-}\gamma^{0}$, $\,{\gamma^{+}}^{\dag}={\gamma^{-}}$, and 
$\,\,\gamma^{+}\gamma^{-}+ 
\gamma^{-}\gamma^{+}=2 I\,$. We may thus introduce the hermitian projection 
operators $\Lambda^{\pm}$

\begin{equation}
\Lambda^{\pm}= {1\over 2} \gamma^{\mp}\gamma^{\pm}= {1\over \sqrt {2}}\gamma^{0}\gamma^{\pm}, 
\quad({\Lambda^{\pm}})^{2}={\Lambda^{\pm}}, 
\quad {\Lambda^{+}}\Lambda^{-}=\Lambda^{-}\Lambda^{+}=0,
\quad \gamma^{0}\Lambda^{+}=\Lambda^{-}\gamma^{0}  
\end{equation}
The corresponding $\,{\pm}\,$ projections of the LF Dirac spinor 
are $\psi_{\pm}=\Lambda^{\pm}\psi$ and  $\psi=\psi_{+}+\psi_{-}$,  
${\bar\psi} = \psi^{\dag}\gamma^{0}=$ 
$\bar\psi_{+}+\bar\psi_{-}$, $\gamma^{\pm}\psi_{\mp}=0$, $ {\Lambda^{\pm}}
\psi_{\pm}= \psi_{\pm}$ etc. The matrix 
$\Sigma_{3}= {\Sigma^{\dag}}_{3}=i\gamma^{1}\gamma^{2}$, 
$ {\Sigma_{3}}^{2}= I$, which commutes with 
$\Lambda^{\pm}$ plays an important role on the LF and   
 we note   :\quad   
$(\Lambda^{+}+\Lambda^{-})= I$,  
$\; (\Lambda^{+}-\Lambda^{-})=\Sigma_{3}\gamma_{5}$,  
$\;\gamma_{5}\psi= \Sigma_{3}(\psi_{+}-\psi_{-})$, and 
$\;\Sigma_{3}\gamma^{\perp}\Sigma_{3}= -\gamma^{\perp}$.

The action  of the free Dirac field is \cite{fno4}  
\begin{eqnarray}
S &=& \int d^{2}x^{\perp}dx^{-}\, {\cal L}\qquad\quad\qquad 
 \mbox{where} \nonumber \\
{\cal L}&=&\bar\psi(i\{\gamma^{+}\partial_{+}+\gamma^{-}\partial_{-}
+\gamma^{\perp}\partial_{\perp}\}-m)\psi
\nonumber \\
&=&i{\sqrt 2}{\psi^{\dag}}_{+}\partial_{+}\psi_{+}+
i{\sqrt 2}{\psi^{\dag}}_{-}\partial_{-}\psi_{-}\nonumber \\
&& -{\psi^{\dag}}_{-}(m+i\gamma^{\perp}\partial_{\perp})\gamma^{0}
\psi_{+}
-{\psi^{\dag}}_{+}(m+i\gamma^{\perp}\partial_{\perp})\gamma^{0}\psi_{-}.
\end{eqnarray}
It shows that only  the component $\psi_{+}$ carries kinetic term and the  
$\psi_{-} $ component is nondynamical.    
 The variation of the action with respect to ${\psi^{\dag}}_{-}$ 
 and $\psi_{-}$ leads to the constraint equation
\begin{eqnarray}
2 i \partial_{-}\psi_{-} &=& (m+i\gamma^{\perp}\partial_{\perp})
\gamma^{+}\psi_{+} 
\end{eqnarray}
 and its conjugate,  while  
 for the dynamical component $\psi_{+}$ we obtain the 
equation of motion
\begin{equation}
4\partial_{+}\psi_{+} = -  (m+ i \gamma^{\perp}\partial_{\perp}) 
\gamma^{-}{1\over{\partial_{-}}}
(m+ i \gamma^{\perp}\partial_{\perp}) 
\gamma^{+}  \psi_{+}, 
\end{equation}
after eliminating the dependent component $\psi_{-}$. Its  right hand side 
  may be simplified to $2 (-m^{2}+\partial^{\perp}\partial^{\perp})
(1/\partial_{-})\psi_{+}$. The canonical  Hamiltonian density is easily seen to be 
 $\,{{\cal H}_{c}}^{lf}= 
{\psi^{\dag}}_{+}(m+i\gamma^{\perp}\partial_{\perp})\gamma^{0}\psi_{-}$  
with $\psi_{-}$ being a dependent field given by the constraint equation 
above. It is straightforward to verify that the equation of motion 
for the dynamical component $\psi_{+}$ in the
quantized theory is recovered as an Heisenberg equation of motion if we 
postulate the following  anticommutation relations, which are {\it  
 local in all the spatial coordiantes}.    
  
 \begin{eqnarray}
 \{\psi_{+}(\tau, x^{-},x^{\perp}),{\psi_{+}}^{\dag}(\tau, y^{-},y^{\perp})\}
 &=&
 {1\over \sqrt {2}} \Lambda^{+} \delta(x^{-}-y^{-})
 \delta^{2}(x^{\perp}-y^{\perp}), \nonumber \\
\{\psi_{+}(\tau, x^{-},x^{\perp}),{\psi_{+}}(\tau, y^{-},y^{\perp})\}=0, 
\quad &&
\{\psi_{+}^{\dag}(\tau, x^{-},x^{\perp}),
{\psi_{+}}^{\dag}(\tau, y^{-},y^{\perp})\}=0.   
 \end{eqnarray}
The  same result is  also  derived if we follow  the straightforward 
Dirac procedure 
as in the case of the scalar theory. No first class constraint, however, 
arises in the present case. The scale dimension of $\psi_{+} $ is 
clearly $[\psi_{+}]=1/(L_{\perp}{\sqrt {L_{\|}}}) $. It follows from (4.3) 
that 
\begin{equation}
 \{\psi_{-}(\tau, x^{-},x^{\perp}),{\psi_{+}}^{\dag}(\tau, y^{-},y^{\perp})\}
 =
 \frac{1}{i4{\sqrt 2}}(m+ i \gamma^{\perp}\partial_{\perp}) 
\gamma^{+}    \epsilon(x^{-}-y^{-})
 \delta^{2}(x^{\perp}-y^{\perp})
\end{equation}

\subsection{LF Spinor in momentum space and its properties}

In order to write the Fourier transform we look for the 
      complete set of linearly independent plane wave solutions of the free 
 Dirac equation  in the {\it front form} 
 theory. For the massive field the signs of $p^{+}$ and   $p^{-}$ are 
 correlated. 
    Choosing, say,   $p^{+}>0$    
 the independent set of the plane wave solutions are 
 $ u(p) e^{-i p\cdot x}$ and 
 $ v(p) e^{i p\cdot x}$ where the  four-spinors $u(p)$ and $v(p)$ satisfy:  
  $(m- \gamma^{\mu}p_{\mu})u(p)=0\,$ and 
  $\,(m+ \gamma^{\mu}p_{\mu})v(p)=0$. We will make the phase convention 
  such that $v(p)=C\gamma^{0T} u(p)^{*}$, the charge conjugate of 
  $u(p)$.

  A very useful form \cite{pre1, pre} of the free {\it LF four-spinor} 
   is given by 
  \begin{equation}
  u^{(r)}(p)= N(p) \left[{\sqrt 2} p^{+}\Lambda^{+} 
  +(m+\gamma^{\perp}p_{\perp})\Lambda^{-}\right] \tilde u^{(r)},
\end{equation}
where the normalization is  chosen 
as $N(p)=1/{({\sqrt 2}\, p^{+}m)}^{1/2}$, with $m>0$ and $p^{+}>0$.    
 The  constant spinors $\tilde u^{(r)}$, which are also the 
 spinors in the rest frame  $\tilde p=(m/\sqrt 2, m/\sqrt 2,  
 0^{\perp})$,   satisfy 
$\;\gamma^{0}\tilde u^{(r)}=\tilde u^{(r)}$, 
    $ \;\Sigma_{3}
\tilde u^{(r)}= r \,\tilde u^{(r)}$ with  $\,r=\pm $. 
 The   charge conjugate rest frame    spinors 
satisfy     
$\gamma^{0}\tilde v^{r}=-\tilde v^{r}$ and  $ \Sigma_{3}
\tilde v^{(r)}= -r \tilde v^{(r)}$ while   
 \begin{equation}
  v^{(r)}(p)
  = N(p) \left[{\sqrt 2} p^{+}\Lambda^{+} +(m-\gamma^{\perp}
p_{{\perp}})\Lambda^{-}\right] \tilde v^{(r)}.  
  \end{equation} 
We note that $\gamma_{5} {u_{}}^{(r)}(p;m)= r\,  {u_{}}^{(r)}(p; -m)$ and  
$\gamma_{5} {v_{}}^{(r)}(p;m)= -r\,  {v_{}}^{(r)}(p; -m)$ 
indicating the mass reversal  property of $\gamma_{5}$ upto a 
phase factor. Also $\Sigma_{3} {u_{}}^{(r)}(p;m)= r\, 
{u_{}}^{(r)}(p^{+}, -p^{\perp};m)$ and  $\Sigma_{3} {v_{}}^{(r)}(p;m)= -r\, 
{v_{}}^{(r)}(p^{+}, -p^{\perp};m)$. We do {\it not} introduce two spinors and 
work only with four-spinors and do not also employ any explicit matrix
representation.  

We recall that the  LF {\it Spin operator} for the massive as well as  
massless particles is defined (Appendix B) by  
${\cal J}_{3}\equiv -W^{+}/P^{+}$ where $W^{\mu}$ is the Pauli-Lubanski four-vector. 
 It contains  solely  the  LF kinematical 
generators and         
the following useful identity can be demonstrated \cite{pre1, pre} 
  \begin{equation}
{\cal J}_{3}(p)= e^{\;(-\frac{i}{p^{+}}){(B_{1}p^{1}+B_{2}p^{2})}}\; J_{3}\; \; 
e^{(\frac{i}{p^{+}}){(B_{1}p^{1}+B_{2}p^{2})}}=J_{3}-\frac{1}{p^{+}}\,
(p^{1}B_{2}-p^{2}B_{1}) 
\end{equation}
where ${\sqrt 2}B_{1}=(K_{1}+J_{2})$ and  ${\sqrt 2}B_{2}=(K_{2}-J_{1})$ 
are the kinemetical boost operators on the LF 
in the standard notation. Applying it to the spin $1/2$ case\footnote{
For spin-$1/2$ case: \quad $J_{j}=\Sigma_{j}/2$, $\;K_{j}=
 i \gamma^{0}\gamma^{j}/2 $ where $ j=1,2,3$. }  
 we derive ($J_{3}=\Sigma_{3}/2$)    
\begin{eqnarray}
{\cal J}_{3}(p)&=&\frac{1}{2}\left[I+
\frac{(\gamma^{\perp}p_{\perp})\, \gamma^{+}}{p^{+}}
\right] \Sigma_{3} \nonumber \\
&=& J_{3}+ \frac{(\gamma^{\perp}p_{\perp})}{2 p^{+}} \gamma^{+}\gamma_{5}
\nonumber \\
{\cal J}_{3}(p)u^{(r)}(p)&=& (r/2) u^{(r)}(p)\nonumber \\
{\cal J}_{3}(p)v^{(r)}(p)&=& -(r/2) v^{(r)}(p
\end{eqnarray}
where  $r/2=\pm(1/2) $ are the projections of $\vec {\cal J}(p)$ on the 
3-axis in the rest frame and we used $i(\gamma^{2}p^{1}- \gamma^{1}p^{2}) 
=(\gamma^{\perp}p_{\perp})\Sigma_{3}$.  
The four-spinors are shown to satisfy 
the following orthogonality  relations:
\begin{equation}
 \bar u^{(r)}(p)u^{(s)}(p)=\delta_{rs}, \qquad \bar v^{(r)}(p)v^{(s)}(p)=-\delta_{rs}, 
\qquad \bar u^{(r)}(p)v^{(s)}(p)= 0. 
\end{equation}
and the  following completeness relations follow easily 
\begin{equation}
\sum_{r=+,-} u^{(r)}(p)\bar u^{(r)}(p)= {(\not p+m)\over {2m}}, \quad 
 \quad \sum_{r=+,-} v^{(r)}(p)\bar v^{(r)}(p)= {(\not p-m)\over {2m}}
\end{equation}
where ${\not p}= \gamma^{\mu}p_{\mu}$.  
We also have  the useful  relations:\quad  $m \bar u^{(r)}(p)
\gamma^{\mu}u^{(s)}(p)=$
$p^{\mu} \bar u^{(r)}(p)u^{(s)}(p)\,$ and 
$m \bar v^{(r)}(p)\gamma^{\mu}v^{(s)}(p)=$
$-p^{\mu} \bar v^{(r)}(p)v^{(s)}(p)$.

\subsection{Fermion propagator }

The  Fourier transform expansion of $\psi(x)$   
 over the complete set of linearly independent plane wave solutions 
 constructed above may be   written as 
\begin {equation}
\psi(x)={1\over {\sqrt {(2\pi)^{3}}}}
              \sum_{r={\pm}}\int d^{2}p^{\perp}
	      dp^{+}\;
	      \theta(p^{+}){\sqrt{m\over p^{+}}}\left[b^{(r)}(p){ u^{(r)}}(p)
	       e^{-ip.x}+
          {d^{\dag (r)}}(p){ v^{(r)}}(p) e^{ip.x}\right]
\end {equation}   
where the $\theta(p^{+})$ is necessarily present.  
For the dynamical component   
$\psi_{+}\equiv \Lambda^{+}\psi$, it follows that   
\begin {equation}
\psi_{+}(x)=\frac{\sqrt{\sqrt 2}} {\sqrt {(2\pi)^{3}}}
              \sum_{r={\pm}}\int d^{2}p^{\perp}
	      dp^{+}\;
	      {\theta(p^{+})}\left[b^{(r)}(p){{\tilde u}_{+}}^{(r)} 
	      e^{-ip\cdot x}+
          d^{\dag {(r)}}(p){{\tilde v}_{+}}^{(r)} e^{ip\cdot x}\right]. 
	  \end{equation}
It is straightforward  to 
verify that the anticommutation relations (4.5)    for the
independent field operator $\psi_{+}$ are in fact satisfied  if we assume the 
standard  canonical anticommutators, with the 
 nonvanishing  ones given by:  
 $\{b^{(r)}(p),{b^{\dag}}^{(s)}(p')\} = \delta_{rs}\delta^{2}(p^{+}-p'^{+}) 
\delta(p^{\perp}-p'^{\perp})$ and 
  $\{d^{(r)}(p),{d^{\dag}}^{(s)}(p')\} = \delta_{rs}\delta(p^{+}-p'^{+})
\delta^{2}(p^{\perp}-p'^{\perp})$. 
 
The $\Lambda^{+} $ projections of our LF spinors are by construction 
very simple,   
$\,{u^{(r)}}_{+}(p)=({\sqrt {2}}p^{+}/m)^{1/2}\Lambda^{+}{\tilde u^{(r)}}$; 
they 
are eigenstates of $\Sigma_{3}$ as well. This is very convenient since 
 on the LF $\psi_{+}$ component is  the      
independent dynamical degrees of freedom while $\psi_{-}$ 
may be eliminated, even in the interacting theory,  making use of the 
 constraint equation. The simplified structure of $\psi_{+}$ gives rise to 
appreciable simplifications  in the context of  LF perturbation theory,   
compensating to some 
extent for the the nonlinearity of  the interaction found along  the 
longituidinal 
direction $x^{-}$.  We have better control \cite{pre3}, say, 
over recovering the 
manifest rotational  and even Lorentz  covariance in the 
perturbation theory calculations if we use the LF four-spinor introduced above.     
 The propagator for the spinor field $\psi_{+}$ also    
  takes a very simple causal form on the LF, 
resembling the one  of the scalar field.

The free propagator for the independent component $\psi_{+}$ in 
momentum space is easily derived using the above Fourier transform   
\begin{eqnarray}
\lefteqn{\left\langle 0\vert \,T(\psi_{+ A}(x)\psi^\dag_{+ B}(0))\, \vert 0
\right\rangle =}\nonumber \\ 
&& 
\left\langle 0\vert \left[\theta(\tau){\psi_{+ A}(x)}\psi^\dag_{+ B}(0)
-\theta(-\tau){\psi}^{\dag}_{+ B}(0){\psi}_{+ A}(x)\right]\vert 0\right\rangle  
\nonumber \\[1ex]
&=& 
\frac{1}{\sqrt 2}\frac{\Lambda^+_{AB}}{(2\pi)^3}\int 
d^2q^{\perp} dq^{+}\theta(q^{+})\left[\theta(\tau) e^{-iqx}- 
\theta({-\tau}) e^{iqx}\right]
\end{eqnarray}
where $A,B=1,2,3,4\,$ label the spinor components.   
The only relevant 
differences, compared with the case of the scalar field, are,  
 apart from the appearance  of the projection operator,  
  the absence of  the factor  
 $(1/2q^{+})$ in the integrand,   and the 
 negative sign of  the second term in  the fermionic case. 
 They, however,   compensate and the standard manipulations to factor out 
 the exponential give rise to the factor 
 $\left[\theta(q^+)+\theta(-q^+)\right]$  which may be interpreted as  
 unity in 
 the distribution theory sense,   parallel to what we find 
 in the derivation of the   scalar field propagator on the LF.    
Hence 
\begin{equation}
<0|T({\psi^{}}_{+}(x){\psi^{\dag }}_{+}(0))|0>  =   
\;{{i}\over {(2\pi)^{4}}} \int d^{4}q \;{{{\sqrt {2}}q^{+}\, 
\Lambda^{+} }\over 
{(q^2-m^{2}+i\epsilon)}}\, e^{-iq.x}. 
\end{equation}
It may 
  also be derived  by functional integral method; we do have to 
 take care of the second class constraint in the measure.  
The  fermionic propagator  here contains   no   instantaneous  
 term usually encountered when doing the  old fashioned perturbation 
 theory  and the integrand factor  may also be expressed as 
$\approx\,\left [\Lambda^{+} (\not q +m)\Lambda^{-}/
{(q^2-m^{2}+i\epsilon)}\right]\gamma^{0} $. 
We verify   that the propagator   satisfies  the equation for the 
Green's function corresponding to the equation of motion 
of $\psi_{+}$.

The 
momentum space representations of the currents and the 
components of the energy-momentum tensor are derived straightforwardly and 
they support the usual interpretation of $b^{\dag (r)}(p)b^{(r)}(p)$ 
and $d^{\dag (r)}(p)d^{(r)}(p)$ as the number operators. 
For example, for the canonical Hamiltonian we find 
\begin{eqnarray}
{H^{lf}}_{c}&=& \frac {1}{\sqrt 2}\int d^{2}x^{\perp}d x^{-}
: {\psi^{\dag}}_{+}(m^{2}-\partial_{\perp}\partial_{\perp})\frac{1}{i\partial_{-}}
\psi_{+} :\nonumber \\
&=& 
\sum_{r,s}\int d^{3}p d^{3}k \theta(p^{+})\theta(k^{+})
: \left[b^{\dag (r)}(p)b^{(s)}(k){{\tilde u}_{+}}^{\dag {(r)}}
{{\tilde u}_{+}}^{(s)} \right. \nonumber \\
& & \left. -d^{(r)}(p)d^{\dag (s)}(k){{\tilde v}_{+}}^{\dag {(r)}}{{\tilde v}_{+}}^{(s)}
\right] : \frac {(m^{2}+p^{\perp}p^{\perp})}{2 p^{+}}\delta^{3}(p-k)
\nonumber \\
&=&
\sum_{r}\int d^{3}p \theta(p^{+})
\left[b^{\dag (r)}(p)b^{(r)}(p)
+d^{\dag (r)}(p)d^{(r)}(p)\right]  \frac {(m^{2}+p^{\perp}p^{\perp})}{2 p^{+}}
\end{eqnarray}
where we use ${{\tilde u}_{+}}^{\dag {(r)}}{{\tilde u}_{+}}^{(s)}
={{\tilde v}_{+}}^{\dag{(r)}}{{\tilde v}_{+}}^{(s)}= \delta_{rs}/2$,  
$d^{3}p\equiv  d^{2}p^{\perp} dp^{+}$, and  $:\;\, :$
indicates the normal ordering.
  
\subsection{$\Gamma_{5}$ Symmetry. Chirality transformation on the LF }

The $\gamma_{5}$ transformation \cite{tiomno}, 
$\psi\to \gamma_{5}\psi$ 
on the spinor field is associated with the mass reversal in the Dirac 
equation. It leaves the Dirac equation form invariant only when 
the mass is vanishing. On the LF we can construct a generalized 
$\Gamma_{5}$ transformation which restores the form invariance  
even for the massive field. 

Consider the  covariant vector and axial current densities 
defined   by 
$j^{\mu}= \bar\psi \gamma^{\mu} \psi\,$ and 
 $\,{j_{5}}^{\mu}= \bar\psi \gamma^{\mu}\gamma_{5} \psi$ respectively. 
The corresponding charge densities are defined on the LF 
by the $+$ components of the currents  
\begin{eqnarray}
j^{+}&=& \; \bar\psi \gamma^{+} \psi ={\sqrt 2}{\psi_{+}}^{\dag} \psi_{+}
\nonumber \\ 
{j_{5}}^{+}&=&\; \bar\psi \gamma^{+} \gamma_{5}\psi ={\sqrt 2}{\psi_{+}}^{\dag} 
\Sigma_{3}\psi_{+}. 
\end{eqnarray} 
The momentum space representations of the charges are easily derived 
\begin{eqnarray}
Q&=&\!\int d^{2}x^{\perp}dx^{-} :j^{+}:=
\sum_{r}\!\int\! d^{3}p \theta(p^{+})
\left[b^{\dag (r)}(p)b^{(r)}(p)-d^{\dag (r)}(p)d^{(r)}(p)\right] 
\nonumber \\
{Q}_{5}&=&\!\! 
 \int d^{2}x^{\perp}dx^{-} :{j_{5}}^{+}:=
 \sum_{r}\!\int\! d^{3}p \theta(p^{+})\,(r)
\left[b^{\dag (r)}(p)b^{(r)}(p)+d^{\dag (r)}(p)d^{(r)}(p)\right]
\end{eqnarray}
The charges $Q$ and $Q_{5}$ commute with  the LF Hamiltonian and are 
thus  constants of motion. The former 
 counts the fermionic number  while the latter the twice the 
projection along the 3-axis of the LF spin operator ${\cal J}_{3}(p)$ discussed
above.   

From the commutation relations of the field $\psi_{+}$ we derive \cite{ken}
 \begin{eqnarray}
\{\psi_{+}, {Q}\}&=&  \psi_{+}, \nonumber \\
\{\psi_{-}, {Q}\}&=&  \psi_{-}, \nonumber \\
\{\psi_{+}, {Q}_{5}\}&=& \gamma_{5} \psi_{+}=
\Lambda^{+}\gamma_{5} \psi_{+}, \nonumber \\
\{\psi_{-}, {Q}_{5}\}&=& \Lambda^{-}
\frac{1}{2i\partial_{-}}(i\gamma^{\perp}\partial_{\perp}
+m)\gamma^{+} (\gamma_{5}\psi_{+}) \neq \gamma_{5} \psi_{-}. 
\end{eqnarray}
The  action of  the infinitesimal generators on $\psi$ is  
\begin{eqnarray}
\delta_{Q}\psi=\{\psi, i\epsilon {Q}\}&=& i\epsilon\, \psi, \nonumber \\
\delta_{Q_{5}}\psi =\{\psi, i\epsilon {Q_{5}}\}&=& i\epsilon\,\gamma_{5}
\left[I-\frac{m}{i\partial_{-}}\gamma^{+}\right] \psi, 
\end{eqnarray}
where we use (4.3) and (4.4). It is well known that  
the infinitesimal transformation with respect to 
$Q$  is associated with  the {\it form invariance} of the 
Dirac equation 
$(i\gamma^{\mu}\partial_{\mu}-m)\psi=0$ and its conjugate 
under the global phase 
transformations. This symmetry gives rise to the 
   on shell conserved Noether vector current $j^{\mu}$. 

The Dirac equation  is  form invariant under
the  $\gamma_{5}$ (or  chiral transformations)  only for the massless
 theory, when  the axial current is also conserved at the classical level. 
Our discussion on the LF in the Hamiltonian formulation indicates  that 
the   Dirac equation   
is also form invaraint  under the following nonlocal  $\Gamma_{5}$  transformation, 
defined  by  ${\Gamma}_{5}$ 
\begin{eqnarray}
\psi & & \to \; \;\Gamma_{5}\psi, \nonumber \\
\Gamma_{5} &=& \gamma_{5}\,\left[I-\frac{m}{i\partial_{-}}\gamma^{+}\right].  
\end{eqnarray}
This can be demonstrated, say,  if we use   of the (on shell)   identity
\begin{equation}
(i\gamma^{\mu}\partial_{\mu}-m)\gamma_{5}
\left[I-\frac{m}{i\partial_{-}}\gamma^{+}\right]= - \gamma_{5}
\left[I+\frac{m}{i\partial_{-}}\gamma^{+}\right]
(i\gamma^{\mu}\partial_{\mu}-m).
\end{equation}
The on shell conserved  current associated with   the 
  $\Gamma_{5}$ symmetry, which holds for both the massive and massless 
fermions, 
 is hence given by   
{
\begin{eqnarray}
{J_{5}}^{\mu}&= &\bar\psi \gamma^{\mu}\Gamma_{5}\psi
={j_{5}}^{\mu}
-m\bar\psi\gamma^{\mu}\gamma_{5}\gamma^{+}\frac{1}{i\partial_{-}}\psi,
\nonumber \\
\partial_{\mu}J^{\mu} &\stackrel{o}{=} &0,  \nonumber \\
{J_{5}}^{+}&=&{j_{5}}^{+}.
\end{eqnarray}
 The chiral charge associated with  the  $\Gamma_{5}$ symmetry  
 coincides with   $Q_{5}$ and the generalized chiral  transformation is 
$\;\psi\to e^{i\alpha \Gamma_{5}} \psi $. 
 
\subsection{Helicity Operator, LF Majorana and Weyl fermions    }

The Fourier transform of the self-charge conjugate 
{\it Majorana spinor field} satisfying,  $\psi_{M}(x)=\psi_{M {c}}(x)$,  
follows easily from (4.13)
\begin{eqnarray}
\psi_{M}(x)&=& \frac{1}{\sqrt 2}(\psi(x)+\psi_{c})\nonumber \\
&=&{1\over {\sqrt {(2\pi)^{3}}}}
              \sum_{r={\pm}}\int d^{2}p^{\perp}
	      dp^{+}\;
	      \theta(p^{+}){\sqrt{m\over p^{+}}}\left[{b_{M}}^{(r)}(p){ u^{(r)}}(p)
	       e^{-ip.x} \right. \nonumber \\
&&    +\left. {{b_{M}}^{\dag (r)}}(p){ v^{(r)}}(p) e^{ip.x}\right]
\end{eqnarray}
where ${b_{M}}^{(r)}(p)=({b}^{(r)}(p)+{d}^{(r)}(p))/{\sqrt 2}$ and the
nonvanishing anti-commutator is given by 
$\{{b_{M}}^{(r)}(p), {{b^{\dag}}_{M}}^{(s)}(k)\}=\delta^{rs}\delta^{3}(p-k)$. 

The {\it  chiral } or  $\gamma_{5}$-{\it projections } of the LF spinor   
  are shown to satisfy the following properties ($\, r\,\gamma_{5}
u^{(r)}(p; m)= u^{(r)}(p; -m) $ and $\, -r\,\gamma_{5}
v^{(r)}(p; m)= v^{(r)}(p; -m) $)

\begin{eqnarray}
 \frac {(I+ r \gamma_{5})}{2}u^{(r)}(p)&=& 
 N(p) \left[{\sqrt 2} p^{+}\Lambda^{+} 
  +(\gamma^{\perp}p_{\perp})\Lambda^{-}\right] \tilde u^{(r)}
  \nonumber \\
 \frac {(I- r \gamma_{5})}{2}u^{(r)}(p)&=& 
 N(p) \;  m\,\Lambda^{-} \tilde u^{(r)} \to 0 \mbox{\qquad for $m\to 0 $}
 \nonumber \\
 \frac {(I- r \gamma_{5})}{2}v^{(r)}(p)&=& 
 N(p) \left[{\sqrt 2} p^{+}\Lambda^{+} 
  -(\gamma^{\perp}p_{\perp})\Lambda^{-}\right] \tilde v^{(r)}
  \nonumber \\
 \frac {(I+ r \gamma_{5})}{2}v^{(r)}(p)&=& 
 N(p) \;  m\,\Lambda^{-} \tilde v^{(r)} \to 0 \mbox{\qquad for $m\to 0 $}
   \end{eqnarray} 
 along with 
 \begin{eqnarray}  
\gamma^{\mu}p_{\mu}\quad \left[\frac {(I+ r \gamma_{5})}{2}u^{(r)}(p)\right]&=& 
 N(p) \;  m^{2}\,\Lambda^{-} \tilde u^{(r)} \to 0 \mbox{\qquad for $m\to 0 $}
 \nonumber \\
 \gamma_{5}\quad \left[\frac {(I+ r \gamma_{5})}{2}u^{(r)}(p)\right]&=&\, r\;
 \left[\frac {(I+ r \gamma_{5})}{2}u^{(r)}(p)\right]
 \end{eqnarray}
etc.,  and we note that $\left[{\cal J}_{3}(p), \gamma_{5}\right]=0$. 
In the {\it massless limit}, $\, m\to 0\,$,  the projections 
$  {(I \mp \gamma_{5})}u^{(\pm)}(p)$,  
$ \, {(I \pm \gamma_{5})}v^{(\pm)}(p)\,$vanish. Also, for example, 
  the nonvanishing  one 
\begin{equation}
\frac {(I+  \gamma_{5})}{2}u^{(+)}(p)
\end{equation}
is an eigenstate of $\,\gamma_{5}$ and  ${\cal J}_{3}(p)$  with the eigenvalues 
$1$ and $1/2$ respectively, while the other one 
\begin{equation}
\frac {(I-  \gamma_{5})}{2}u^{(-)}(p)
\end{equation}
has the corresponding eigenvalues given by $-1$ and $-1/2$. 
The explicit discussion here shows  that on 
the LF the definition of the spin operator for the massive 
and massless cases gets unified. 

The {\it Helicity  operator } $\;{\hat {h}}\;$ is defined  by 
\begin{equation}
{\hat{ h}}\; =\;{\frac{1}{2}} \frac{{\vec \Sigma}\cdot \hat{\vec {P}}} 
{\vert\hat{\vec {P}}\vert}  \; 
= \frac{\left[ \Sigma_{3} {\hat {P}}^{3}
     +\gamma^{\perp}{\hat {P}}_{\perp}\gamma^{0}\gamma_{5}\right] }	
     {2 \vert\hat {\vec P }\vert}	 
\end{equation}
which  is {\it not} the same as the LF spin operator.

For {\it massless fermions } it  is easily shown that 
\begin{eqnarray}
\hat{h}(p)\; u^{(r)}(p) & = &  (\frac {r}{2})\; u^{(r)}(p) \nonumber\\
\hat{h}(p)\ v^{(r)}(p) & = & - (\frac {r}{2})\; v^{(r)}(p)
\end{eqnarray}
Experimental observations show that only the negative chirality, 
$ {(I-\gamma_{5})}u^{(-)}(p)/{2}$, 
 neutrinos exist. Neutrinos have helicity $-1/2$, antineutrinos 
helicity $1/2$. There is no charge conjugation invariance if neutrinos have 
a definite chirality. The CP transformations of these spinors can be discussed 
as usual (Appendix B). The normalization factor in the massless case has to be
redefined. The massive particle  does not have Lorentz invariant helicity; in the rest
frame of the particle there is no preffered direction in what to measure spin.   
          
\subsection{Bilocal operators} 

From the anticommutators in Sec. 4.1 we may derive the (free theory) 
equal-$\tau$ current commutation relations, for example, $\left[j^{+}(x),
 j^{+}(y)\right]_{\tau}=0$. The commutators among the other components 
 are  derived straightforwardly.  They  involve 
 bilocal operators \cite{jacc} of the form $\bar\psi(x) \Gamma \psi(y)$, 
  with the nonlocality {\it only} along the longitudinal direction. 
   In the context of the deep 
 inelastic scattering limit they  are found relevant in the hadron  
 tensor $W^{\mu\nu}$ and the explanation of the Bjorken scaling and the 
 introduction of the parton model of Feynman.   
 Similar bilocal operators 
 appear also in bosonic theories,  for example, in the LF 
 quantization \cite {prech} of  Chern-Simons systems. We recall (Sec. 1) 
 that on the LF  nonlocality in the $x^{-}$ direction does not 
 conflict with the 
microcausality principle. The bilocals have also been shown useful recently, 
for example,   in the context \cite {dhar}
of the dynamics of hadrons in two dimensions and in revealing the string 
 like structure in $QCD_{2}$. 

\section{LF quantization of Gauge theory}
\setcounter{equation}{0}
\renewcommand{\theequation}{5.\arabic{equation}}

In perturbative QCD we employ, in the interaction representation,  
the free  abelian gauge theory propagator.  It is customary on the LF to 
adopt  the light-cone gauge\footnote{See the discussion below on
the LF quantized two dimensional SM where this gauge 
is not convenient to employ if  we are seeking for nonperturbative effects in
the theory. }
 $A_{-}=0$ which results in 
a simplified interaction 
Hamiltonian.   The noncovariant gauge, however, introduces in the theory 
undesireable features. The rotational invariance becomes very difficult to 
track down making  the comparasion with the conventional theory results
sometimes extremely difficult. 
In the frequently employed  old fashioned perturbation theory computations      
it is sometimes not easy to see  
if the conventional and the {\it front form} theories are really in 
agreement \cite{broo}. The LF quantized QCD  was recently 
studied \cite{pre3} in covariant gauges  in the context of the 
Dyson-Wick perturbation theory expansion based on the LF-time ordered 
Wick products. Here all the relevant propagators become causal and the 
rotational invariance is easily recovered, when the 
 LF spinor (4.7) introduced in the Sec. 4 is employed.  
 The loop integrals
can also be converted \cite {pre4} to the Euclidean space integrals 
and  the dimensional regularization may be used.

   The  Lagrangian density for the Abelian gauge theory  
written in LF coordinates is  
\begin{equation}
{1\over 2}\left[(F_{+-})^{2}-(F_{12})^2 +2F_{+\perp}
F_{-\perp}\right]+B(\partial_{+}A_{-}+\partial_{-}A_{+}+\partial_{\perp}A^{\perp})
+{\xi\over2}B^{2},
\end{equation}
where   
$F_{\mu\nu}\equiv (\partial_{\mu}A_{\nu}-\partial_{\nu}A_{\mu})$. The covariant
gauge-fixing is introduced by adding to  the  Lagrangian the linear gauge-fixing
term $B\partial_{\mu}A^{\mu}+(\xi/2) B^{2}\, $ where $B$ is the
Nakanishi-Lautrup auxiliary field and $\xi$ is a parameter.  
The canonical momenta 
are  $\pi^{+}=0$, $\pi_{B}=0$, 
$\pi^{\perp}=F_{-\perp}$, $\pi^{-}=F_{+-}+B$ and the canonical 
 Hamiltonian density is found to be 
\begin{equation}
{\cal H}_{c}= {1\over 2} ({\pi}^{-})^{2}+{1\over2}(F_{12})^{2}-
A_{+}(\partial_{-}\pi^{-}+\partial_{\perp}\pi^{\perp}-2\partial_{-}B)
-B(\pi^{-}+\partial_{\perp}A^{\perp})+{1\over2}(1-\xi)B^2
\end{equation}
Following  the Dirac procedure,  
the primary constraints  are $\pi^{+}\approx 0$, 
$\,\pi_{B}\approx 0$ and $\,\eta\equiv\pi^{\perp}-\partial_{-}A_{\perp}+
\partial_{\perp}A_{-}\approx 0$, where $\perp=1,2$ and $\,\approx\,$ 
stands for {\it weak equality} relation. We now require 
the persistency in $\tau$ of these constraints employing  the 
preliminary Hamiltonian, which is obtained by adding to the canonical 
Hamiltonian the primary constraints multiplied by the Lagrange multiplier 
fields. We assume the standard Poisson brackets for the dynamical variables 
in the  computation for  obtaining the Hamilton's equations of motion. 
We are  led to the following two  
secondary constraints 
\begin{eqnarray}
\Phi\equiv \partial_{-}\pi^{-}+\partial_{\perp}\pi^{\perp}-2\partial_{-}B 
& \approx & 0,  \nonumber \\
\Psi\equiv \pi^{-}+ 2 \partial_{-}A_{+}+\partial_{\perp}A^{\perp}
-(1-\xi)B & \approx & 0. 
\end{eqnarray}
The Hamiltonian is next enlarged by  including  these additional 
constraints as well.   The procedure is repeated.  
No more constraints are seen to arise.  
we now go over from the standard Poisson brackets to the  
  Dirac brackets,  such that inside 
 them we are  able to substitute 
the above constraints as  {\it strong} equality.  
The equal-$\tau$ Dirac bracket $\{f(x),g(y)\}_{D}$ which carries this property 
is constructed straightforwardly. 
 Hamilton's equations now employ the Dirac brackets and    
the   phase space   
constraints $\pi^{+}= 0$, 
$\pi_{B}= 0$, $\eta= 0$, $
\Phi=0$, and $\Psi= 0$ then  effectively reduce the (extended)
Hamiltonian. In the covariant {\it Feynman gauge} with $\xi=1$ 
the free Hamiltonian takes the simple form 
\begin{equation}
{H_{0}}^{LF} = 
-{1\over2}\int {d^{2}x^{\perp}}dx^{-}\; {g}^{\mu\nu} 
A_{\mu}\,\partial^{\perp}\partial_{\perp}\,A_{\nu}.  
\end{equation}
 The theory is canonically quantized 
through the correspondence $i\{f(x),g(y)\}_{D} \to 
\left[f(x),g(y)\right]$,  the  commutator among the 
corresponding operators.

 The equal-$\tau$ commutators of the gauge field are found to be 
\begin{equation}
\left[A_{\mu}(x),A_{\nu}(y)\right]_{x^{+}=y^{+}=\tau}
=-ig_{\mu\nu} K(x,y)
\end{equation}
 where $K(x,y)=-(1/4)\epsilon(x^{-}-y^{-})\delta^{2}(x^{\perp}-y^{\perp})$ is 
 nonlocal in the longitudinal coordinate.    
The transverse components of the gauge field have the physical LF 
commutators $ \left[A_{\perp}(x),A_{{\perp}'}(y)\right]_{\tau}
= i \delta_{\perp,\perp'}\,K(x,y)\;$,
while for the $\pm $ components we have only the  mixed 
 commutator nonvanishing $\left[A_{+}(x),A_{-}(y)\right]_{\tau}
 =- i K(x,y)$,  it has a  negative  sign 
 which   indicates  the presence of unphysical 
degrees of freedom in Feynman  gauge.   For        
  $\xi\neq 1$ 
the commutator, for example, 
of $A_{\pm}$ with $A_{\perp}$ is  found  to be nonvanishing. We note that 
the dimension of the gauge field is $[A_{\mu}]=1/L_{\perp}$.   

From the discussion analogous to that given in Sec. (2.6) for the scalar field 
it is clear, from the primary constraints, e.g., $\chi^{\perp}\approx 0$, 
 in the discussion  here, that there are 
also first class constraints present in the gauge theory. They may be taken 
care of like in the case of the scalar theory. In the context of perturbation theory 
we may possibly ignore the zero-longitudinal-mode of the components of the gauge field. 
However, when dealing with nonperturbative effects they may not be ignored. 
For example, in  the discussion of the (nonperturbative) vacuum structure of the 
completely soluble $QED_{2}$ (SM ) theory the  zero-momentum mode of $A_{-}$ plays a 
crucial role together  with the bosonic condensate variable (Sec. 6).

The Heisenberg
equations of motion lead to 
${\Box A_{\mu}}=0$  
for all the components, and consequently  the Fourier transform  
of the free gauge field
over the complete set of plane wave solutions   takes 
 the following form on the LF 
\begin{equation}
A^{\mu}(x)={1\over {\sqrt {(2\pi)^{3}}}}
\int d^{2}k^{\perp}dk^{+}\,
{\theta(k^{+})\over {\sqrt {2k^{+}}}}\,  
e^{\mu (\lambda)}(k)\left[a_{(\lambda)}(k^{+},k^{\perp})
 e^{-ik.x}
+a^{\dag}_{(\lambda)}(k^{+},k^{\perp})
 e^{ik.x} \right ] 
\end{equation}
where  $e^{\mu (\lambda)}(k)$, $\lambda=-,+,1,2\,$ label  
the  set of four linearly independent 
polarization four-vectors. 

In the {\it front form} theory 
 the two transverse (physical) polarization vector
are space-like as usual while\footnote{ $e^{(-)}(k)$ is called the dual  
of $e^{(+)}(k)$. Such a pair of null vectors is employed also in the well known 
ML prescription in the light-cone gauge and in the context of CNPA \cite{mit2, carbonel}.}
{\it the other two  are null four-vectors}.  
For a fixed  $k^{\mu}=(k^{0},{\vec k})$, where $k^{0}=|{\vec k}|$, 
we may construct them as follows
\begin{equation}  
e^{(+)}=(1,{\vec k}/k^{0})/{\sqrt 2}, \quad 
e^{(-)}=(1,-{\vec k}/k^{0})/{\sqrt 2}, \quad 
e^{(1)}=(0, {\vec \epsilon }(k;1)), \quad 
e^{(2)}=(0, {\vec \epsilon }(k;2)). 
\end{equation}
 Here $(0,1,2,3)$ components are 
specified for convenience while  ${\vec \epsilon }(k;1)$, 
${\vec \epsilon }(k;2)$ and ${\vec k}/|{\vec k}|$ constitute the 
usual orthonormal set of 3-vectors with the associated completeness 
relation.  
The polarization vectors are orthonormal: 
$g_{\mu\nu}e^{(\lambda){\mu}}(k)e^{(\sigma){\nu}}(k)= 
g^{\lambda\sigma}$ and  satisfy the completeness relation: 
$g_{\lambda\sigma}{e^{(\lambda)}}_{\mu}(k){e^{(\sigma)}}_{\nu}(k)
=  g_{\mu\nu}$. 

The field commutation relations for the gauge field found above   
are  shown to be satisfied  if we assume, parallel to the 
discussion in the fermionic case, the canonical commutation relations:      
$\,\left[a_{(\lambda)}(k^{+},k^{\perp}),{a^{\dag}}_{(\sigma)}(k'^{+},k'^{\perp})
\right]$ $=-g_{\lambda\sigma}$
$\delta(k^{+}-k'^{+})$ $\delta^{2}(k^{\perp}-k'^{\perp})$. 
We note that   the operators $a_{(0)}=(a_{(+)}+a_{(-)})/{\sqrt 2}$ and 
$a_{(3)}=(a_{(+)}-a_{(-)})/{\sqrt 2}$ obey 
the usual canonical  commutation relations   except that  in the 
case of $a_{(0)}$  a negative 
sign is obtained. The discussion of the 
  Gupta-Bleuler consistency condition  then becomes parallel 
to that in the 
 conventional equal-time treatment  of the theory.
   
 The Feynman gauge free gauge field propagator  on the LF 
 can be derived straightforwardly   
\begin{equation}
 <0| T({A^{}}_{\mu}(x){A^{}}_{\nu}(0))|0>= {{{i}}
 \over {(2\pi)^{4}}} 
 \int d^{4}k \;e^{-ik.x}\; 
 {-g_{\mu\nu}\over {k^{2}+i\epsilon}}
 \end{equation}
The momentum space representations of the components of the energy-momentum 
tensor are straightforward to derive as in the fermionic case. 
 The canonical Hamiltonian, 
for example,  gets  contributions  from the physical transversly polarized 
photons  as well as from the longitudinally polarized ones. The 
Gupta-Bleuler  consistency condition is required \cite{pre3} to be 
imposed in order to define the physical Hilbert space.

The computations done \cite{pre3}, 
employing the covariant gauge on the LF,  for the 
electron self-energy, electron-muon scattering, and the 
Compton scattering demonstrate complete 
 agreement with the results known in the conventional equal-time 
theory.  We find that on the LF the tree level {\it seagull} 
term dominates the (classical) Thomas formula for the scattering 
at vanishingly small photon energies. It is suggestive that on the LF 
the (conventional theory) semi-classical approximation may  reveal itself 
already  at the tree level (after having  removed  the constraints). 
  We will consider the  
LF quantized QCD after  the study  in the {\it front 
form } theory of the  
 nonperturbative vacuum 
structures in some two dimensional completely solvable gauge theories.

\section{Vacuum Structures in Schwinger and Chiral Schwinger Models}
\label{bosmod}
\setcounter{equation}{0}
\renewcommand{\theequation}{6.\arabic{equation}}

It is pertinent   to study two dimensional gauge theories on the LF.  
The  models like SM and CSM  can be solved completely. They may give 
 clues, for example, on the accessibility or not, 
 in the fully interacting theory,  
of certain gauge-fixing condition, found practical  in the context of 
perturbation theory.  The study \cite{pre1} of the SM, for example, 
shows that the light-cone gauge, $A_{-}=0$, is {\it not} convenient on the LF; it
would subtract out  the gauge invariant information from the theory itself,   
which is needed for describing the nonperturbative vacuum structure in the 
theory. 
 
The models mentioned above are known to have  non-trivial 
vacuum structure, a non-perturbative effect, 
 from the studies \cite {abda} in the conventional framework. 
Their study would indicate as to how to look for such and other 
nonperturbative effects in the LF quantized QCD in $3+1$ dimensions.  

The  massless $QED_{2}$ or SM is describe by   
\begin{equation}
{\cal L}=\bar\psi\,i\gamma^{\mu}\partial_{\mu}\psi-{1\over 4}F^{\mu\nu}
F_{\mu\nu}-e\bar\psi\,\gamma^{\mu}\psi A_{\mu}.
\end{equation} 
Its   exact solvability \cite {sch} 
derives from the remarkable 
property of one-dimensional fermion systems, viz, that they can 
equivalently be described in terms of canonical one-dimensional boson
fields.   
 Some of the 
correspondences in the abelian bosonization 
are \quad  $\bar\psi\psi=K :\cos 2\sqrt{\pi}\,
\phi:,\,\bar\psi\gamma_{5}\psi=K :\sin 2\sqrt{\pi}\,\phi:,\,
\bar\psi\gamma_{5}\gamma_{\mu}\psi=\partial_{\mu}\phi/\sqrt{\pi},\,
\bar\psi\gamma_{\mu}\psi=\epsilon_{\mu\nu}\partial^{\nu}\phi/\sqrt{\pi},\,
\bar\psi\,i\gamma.\partial\psi=
{1\over2}\partial_{\mu}\phi\partial^{\mu}\phi\,$ where  $\phi$ is a
bosonic scalar field and   $K$ is a constant. The fermionic
condensate $<\bar\psi\psi>_{0}$, for example,  may then be expressed 
in terms of the value of the bosonic condensate. The bosonized 
theory can also be constructed 
with the use of the functional integral method.   
The original fermionic and the bosonized theories 
are {\it equivalent} in the sense 
that they have the same current 
commutation relations and the energy-momentum tensor is the same when
expressed in terms of the currents. For studying  nonperturbative vacuum structure 
the bosonized theory is convenient to use. 
The {\it bosonized} version of $QED_{2}$ is found to be 
\begin{equation}
{\cal L}= {1\over2}\partial_{\mu}\phi\partial^{\mu}\phi-g
A_{\mu}\epsilon^{\mu\nu}\partial_{\nu}\phi-
{1\over4}F^{\mu\nu}F_{\mu\nu},
\end{equation}
 where $g={e/\sqrt{\pi}}$. It carries in it
all the symmetries of the original fermionic model including the 
information on the dynamical mass generation \cite{sch} for the gauge field.   
Under  the $U(1)$ gauge field transforamtion the scalar field 
is  invariant (or neutral) 
while under the 
 chiral transformations, $U_{5}(1)$, in view of the 
correspondences above, the field suffers a translation  by a 
constant.  

Following the procedure of Secs. 2 and 3   we  make  the {\it separation},   
of the condensate variable in  the scalar field 
:\quad  $\phi(\tau,x^{-})= \omega (\tau)+\varphi(
\tau,x^{-})$. 
The {\it chiral transformation} is now  defined  by:\quad
$\omega\to \omega+const., \; \varphi\to \varphi$,  and $A_{\mu}\to A_{\mu}$ 
so that  the {\it boundary conditions} at infinity on the quantum
fluctuation field $\varphi$ are kept   
unaltered under these transformations and the mathematical framework 
be considered  {\it well posed}. 
The bosonized Lagrangian written in the LF coordinates reads as is rewritten as
\begin{equation}
L= \int\, dx^{-}\;\Bigl[{\dot\varphi}\varphi' +g(A_{+}\varphi'
-A_{-}{\dot\varphi})+{1\over 2}({\dot A}_{-}-{A_{+}}')^{2}\Bigr]-
g{\dot\omega}h(\tau)
\end{equation}
 where $h(\tau)=\int dx^{-} A_{-}(\tau,x^{-})$, 
an overdot (a prime) indicates the partial derivative 
with respect to $\tau \; (x^{-})$. We work in the {\it continuum} and 
require (on  physical
considerations)  that the relevant fields   satisfy the
necessary conditions such that their Fourier transforms 
with respect to the spatial longitudinal coordinate  $x^{-}$ exist.
 
The last term in the Lagrangian density  shows that the  
{\it light-cone gauge},  $A_{-}=0$, employed often in    perturbation 
theory computations, {\it may not be appropriate to use 
in the fully interacting theory}\footnote{ Similar considerations 
are clearly pertinent  to  $3+1$ dimensional QCD as well.},  if we are seeking  
to study also the nonperturbative effects in the theory. Also the zero-momentum 
 mode of $A_{-}$ is a gauge invariant quantity under the boundary conditions 
assumed. We may, of course,  impose 
 different boundary conditions on the  fields or add 
  new ingredients in the theory so as to compensate for the 
elimination of the physical  dynamical  variable $h(\tau)$.
A convenient alternative is the local gauge-fixing  condition 
 $\partial_{-} A_{-}=0$,   which is 
accessible on the phase space.  We remove only the nonzero modes of $A_{-}$. 

Following the Dirac method to eliminate the constraints in the 
{\it front form} theory  only  the three linearly independent 
operators survive:\quad    
 the {\it condensate }
$\omega$, $\;h(\tau)$, the zero-momentum-mode of $A_{-}$ and 
  canonically conjugate to $\omega$ as well,    
and $\varphi$ which satisfies the LF commutator while it commutes with the 
others.    
The $H^{lf}$ contains  in it only 
the field $\varphi$. The Hilbert space can thus  be described in two 
different  fashions. 
Selecting  $\varphi$ and  $ h$ as forming the complete set of 
mutually commuting operators 
leads to the {\it chiral vacua} while selecting $\varphi$ together 
with $\omega$ leads to 
the description built on  the {\it condensate} or $\theta$-vacua.  
In the $QED_{2}$ the $\omega$  is {\it not} a background field rather it is 
shown \cite{pre1} to be an   operator and its eigenvalues, with continuous
spectrum,  
label the {\it condensate}  vacua of the theory. The 
cluster decomposition property  requirement \cite{weif} indicates the 
preference in favor of the {\it condensate} vacua.

The other related gauge theory model is the {\it chiral} $QED_{2}$ or 
CSM  described by 
\begin{equation}
{\cal L}= -{1\over 4}F^{\mu\nu}F_{\mu\nu} 
+  {\bar\psi}_{R}\,i\gamma^{\mu}\partial_{\mu}
\psi_{R}+ {\bar\psi}_{L}\,\gamma^{\mu}(i\partial_{\mu}+2e\sqrt{\pi}
 A_{\mu})\psi_{L},
 \end{equation}
where\footnote{\baselineskip=12pt  In two dimensions the $\pm$ 
projections of the spinor coincide with the chiral or $\gamma_{5}$ 
projections. We define 
$\gamma^{0}=\sigma_{1}$, $ \gamma^{1}=i\sigma_{2}$, 
$\gamma_{5}=\gamma^{0}\gamma^{1}=-\sigma_{3}$, $\Lambda^{-}=
\gamma^{0}\gamma^{-}/\sqrt{2}=(1-\gamma_{5})/2\equiv P_{L}$,  
$\Lambda^{+}=\gamma^{0}\gamma^{+}/\sqrt{2}=(1+\gamma_{5})/2\equiv P_{R}$, 
  $x^{\mu}:\,(x^{+}\equiv \tau,x^{-}\equiv x)$ 
with ${\sqrt 2}x^{\pm}={\sqrt 2}x_{\mp}=(x^{0}{\pm} x^{1})$, 
$A^{\pm}=A_{\mp}=(A^{0}\pm A^{1})/
{\sqrt 2}$,  
$\psi_{L,R}=P_{L,R}\;\psi $, 
$\bar\psi= \psi^{\dag}\gamma^{0}$. 
} 
 $\psi= \psi_{R}+\psi_{L}$ is a two-component spinor field and
$A_{\mu}$ is the abelian gauge field, 
$\gamma_{5}\psi_{L}=-\psi_{L}$, and $\gamma_{5}\psi_{R}=\psi_{R}$. 
The classical Lagrangian  is  
 invariant under the local $U(1)$ gauge transformations $A_{\mu}\to 
A_{\mu}+\partial_{\mu}\alpha/(2\sqrt{\pi}e)$, $\psi\to [P_{R}+
e^{i\alpha}P_{L}] \psi $ and under the global 
$U(1)_{5}$ chiral transformations $\psi\to exp(i\gamma_{5}
\alpha)\,\psi $. 

The bosonized  version is convenient to study the vacuum structure; 
it is shown to be   given by 
\begin{equation} 
    S = \int d^2x \left[ -{{1}\over {4}}F_{\mu \nu}F^{\mu \nu} 
                          +{{1}\over{2}}\partial_{\mu}\phi\partial^{\mu}\phi
                          +eA_{\nu}(\eta^{\mu \nu}
                          -\epsilon^{\mu \nu})\partial_{\mu}\phi
              +{{1}\over{2}}ae^{2}A_{\mu}A^{\mu}\right] 
\end{equation}
Here the explicit mass term for the gauge field 
parametrized by the constant parameter $a$ represents a 
regularization ambiguity \cite{jacraj} 
and the breakdown of $U(1)$ gauge symmetry. 
The model has received  much attention since Jackiw and Rajaraman \cite{jacraj} 
pointed out that, despite the gauge anomaly the theory can be shown to be unitary
and consistently quantized. 
In the LF coordinates it reads as
\begin{equation} 
 S = \int\, d^2x \;\left[{\dot\varphi}\varphi'+
{1\over 2}({\dot A}_{-}-{A'_{+}})^{2}+ a e^{2}[A_{+}+{2\over {ae}}(
\dot\omega+\dot\varphi)] A_{-} \right]. 
\end{equation}

We note now that  $A_{+}$ appears in the action 
as an {\it auxiliary} field, 
without a kinetic term. It is clear that 
the condensate variable may thus be subtracted out 
from the theory using the frequently adopted 
procedure of  
{\it field redefinition} \cite{pre2} on it:  
$\;\;A_{+}\to A_{+}-2\dot\omega/(ae)$, obtaining thereby 
\begin{equation}
{\cal L}_{CSM}={\dot\varphi}\varphi'+
{1\over 2}({\dot A}_{-}-{A'_{+}})^{2}+ 2e {\dot\varphi} A_{-}+
a e^{2} A_{+} A_{-},   
\end{equation}
which signals the emergence of a  {\it different structure} of the
Hilbert space compared to that of the SM.  
 
The  Lagrange equations in the CSM   follow to be  

\begin{eqnarray}
    \partial_{+}\partial_{-}\varphi~&=&~-e \partial_{+}A_{-}, \nonumber \\
    \partial_{+}\partial_{+}A_{-}-\partial_{+}\partial_{-}A_{+} ~&=&~
a e^{2} A_{+} +2 e \partial_{+}\varphi, \nonumber \\
\partial_{-}\partial_{-}A_{+}-\partial_{+}\partial_{-}A_{-}
  ~&=&~ a e^{2} A_{-}. 
\end{eqnarray}
and for $ a\neq 1 $ they lead to: 
\begin{eqnarray}
       \Box G(\tau,x)& = & 0 \nonumber\\
\left[ \Box + {e^2a^2\over (a-1) }\right] E(\tau,x) & =& 0, 
\end{eqnarray}  
where $E=(\partial_{+}A_{-}-\partial_{-}A_{+})$ and 
$G= (E-ae\varphi)$. 
Both the massive and massless scalar  excitations 
are present in the theory and  the tachyons
would be absent in the specrtum if   $a>1$; the case  considered 
in this paper. 
 We will confirm in the Hamiltonian framework below 
that the $E$ and $G$ represent, in fact, the two 
independent field operators on the  LF phase space. 

The  Dirac procedure  as applied to  
the very simple action (6.7) of the CSM 
is straightforward. The canonical momenta  are 
$    \pi^{+}\approx 0,     \pi^{-}\equiv E=
 \dot{A}_{-}- A'_{+},   \pi_{\varphi}= {\varphi}'+2e A_{-}$ which
result in 
two primary  weak constraints  
$    \pi^{+} \approx 0 $ and 
$ \Omega_1 \equiv (\pi_{\varphi}-\varphi'-2eA_{-})\approx 0$. A 
secondary constraint 
$    \Omega_2 \equiv \partial_{-} E + 
a e^{2} A_{-} \approx 0$ is shown to emerge 
when we require the $\tau$ independence 
(persistency) of 
$\pi^{+}\approx 0$ employing the preliminary  Hamiltonian
\begin{equation}
    H' = {H_c}^{lf} + \int dx~u_{+}\pi^{+}+\int dx~u_{1}\Omega_1 ,
\end{equation}
where $u_{+}$ and $u_{1}$ are the Lagrange multiplier fields and 
${H_c}^{lf}$ is the canonical Hamiltonian  
\begin{equation}
    {H_c}^{lf} = \int\; dx~\left[~
            \frac{1}{2}{E}^{2} + E A_{+}'
-ae^{2} A_{+}A_{-}
               \right]. 
\end{equation}
and we assume initially the  standard   
equal-$\tau$ Poisson  
brackets : 
 $\{E^{\mu}(\tau,x^{-}),A_{\nu}(\tau,
y^{-}) \}=-\delta_{\nu}^{\mu}\delta (x^{-}-y^{-})$,  
$\{\pi_{\varphi}(\tau,x^{-}),\varphi(\tau,y^{-}) \}
=-\delta(x^{-}-y^{-})$ etc.. 
 The persistency requirement for $\Omega_{1} $ results in an equation for 
determining  $u_{1}$. The procedure is repeated with the 
following extended 
Hamiltonian which includes in it also   the secondary constraint  
\begin{equation}
    {H_e}^{lf} = {H_c}^{lf} + \int dx~u_{+}\pi^{+}+\int dx~u_{1}\Omega_1 
+\int dx~u_{2}\Omega_{2}. 
\end{equation}
No more secondary constraints are seen 
to arise; we are left with the persistency conditions which
determine the multiplier fields  $u_{1}$ and $u_{2}$ while $u_{+}$
remains undetermined. We also  
find\footnote{\baselineskip=12pt 
We make the convention that the first variable in an equal-
$\tau$ bracket refers to the longitudinal coordinate $x^{-}\equiv x$
while the second one to $y^{-}\equiv y$ while  $\tau$ is suppressed.} 
 $(C)_{ij}=\{\Omega_{i},\Omega_{j}\}~=~D_{ij}~$ $ (-2
\partial_{x} \delta(x-y))$ where $i,j =1,2$ and $~D_{11}=1,~D_{22}=a
e^2, ~D_{12}=~D_{21}= -e$ and  that $\pi^{+}$ has 
vanishing brackets with $\Omega_{1,2}$. 
The $\pi^{+}\approx 0 $ is     first class weak 
constraint while 
 $\Omega_1$ and $\Omega_2 $, which does not depend on  $A_{+}$ or 
$\pi^{+}$,  are second class ones.

We  go over from the Poisson bracket  
to the  Dirac bracket $\{,\}_{D}$ 
 constructed in relation to the pair,    
 $\Omega_1\approx 0$ and   $\Omega_2\approx 0 $ 

\begin{equation} 
\{f(x),g(y)\}_{D}=\{f(x),g(y)\}-\int\int du dv\;\{f(x),\Omega_{i}(u)\}
(C^{-1}(u,v))_{ij}\{\Omega_{j}(v), g(y)\}. 
\end{equation}
Here $C^{-1}$ is the inverse of $C$ and we find 
$(C^{-1}(x,y))_{ij}=B_{ij}$ $K(x,y)$ with  $~B_{11}=a/(a-1)$,$
~B_{22}=1/[(a-1)e^2]$, $~B_{12}=B_{21}$$= 1/[(a-1)e],$ and 
$K(x,y)=-\epsilon(x-y)/4$.  Some of the Dirac brackets are 
$\{\varphi,\varphi\}_{D}= B_{11} ~K(x,y); 
~\{\varphi,E\}_{D} = e B_{11} ~K(x,y); 
~\{E,E\}_{D} = ae^{2} B_{11} ~K(x,y); 
~\{\varphi, A_{-}\}_{D}=-B_{12}~\delta(x-y)/2; 
~\{A_{-},E\}_{D}=B_{11}~\delta (x-y)/2;
~\{A_{-},A_{-}\}_{D}=B_{12}\partial_{x}~\delta(x-y)/2$  
and the only nonvanishing one involving $A_{+}$ or $\pi^{+}$ is 
$\{A_{+},\pi^{+}\}_{D}= \delta(x-y)$.

The equations  of motion employ now the Dirac brackets and 
inside them, in view of their very construction, we may set 
$\Omega_1=0$ and $\Omega_2=0 $  as strong relations. The 
Hamiltonian is therefore  effectively given by $H_{e}$ 
with the terms involving the multipliers $u_{1}$ and $u_{2}$ 
dropped. The multiplier $u_{+}$ is not determined since the constraint 
$\pi^{+}\approx 0$ 
continues to be first class even when the above 
Dirac bracket is employed. 
The variables $\pi_{\varphi}$ and $A_{-}$ are then removed from the 
theory leaving behind $\varphi$, $E$, $A_{+}$, and $\pi^{+}$  as 
the remaining independent variables. 
The canonical Hamiltonian density reduces to 
${\cal H}_{c}^{lf}= E^2/2 +\partial_{-}(A_{+}E)$ while $\dot A_{+}=
\{A_{+}, H_{e}^{lf}\}_{D}=u_{+}$. The surface term in 
the canonical LF Hamiltonian may be ignored if, 
say, $E (=F_{+-})$ vanishes at 
infinity. The variables $\pi^{+}$ and $A_{+}$ are then seen to describe 
a decoupled (from $\varphi$ and $E$) free theory 
and we may hence drop these variables as well. 
The effective LF  Hamiltonian thus takes the simple form 

\begin{equation}
H_{CSM}^{lf} = {1\over {2}} \int dx \; E^{2}, 
\end{equation}
which is to be contrasted with the one found 
in the conventional treatment \cite{abda}. 
 $E$ and $G$ (or $E$ and $\varphi$) are now the independent variables
on the phase space and the Lagrange equations are verified to be
recovered for them, which assures us of the selfconsistency
\cite{dir1}. 
We  stress that in our 
discussion we do {\it not} employ any gauge-fixing.  The same result
for the Hamiltonian could be alternatively  
obtained\footnote{\baselineskip=12pt  A similar discussion is
encountered also in the LF quantized 
Chern-Simons-Higgs system \cite{prech}.}, 
however,  
if we did introduce  the gauge-fixing constraint $A_{+}\approx 0$ 
and made further 
modification on $\{,\}_{D}$ in order to implement  $A_{+}\approx 0, 
\pi^{+}\approx0$ as well.  
That it is accessible  on the phase space to take care of the 
remaining first class constraint, but not in the bosonized Lagrangian,  
follows from the Hamiltons 
eqns. of motion. We recall \cite{pre1} that in the SM
 $\varphi$, $\omega$, and $\pi_{\omega}=
(e/\sqrt {\pi})\int dx A_{-}$ were shown to be the 
independent operators 
 and that the matter field $\varphi$ appeared instead in the LF 
Hamiltonian.
The  {\it canonical quantization} is peformed 
via the correspondence $i\{f,g\}_{D}\to [f,g] $ and we find the
following equal-$\tau$ commutators 
\begin{eqnarray}
\left[ E(x),E(y) \right]&=& i K(x,y){ a^2 e^2/ (a-1)},\nonumber\\ 
\left[ G(x),E(y)\right]& =& 0,   \nonumber\\
\left[ G(x),G(y)\right]& =& {ia^2 e^2} K(x,y).
\end{eqnarray} 
For $a>1$, when the tachyons are absent as seen from (6), 
these commutators  are also 
physical and  the independent field operators $E$ and $G$  generate 
the Hilbert space with a tensor product structure 
of the Fock spaces 
$F_{E}$ and $F_{G}$ of these fields
 with  the positive definite metric. 

The commutators obtained can  be realized in the momentum space through 
the following Fourier transforms   
\begin{eqnarray}
E(x,\tau)&=&{ae\over {{\sqrt{(a-1)}}{\sqrt{2\pi}}}} \int_{-\infty}^{\infty} dk
\;{{\theta(k)}\over{\sqrt{2k}}}
\left[d(k,\tau)e^{-ikx}~+~d^{\dag}(k,\tau)e^{ikx}\right],\nonumber\\ 
G(x,\tau) &=& {{ae}\over {\sqrt{2\pi}}} 
\int_{-\infty}^{\infty} dk\;{{\theta(k)}\over{\sqrt{2k}}}
\left[g(k,\tau)e^{-ikx}~+~g^{\dag}(k,\tau)e^{ikx}\right], 
\end{eqnarray}
if  the operators ($d, g, d^{\dag},g^{\dag}$) satisfy the well known 
canonical commutation relations of two independent harmonic
oscillators;   
the well known set of Schwinger's bosonic oscillators, often employed
in the angular momentum theory. The
expression for the Hamiltonian becomes 
\begin{equation}
H_{CSM}^{lf}= \delta(0){{a^2e^2}\over {2(a-1)}}~\int 
{{dk}\over {2k}}\, \theta(k)\,
\; N_{d}(k,\tau)
\end{equation}
where we have dropped the infinite zero-point energy term and 
note that \cite{ryd} $\left[d^{\dag}(k,\tau),d(l,\tau)\right]=
-\delta(k-l)$, $d^{\dag}(k,\tau)d(k,\tau)= 
\delta(0) N_{d}(k,\tau)$ etc. 
 with similar expressions 
for the independent g-oscillators. 
We verify that $\left [N_{d}(k,\tau),N_{d}(l,\tau)\right]=0$, 
$\left [N_{d}(k,\tau),N_{g}(l,\tau)\right]=0$,  
$\left [N_{d}(k,\tau),d^{\dag}(k,\tau)\right]=  d^{\dag}(k,\tau) $
etc.. 

The Fock 
space can hence be built on a basis of eigenstates of the 
hermitian number operators $N_{d}$ and  $N_{g}$. 
The   ground state of CSM is  degenerate and 
described by $\vert 0> = \vert E=0)\otimes \vert G\}$ and it carries 
vanishing LF energy in agreement with the conventional theory discusion 
\cite{abda}.   
For a fixed $k$ these states,  $ \vert E=0)\otimes 
{({g^{\dag}(k,\tau)}^{n}/\sqrt {n!})}\vert 0\}$,  
are labelled by the integers $n=0,1,2,\cdots $. The $\theta$-vacua 
are absent in the CSM. However, we 
recall \cite {pre1} 
that in the SM the degenerate {\it chiral vacua} are also 
labelled  by 
such integers. We remark also that on the LF we work in the Minkowski 
space and that in our discussion we do {\it not} make use of the 
Euclidean space theory action, where the (classical) 
vacuum configurations of
the Euclidean theory gauge field, belonging  to the  
distinct topological sectors, are useful, for example, in the 
functional integral quantization of the gauge theory.   

On the LF both the bosonized SM and CSM are described in terms of 
a minimum number of 
dynamical variables,  which survive after the elimination of  the phase space 
constraints. We recall that  the introduction of the {\it bosonic condensate} 
variable $\omega(\tau)$ (or in general  
$\omega(\tau, x^{\perp})\,$)  corresponds to the gauge-fixing required 
in order to deal with the first class constraint 
$\int (\pi-\partial_{-}\phi) \approx 0$. On the other hand we  have the gauge invariant 
zero-momentum mode $h(\tau)$ of the gauge field $A_{-}$, apart from the quantum 
fluctuation field $\varphi$. They are in a sense the minimal set of operators 
which survive in the {\it front form} theory. With their help the 
vacuum structures of both the SM and CSM are described in a very economical 
and transparent fashion on the LF, 
which agree with the conventional theory conclusions. 
In the latter, however, we have to go through quite an elaborate and extensive 
discussion \cite{abda}. 
Finally,  if we did adopt the light-cone gauge we must compensate 
for the loss of the gauge invariant information by some other 
ingredient, say, by imposing more complicated boundary conditions on the 
fields involved or by introducing new fields.

\section{QCD in Covariant gauges}
\setcounter{equation}{0}
\renewcommand{\theequation}{7.\arabic{equation}}

We describe briefly the recent study \cite{pre3} done  on the 
 {\it front form} QCD  in covariant gauges. 
The Lagrangian density corresponding to the quantum action 
 of  QCD is described in  standard notation  by 
\begin{equation}
{\cal L}_{QCD}=-{1\over 4}F^{a\mu\nu}{F^{a}}_{\mu\nu}+B^{a}\partial_{\mu}A^{a\mu}+
{{\xi} \over 2}B^{a}B^{a}+i\partial^{\mu}{\chi_{1}}^{a}{{\cal D}^{ac}}_{\mu}
{\chi_{2}}^{c}+ {\bar\psi}^{i}
(i\gamma^{\mu}{D^{ij}}_{\mu}-m\delta^{ij})\psi^{j}
\end{equation}
Here $\psi^{j}$ is the quark field with color index $j=1..N_{c}$ for an  
$SU(N_{c})$ color group, ${A^{a}}_{\mu}$ the gluon field,   
 $F^{a}_{\mu\nu}=\partial_{\mu}A^{a}_{\nu}
-\partial_{\nu}{A^{a}}_{\mu}
+g f^{abc}{A^{b}}_{\mu}{A^{c}}_{\nu}$ the  
field strength, ${{\cal D}^{ac}}_{\mu}=
(\delta^{ac}\partial_{\mu}+g f^{abc}{A^{b}}_{\mu})$,  
${D^{ij}}_{\mu}\psi^{j}=
(\delta^{ij}\partial_{\mu}-ig {A^{a}}_{\mu}(\lambda^{a}/2)^{ij})\psi^{j} $, 
$a=1..({N_{c}}^{2}-1)$  the gauge group 
index etc. The covariant  
gauge-fixing is introduced  by adding to the Lagrangian 
the linear gauge-fixing term $(B^{a}\partial_{\mu}A^{a\mu}+
{{\xi} \over 2}B^{a}B^{a})$ where $B^{a}$ is 
the  Nakanishi-Lautrup  auxiliary field 
and $\xi$ is a parameter.  
The ${\chi_{1}}^{a}$ and 
${ \chi_{2}}^{a}$ are the (hermitian) anticommuting Faddeev-Popov 
ghost fields, and the   action is 
invariant under the BRS  transformation.
 
The quark field term in  LF coordinates reads 
\begin{eqnarray}
{\bar\psi}^{i}
(i\gamma^{\mu}{D^{ij}}_{\mu}-m\delta^{ij})\psi^{j}
&=& i{ \sqrt {2}}{\bar\psi_{+}}^{i}
\gamma^{0}{D^{ij}}_{+}{\psi_{+}}^{j}+{\bar\psi_{+}}^{i}
(i\gamma^{\perp}{D^{ij}}_{\perp}-m\delta^{ij}){\psi_{-}}^{j} \nonumber \\
&+&{\bar\psi_{-}}^{i}\left[{i \sqrt {2}}\gamma^{0}
{D^{ij}}_{-}{\psi_{-}}^{j}+ (i\gamma^{\perp}{D^{ij}}_{\perp}-m\delta^{ij})
{\psi_{+}}^{j}\right]
\end{eqnarray}
where ${D^{ij}}_{\pm}=
(\delta^{ij}\partial_{\pm}-ig {A^{a}}_{\pm}(\lambda^{a}/2)^{ij}) $.                              
It shows that the  minus components ${\psi_{-}}^{j}$  are in fact nondynamical 
( Lagrange multiplier ) fields 
without  kinetic terms. 
The variation of the action with respect to  
${{\bar{\psi^{j}}}_{-}}$ and ${{{\psi^{j}}}_{-}}$ leads to 
the following gauge covariant constraint equation 
\begin{equation}
i{\sqrt 2} {D^{ij}}_{-}{\psi_{-}}^{j}= -(i\gamma^{0}\gamma^{\perp}{D^{ij}}_{\perp}-m\gamma^{0}\delta^{ij})
{\psi_{+}}^{j}, 
\end{equation}
and its conjugate.  
The   ${\psi^{j}}_{-}$ components may thus be eliminated in favor of 
 the dynamical components $\psi_{+}^{j}$
\begin{equation}
{\psi_{-}}^{j}(x)
= \frac{i}{\sqrt 2}  \left[U^{-1}(x|A_{-})
\frac{1}{\partial_{-}}
U(x|A_{-})\right]^{jk}
(i\gamma^{0}\gamma^{\perp}{D^{kl}}_{\perp}-m\gamma^{0}\delta^{kl})
{\psi_{+}}^{l}(x). 
\end{equation}
Here, for a fixed $\tau$,  $U\equiv U(x|A_{-})$ is an $N_{c}\times N_{c}$ 
gauge matrix  in the fundamental representation of $SU(N_{c})$ and 
it satisfies   
\begin{equation}
\partial_{-}U(x|A_{-})= -ig \,U(x|A_{-})\, A_{-}(x)
\end{equation}
with  $A_{-}={A^{a}}_{-}\lambda^{a}/2$. It has  
 the formal solution 
\begin{equation}
U(x^{-},x^{\perp}|A_{-})=U({x^{-}}_{0},x^{\perp}|A_{-})\,\tilde {\cal P} 
\,exp\left\{-ig \int_{{x^{-}}_{0}}^{x^{-}} dy^{-}
 A_{-}(y^{-},x^{\perp})\right\}
\end{equation}
where $\tilde {\cal P}$ indicates the anti-path-ordering along the 
longitudinal direction $x^{-}$.  $U$ has  a series expansion 
in the powers of the coupling constant.

The  Hamiltonian density  in Feynman gauge is  
\begin{eqnarray}
{\cal H}^{LF}
&=& {\cal H}_{0}+{\cal H}_{int }\nonumber \\
&=&  -{1\over2} g^{\mu\nu} 
{A^{a}}_{\mu}\,\partial^{\perp}\partial_{\perp}\,{A^{a}}_{\nu}
  -g { \sqrt {2}}{\bar\psi_{+}}^{i}
\gamma^{0}{A_{+}}^{ij}{\psi_{+}}^{j} \nonumber \\
&-& {\bar\psi_{+}}^{i}\left[\delta^{ij}
(i\gamma^{\perp}\partial_{\perp}-m)+g\gamma^{\perp}{A^{ij}}_{\perp}
\right] {\psi_{-}}^{j} 
 +{g\over2}f^{abc}(\partial_{\mu}{A^{a}}_{\nu}-
\partial_{\nu}{A^{a}}_{\mu}) A^{b\mu} A^{c\nu} \nonumber \\
&+&{{g^2}\over 4} 
f^{abe}f^{cde} {A^{a}}_{\mu} {A^{b}}_{\nu} A^{c\mu} A^{d\nu} 
+\partial^{\mu}({\bar\chi}^{a})\partial_{\mu}\chi^{a}
 +g f^{abc} (\partial^{\mu}{\bar\chi^{a}})\chi^{b}{A^{c}}_{\mu}
\end{eqnarray}
where ${\psi_{-}}^{j}$ is given above,  we have 
set ${\sqrt2}\chi=({\chi_{1}}+i {\chi_{2}}),\,   
{\sqrt 2}\bar\chi=({\chi_{1}}-i {\chi_{2}})$, and in ${\cal H}^{LF}$ 
the cubic and higher
order  terms belong to  ${\cal H}_{int}$ which is also understood to be normal
ordered. It is worth remarking that despite the presence of the 
longitudinal operators  $a_{\pm}$ and ${a^{\dag}}_{\pm}$ 
in the fields $A_{\mu}$, 
there are no non-zero matrix elements involving these quanta
as external lines in view of the commmutation relations of these operators 
as discussed in the previous section.

The  perturbation theory expansion in the interaction representation 
where we time order with respect to the LF time $\tau$ is 
built following the Dyson-Wick  procedure. We will illustrate 
it in our context through   some explicit  computations,  
for simplicity, in QED where $U(x|A_{-})=exp \{-ie\int_{}^{x^{-}} du^{-}
A_{-}(\tau,u^{-},x^{\perp})\}$ and $D_{\mu}=(\partial_{\mu}-ieA_{\mu})$. 
We observe    that a {\it seagull} term of the order $e^{2}$ 
is present in the interaction Hamiltonian at the tree level; like that 
found also in  the scalar field QED. 

Towards an illustration consider  the  computation of  {\it  Electron Self-Energy}.  
The contribution  from the longitudinal components arises from 
\begin{eqnarray}
e^{2}{\int }d^{4}x_{1} d^{4}x_{2}&&  
: {\psi_{+}}^{\dag}(x_{2})(m+i{\not{\partial}_{2}}^{T})\int_{-\infty}^{\infty}
{1\over2}d{y_{2}}^{-}\epsilon({x_{2}}^{-}-{y_{2}}^{-}) \nonumber \\
&& \{\int_{{y_{2}}^{-}}^{{x_{2}}^{-}}d{u_{2}}^{-} \dot{A}_{-}(u_{2})\}
(m-i\not{\partial_{2}}^{T}) {\ddot\psi}_{+}(y_{2}){{\ddot\psi}_{+}}^{\dag}(x_{1})
\psi_{+}(x_{1})\dot {A}_{+}(x_{1}):
\end {eqnarray}
leading to 
\begin{equation}
e^{2}\int d^{4}q \;{{{\bar u}^{(r)}(p)[\gamma^{-}(m+\not q^{T})
\gamma^{+}] u^{(s)}(p)}\over
 {[(p-q)^{2}+i\epsilon ]}(q^{2}-m^{2}+i\epsilon)}\; (-g_{-+})
\end{equation}
The  graph with the $A_{+}$ and $A_{-}$ interchanged gives rise to 
a similar expression with  $g_{+-}\to g_{-+}$ while $\gamma ^{\pm}\to 
\gamma ^{\mp}$. 
The matrix elements following from  the  four 
graphs corresponding to the exchange of the 
 ( photon ) fields $A_{1}$ and $A_{2}$  is also
written down by simple inspection.  As in the earlier case  
the expressions get  simplified in virtue of (10) and 
acquire the covariant form encountered in the conventional covariant perturbation 
theory. The   complete matrix element 
  is found to be 
\begin{equation}
e^{2}\int d^{4}q \;{{{\bar u}^{(r)}(p)[\gamma^{\mu}(m+\not{\tilde q})
\gamma^{\nu}]
 u^{(s)}(p)}\over
 {[(p-q)^{2}+i\epsilon ](q^{2}-m^{2}+i\epsilon)}} \;(-g_{\mu\nu})
\end{equation}
where ${{\tilde q}^{\mu}}\equiv( (m^{2}+q^{\perp}q^{\perp})/{(2q^{+})}, 
q^{+}, q^{\perp})$
  and the integration measure is $d^{4}q= d^{2}q^{\perp}dq^{+}dq^{-}$. 
  Considering that the integrand has  the pole at $q^{2}-m^{2}\approx 0$ we may 
regard the expression obtained \cite{pre3} on the LF to be effectively identical 
to the one obtained in the  conventional covariant theory  framework. 
The discussion parallel to that given here may be followed also 
in the context of
the light-cone gauge. The latter, however, demands the further introduction
\cite{mandelstam} of 
a light-like vector $n^{\mu}=(n^{0}, {\vec n} )$ and its dual 
${\tilde n}^{\mu}=(n^{0}, -{\vec n} )$ in order to evaluate the corresponding
Feynman integrals in a consistent manner. 

\section{Conclusions}

 Collected  below are  some of the interesting conclusions   
we seem to  reach.     
 
\begin{itemize}
\item 
The LF hyperplane is {\it equally valid and appropriate} 
as the conventional equal-time 
one for the field theory quantization. 

\item The appearence of the nonlocality 
along the longitudinal direction in the {\it front form} quantized  theory 
is not unexpected; it does 
not conflict with the the microcausality (or 
cluster decomposition)  principle.

\item The covariant  phase space and Fourier expansion considerations based 
on the description of the relativistic theory 
using light-cone coordinates lead to   
 the  LF commutator for the free scalar field,  which is nonlocal 
in the longitudinal direction.  

\item 
The  
hyperplanes $x^{\pm}=0$ define the characteristic surfaces of a 
hyperbolic partial differential equation. From the mathematical theory of 
classical partial differential equations \cite{sne}  
it is known that the Cauchy initial value problem 
would require us to specify the  data on both the hyperplanes. 
From our studies  
we conclude \cite{pre2} that 
it is sufficient in the {\it front form} theory  to choose  
  one of the two LF hyperplanes 
for canonically quantizing the theory. 

- In the quantized theory the equal-$\tau$  commutators of 
the field operators,   at a fixed 
initial LF-time, form now  a part of the initial data instead and we 
deal with operator differential equations. 

- The information on the commutators on 
the other characteristic 
hyperplane seems already to be contained \cite{pre1} 
in the quantized theory; it  may  not, in general, be required 
to specify it  separately. 

\item The constrained phase space dynamics  in the LF theory 
with  one more kinematical 
generator and  the  inherent symmetry with regard to $x^{\pm}$   
result in a reduced number of independent 
field operators.   
The discussion of the Hilbert space  becomes  more transparent 
compared to that in the conventional treatment.    
The lack of manifest covariance which appears problematic can  
be handled\footnote{ See also, \cite{prech}}
 by employing, for example, the {\it LF  four-spinor} \cite{pre1} and the 
Fourier transform of the spinor field as defined \cite{pre3} in Sec. 4.

\item On the LF the $\gamma_{5}$ symmetry of 
free massless Dirac equation can be 
generalized to a nonlocal (chiral) $\Gamma_{5}$ symmetry valid also  in the massive case. 
 The Weyl and Majorana spinors and the helicity operator 
  may be defined on the LF in straightforward 
 fashion.

 \item 
The zero-longitudinal-momentum  modes of the fields are important for describing
the nonperturbative effects on the LF. In the scalar and gauge theories 
 they are   dynamical variables 
in the frame work of the {\it standard} Dirac procedure.   
The {\it separation} $\;\phi(\tau, x^{-}, x^{\perp})= 
\omega(\tau, x^{\perp})+\varphi(\tau, x^{-}, x^{\perp})$ 
introduced in Secs. 2.6, 5  correspond to 
the  gauge-fixing conditions \cite{dir1} 
required to be introduced in the theory for  handling  
 first class   constraints.   
 
- In the case of the scalar theory we obtain  constraint equations 
which enable us to describe SSB and the (tree level) Higgs mechanism. 
Associated to the local theory in the conventional coordinates we find 
a nonlocal  LF Hamiltonian. 

- The gauge field zero modes play a crucial role in the 
description of the nonperturbative vacuum structures in the LF quantized 
SM and CSM. 
 They also indicate that the  (popular) 
light-cone gauge may not be accessible  in the {\it front form } theory  
if we are concerned with the study of nonperturbative effects.

\item   The physical content following from the {\it front form} theory 
is the same, even though 
arrived at through different description on the LF, when compared with the 
one in the {\it instant form} case.

\item Not all the constraints in the LF theory need to be solved first 
before considering its  renormalization; it is sometimes 
convenient to obtain some of them as renormalized constraint 
equations \cite{pre4} instead.

\item In the conventional treatment we may be required to 
introduce external 
constraints in the quantized theory based on physical considerations, say, 
while describing the SSB.  The analogous relevant 
constraints in the {\it front form} theory appear to be already contained
    in the quantized theory.  

\item On the   LF the quantized theory of {\it chiral boson}   
appears  straightforward (Sec. 3.4).   The field commutator does not  
 conflict with the microcausality principle.  

\item A theoretical demonstration of the well accepted notion 
that a  classical model field 
theory must be upgraded first through its quantization 
before we confront it  with the 
experimental data, seems to emerge.

\item The LF quantized QCD employing covariant gauges \cite{pre3} 
looks  promising. All of the propagators become causal and 
 the covariance of the 
theory is tractable.   The semiclassical theory is found  
revealed at the tree level.
  The algebra of bilocals 
 in the LF quantized theory  
may help reveal the string like structure as seems to be found \cite{dhar}, 
for example, in $QCD_{2}$.

\item The recently proposed BRS-BFT \cite{bata} quantization procedure is 
extended straightforwardly 
on the LF  (Appendix C).

\item It is well known that topological considerations are often 
required  in the field theory quantization employing the functional  
integral method, where the Euclidean theory action is  
employed.  
The corresponding ingredients seem to arise in the 
  canonically quantized {\it front form} theory  as well but with 
different  interpretation. 
  This is suggested, for example, from the studies of the 
SM,   CSM, and  the study of the {\it kink} solutions.  
  
\item In connection with the relativistic bound state problem, not touched 
upon in this article,   the 
LF Tamm-Dancoff method \cite{ken,broo} and Bethe-Salpeter dynamics on 
the covariant null plane \cite{mit1, mit2, carbonel} 
seem to be promising alternatives to lattice gauge theory 
approach.  

\end{itemize}

\section*{Acknowledgements}
The author acknowledges with thanks the helpful 
comments from Stan Brodsky, Richard Blankenbecler and   Sidney Drell.  
The hospitality offered to him at the SLAC and a financial grant 
of Proci\^encia 
program of the UERJ, Rio de Janeiro, Brasil, are gratefully acknowledged.




\vspace{1cm}

\noindent {\bf Appendix A:  \quad Poincar\'e Generators on the LF} 
\vspace{0.5cm}

The Poincar\'e generators in coordinate system 
$\,(x^{0},x^{1},x^{2},x^{3})$,  satisfy 
$[M_{\mu\nu},P_{\sigma}]=-i(P_{\mu}g_{\nu\sigma}-P_{\nu}g_{\mu\sigma})$ and 
$[M_{\mu\nu},M_{\rho\sigma}]=i(M_{\mu\rho}g_{\nu\sigma}+
M_{\nu\sigma}g_{\mu\rho}-M_{\nu\rho}g_{\mu\sigma}-M_{\mu\sigma}g_{\nu\rho})
\,$  where the metric is 
 $g_{\mu\nu}=diag\,(1,-1,-1,-1)$, $\mu=(0,1,2,3)$ and we take 
 $\epsilon_{0123}=\epsilon_{-+12}=1$. If we define 
 $J_{i}=-(1/2)\epsilon_{ikl}M^{kl}$
and  $K_{i}=M_{0i}$, where   $i,j,k,l=1,2,3$, we find 
$[J_{i},F_{j}]=i\epsilon_{ijk}F_{k}\,$ 
for $\,F_{l}=J_{l},P_{l}$ or  $ K_{l}$ while 
$[K_{i},K_{j}]=-i \epsilon_{ijk}J_{k}, \, [K_{i},P_{l}]=-iP_{0}g_{il},
\, [K_{i},P_{0}]=iP_{i},$ and  $[J_{i},P_{0}]=0$.

 The LF  generators are  
$P_{+}$, $P_{-}$, $P_{1} $, $P_{2}$, $M_{12}=-J_{3}$, $M_{+-}=-K_{3}$, $M_{1-}=
-(K_{1}+J_{2})/{\sqrt 2}\,
\equiv {-B_{1}}$, $M_{2-}=-(K_{2}-J_{1})/{\sqrt 2}
\equiv {-B_{2}}$, 
$M_{1+}=-(K_{1}-J_{2})/{\sqrt 2}\equiv -S_{1}$ and 
$M_{2+}=-(K_{2}+J_{1})/{\sqrt 2}\equiv -S_{2}$. We find 
$[B_{1},B_{2}]=0$, $[B_{a},J_{3}]=-i\epsilon_{ab} B_{b}$, 
$[B_{a},K_{3}]=iB_{a}$, $[J_{3},K_{3}]=0$, 
$[S_{1},S_{2}]=0$, $[S_{a},J_{3}]=-i\epsilon_{ab} S_{b}$, 
$[S_{a},K_{3}]=-iS_{a}$ where $a,b=1,2$ and $\epsilon_{12}=-\epsilon_{21}=1$. 
Also $[B_{1},P_{1}]=[B_{2},P_{2}]=i P^{+}$, $[B_{1},P_{2}]=[B_{2},P_{1}]=
0$,  $[B_{a},P^{-}]=iP_{a}$,  $[B_{a},P^{+}]=0$, 
$[S_{1},P_{1}]=[S_{2},P_{2}]=i P^{-}$, 
$[S_{1},P_{2}]=[S_{2},P_{1}]=
0$, $[S_{a},P^{+}]=iP_{a}$,  $[S_{a},P^{-}]=0$,  
$[B_{1},S_{2}]= - [B_{2},S_{2}]=-iJ_{3}$, 
$[B_{1},S_{1}]=[B_{2},S_{2}]=-iK_{3} $. 
For $ P_{\mu}=i\partial_{\mu}$,  and 
$M_{\mu\nu}\to L_{\mu\nu}= i(x_{\mu}\partial_{\nu}-x_{\nu}\partial_{\mu})$ 
we find 
$B_{a}=(x^{+}P^{a}-x^{a}P^{+})$, $S_{a}=(x^{-}P^{a}-x^{a}P^{-})$,  
 $K_{3}=(x^{-}P^{+}-x^{+}P^{-})$  and 
$  \quad J_{3}=(x^{1}P^{2}-x^{2}P^{1})$. 
Under the conventional {\it parity} operation  ${\cal P}$: 
($\;x^{\pm}\leftrightarrow x^{\mp}, x^{1,2}\to 
-x^{1,2}$) and $(  p^{\pm}\leftrightarrow p^{\mp}, p^{1,2}\to 
-p^{1,2}),$ we find  $\vec J\to \vec J,\, \vec K \to -\vec K $, $B_{a}\to -S_{a}$  
etc..
The  six generators 
$\,P_{l}, \, M_{kl}\,$  leave  $x^{0}=0$ hyperplane 
invariant and are  called    
{\it kinematical}  while the remaining $P_{0},\,M_{0k}$ 
the  {\it dynamical } ones. On the LF 
there are {\it seven}  kinematical generators :  $P^{+},P^{1},
P^{2}, B_{1}, B_{2}, J_{3}$ and  $K_{3} $ which leave the LF hyperplane, 
$x^{0}+x^{3}=0$,  invariant and the three {\it dynamical} 
ones $S_{1},S_{2}$ and  $P^{-}$ 
form a mutually commuting set. The $K_{3}$ which was dynamical becomes now 
a kinematical; it generates scale transformations of the LF components of 
$x^{\mu}$, $P^{\mu}$ and $M^{\mu\nu}$.  We note that each of the set 
 $\{B_{1},B_{2},J_{3}\}$ and   $\{S_{1},S_{2},J_{3}\}$ generates 
an  $E_{2}\simeq SO(2)\otimes T_{2} $ algebra; this will be shown below to be 
relevant for 
defining the {\it spin} for massless particle. Including $K_{3}$ in each set 
we find two subalgebras each with four elements. Some useful identities are 
$e^{i\omega K_{3}}\,P^{\pm}\,e^{-i\omega K_{3}}= e^{\pm \omega}\,P^{\pm},
\, e^{i\omega K_{3}}\,P^{\perp}\,e^{-i\omega K_{3}}= P^{\perp}, 
 e^{i\bar v.\bar B}\,P^{-}\,e^{-i\bar v.\bar B}= 
 P^{-}+\bar v.\bar P + {1\over 2}{\bar v}^{2}P^{+},  
 e^{i\bar v.\bar B}\,P^{+}\,e^{-i\bar v.\bar B}= P^{+}, 
 e^{i\bar v.\bar B}\,P^{\perp}\,e^{-i\bar v.\bar B}= 
 P^{\perp}+v^{\perp} P^{+}, 
 e^{i\bar u.\bar S}\,P^{+}\,e^{-i\bar u.\bar S}= 
 P^{+}+\bar u.\bar P + {1\over 2}{\bar u}^{2}P^{-}, 
  e^{i\bar u.\bar S}\,P^{-}\,e^{-i\bar u.\bar S}= P^{-}, 
 e^{i\bar u.\bar S}\,P^{\perp}\,e^{-i\bar u.\bar S}= 
 P^{\perp}+u^{\perp} P^{-}$ 
  where  $P^{\perp}\equiv\bar P=(P^{1},P^{2}), \, v^{\perp}\equiv {\bar v}
= (v_{1},v_{2})\, $ 
and  $(v^{\perp}. P^{\perp})\equiv(\bar v.\bar P)=v_{1}P^{1}+v_{2}P^{2}$ etc. Analogous expressions with 
$P^{\mu}$  replaced by $X^{\mu}$ can be obtained if we use 
$\,[P^{\mu},X_{\nu}]\equiv [i\partial^{\mu},x_{\nu}]= i\delta^{\mu}_{\nu}\,$. 

\vspace{0.5cm}


\nl {\bf Appendix B\footnote{See, P.P. Srivastava, {\sl Lightfront quantization of 
field theory} in {\it Topics in Theoretical Physics}, 
{\sl Festschrift for Paulo Leal Ferreira}, eds., V.C. Aguilera-Navarro et 
al., pgs. 206-217,  IFT-S\~ao Paulo, SP, Brasil (1995); 
hep-th/9610044; 9610149. }:\quad   LF Spin Operator. Hadrons in LF Fock basis}

\vspace{0.5cm}

The Casimir generators of the Poincar\'e group are : $P^{2}\equiv 
P^{\mu}P_{\mu}$ and  $W^{2}$, where  
$W_{\mu}=(-1/2)\epsilon_{\lambda\rho\nu\mu}
M^{\lambda\rho}P^{\nu}$ defines the  Pauli-Lubanski pseudovector. It follows
from $[W_{\mu},W_{\nu}]=i\epsilon_{\mu\nu\lambda\rho} W^{\lambda}P^{\rho}, 
\quad [W_{\mu},P_{\rho}]=0\;$  and  $\;W.P=0$ that in a   
representation charactarized by particular eigenvalues of 
the two Casimir operators we 
may simultaneously diagonalize  $P^{\mu}$ along with just one 
component of  $W^{\mu}$. We have 
$ W^{+} =-[J_{3} P^{+}+B_{1} P^{2}-B_{2} P^{1}], 
W^{-} =J_{3} P^{-}+S_{1} P^{2}-S_{2} P^{1}, 
W^{1} =K_{3} P^{2}+ B_{2} P^{-}- S_{2} P^{+},$ and 
$W^{2} =-[K_{3} P^{1}+ B_{1} P^{-}- S_{1} P^{+}]$ and it shows 
that  $W^{+}$ {\it has a  special place} since it 
contains only the kinematical generators \cite {pre1}. On the LF we define  
  ${\cal J}_{3}= -W^{+}/P^{+}$ as the   {\it spin operator}. 
It may be shown  to 
commute with  $P_{\mu}, B_1,B_2,J_3,$ and  $K_3$. 
For $m\ne 0$ we may use the parametrizations $p^{\mu}:( p^{-}=(m^{2}+
{p^{\perp}}^{2})/(2p^{+}), 
p^{+}=(m/{\sqrt 2})e^{\omega}, 
p^{1}=-v_{1}p^{+},  p^{2}=-v_{2}p^{+})$ and 
${\tilde p}^{\mu}: (1,1,0,0)(m/{\sqrt 2})$ in the rest frame. We have 
$P^{2}(p)= m^{2} I$ and  $W(p)^{2}=
W(\tilde p)^{2}= -m^{2} [J_{1}^2+J_{2}^2+J_{3}^2] = -m^{2} s(s+1) I$ 
where $s$ assumes half-integer values.  
Starting from the rest  state $\vert \tilde p; m,s,\lambda, ..\rangle $
with ${J}_{3}\, \vert\tilde p; m,s,\lambda, ..\rangle
= \lambda \,\vert \tilde p; m,s, \lambda, ..\rangle $ we may build  an 
arbitrary eigenstate of $P^{+}, P^{\perp}, {\cal J}_{3} $  (and $ P^{-}$ ) 
on the LF by 

$$\vert p^{+},p^{\perp}; m,s,\lambda, ..\rangle= e^{i(\bar v. \bar B)} 
e^{-i\omega K_{3}} \vert \tilde p; m,s,\lambda, ..\rangle 
$$ 

\noindent If we make use of the following  {\it identity} \cite{pre}

$${\cal J}_{3}(p)=\;J_{3}+v_{1}B_{2}-v_{2}B_{1}=\quad 
e^{i(\bar v. \bar B)} \;
J_{3} \;e^{-i(\bar v. \bar B)}  $$

\noindent we find ${\cal J}_{3}\, \vert p^{+},p^{\perp}; m,s,\lambda, ..\rangle
= \lambda \,\vert p^{+},p^{\perp};m,s,\lambda, ..\rangle $. 
Introducing  also  the operators ${\cal J}_{a}= 
-({\cal J}_{3} P^{a} + W^{a})/{\sqrt{P^{\mu}P_{\mu}}}$, $a=1,2$, which
do, however, contain 
dynamical generators, we verify that 
$\;[{\cal J}_{i},{\cal J}_{j}]=i\epsilon_{ijk} {\cal J}_{k}$.

For $m=0$ case when $p^{+}\ne0$ 
a convenient parametrization is 
$p^{\mu}:( p^{-}=p^{+} {v^{\perp}}^{2}/2, \,p^{+}, 
p^{1}=-v_{1}p^{+}, p^{2}=-v_{2}p^{+})$ and $\tilde p: 
(0, p^{+}, 0^{\perp})$. We have 
$W^{2}(\tilde p) = -(S_{1}^{2}+S_{2}^{2}){p^{+}}^{2}$ 
and  $[W_{1},W_{2}](\tilde p)=0, \,
[W^{+},W_{1}](\tilde p)=-ip^{+}W_{2}(\tilde p), 
\, [W^{+}, W_{2}](\tilde p)=ip^{+}W_{1}(\tilde p)$ showing that 
$W_{1}, W_{2}$ and  $W^{+}$  generate the algebra 
$SO(2)\otimes T_{2}$. The eigenvalues of 
$W^{2}$ are hence not quantized and they vary continuously. 
This is contrary to the experience so we impose that the physical states 
satisfy in addition  $W_{1,2}
\vert \, \tilde p;\,m=0,..\rangle=0$. Hence 
$ W_{\mu}=-\lambda P_{\mu}$ and  the invariant parameter 
$\lambda $ is taken to define as the {\it spin} of the massless particle.  
From  $-W^{+}(\tilde p)/{\tilde p}^{+}=J_{3}$ we conclude that 
$\lambda$ assumes half-integer values as well.  
We note that   $W^{\mu}W_{\mu}=\lambda^{2} P^{\mu}P_{\mu}=0$  and that 
on the LF the definition of the spin operator  appears  unified 
for massless and massive particles. A parallel discussion based on 
$p^{-}\ne0$ may also be given. 

As an illustration consider the three particle state on the LF with the 
total eigenvalues 
$p^{+}$,  $\lambda$  and $p^{\perp}$. In the {\it standard frame } 
with  $p^{\perp}=0\;$ it may be written as 
($\vert x_{1}p^{+},k^{\perp}_{1}; \lambda_{1}\rangle
\vert x_{2}p^{+},k^{\perp}_{2}; \lambda_{2}\rangle
\vert x_{3}p^{+},k^{\perp}_{3}; \lambda_{3}\rangle$ ) 
with  $\sum_{i=1}^{3} \,x_{i}=1$, $\sum_{i=1}^{3}\,k^{\perp}_{i}=0$, and 
$\lambda=\sum_{i=1}^{3}\,\lambda_{i}$. Aplying 
$e^{-i{(\bar p.\bar B)/p^{+}}}$ on it we obtain 
($\vert x_{1}p^{+},k^{\perp}_{1}+x_{1}p^{\perp}; \lambda_{1}\rangle
\vert x_{2}p^{+},k^{\perp}_{2}+x_{2}p^{\perp}; \lambda_{2}\rangle
\vert x_{3}p^{+},k^{\perp}_{3}+x_{3}p^{\perp}; \lambda_{3}\rangle $ ) 
now with  $p^{\perp}\ne0$. The  $x_{i}$ and $k^{\perp}_{i}$ indicate 
relative (invariant) parameters\footnote{We note ${p_{i}}^{+}=x_{i}p^{+}$, 
$ \;{p_{i}}^{\perp} =\,x_{i} {p}^{\perp} +  {k_{i}}^{\perp}$, and 
$(p\cdot p)= (2p^{+}p^{-}- {p}^{\perp} {p}^{\perp})=
\sum_{i}\left[({m_{i}}^{2}+{k_{i}}^{\perp} {k_{i}}^{\perp})/x_{i}\right]$ where 
$(p_{i}\cdot p_{i})= {m_{i}}^{2}$ and $\sum {p_{i}}^{\mu}= p^{\mu}$.                                 }
and do not depend upon the reference frame. The  $x_{i}$ is 
the fraction of the total longitudinal momentum 
carried by the  $i^{th} $ particle while 
 $k^{\perp}_{i}$ its transverse momentum. The state of a pion with 
 momentum ($p^{+},p^{\perp}$),  for example, 
 may be expressed  as 
an expansion over the LF Fock states constituted by the different 
number of partons 

$$
\vert \pi : p^{+},p^{\perp} \rangle=
\sum_{n,\lambda}\int {\bar \Pi}_{i}{{dx_{i}d^{2}{k^{\perp}}_{i}}\over
{{\sqrt{x_{i}}\,16\pi^{3}}}} \vert n:\,x_{i}p^{+},x_{i}p^{\perp}+
{k^{\perp}}_{i},
\lambda_{i}\rangle\;\psi_{n/\pi}(x_{1},{k^{\perp}}_{1},
\lambda_{1}; x_{2},...) 
$$

\noindent where \cite{bro} the summation is over all the Fock states 
$n$ and spin projections  $\lambda_{i}$, with 
${\bar\Pi}_{i}dx_{i}={\Pi}_{i} dx_{i}\; \delta(\sum x_{i}-1), $ 
and ${\bar\Pi}_{i}d^{2}k^{\perp}_{i}={\Pi}_{i} dk^{\perp}_{i} \;
 \delta^{2}(\sum k^{\perp}_{i})$.  The wave function of the 
parton $\psi_{n/\pi}(x,k^{\perp})$ 
indicates the probability amplitude for finding inside the pion 
the partons in the Fock state $n$ carrying 
the 3-momenta 
$(x_{i}p^{+}, x_{i}p^{\perp}+ k^{\perp}_{i}) $.

The {\it discrete symmetry} transformations may also be defined on the 
LF Fock states \cite{bro, pre1}
For example, under the conventional parity ${\cal P} $ 
the  spin operator  ${\cal J}_{3}$ is not invariant. 
We may rectify this by defining {\it LF Parity operation} by 
${\cal P}^{lf}=e^{-i\pi J_{1}}{\cal P}$. We find 
then  $B_{1}\to -B_{1}, B_{2}\to B_{2}, P^{\pm}\to P^{\pm}, P^{1}\to 
-P^{1}, P^{2}\to P^{2}$ etc. such that 
${\cal P}^{lf}\vert p^{+},p^{\perp}; m,s,\lambda, ..\rangle
\simeq \vert p^{+},-p^{1}, p^{2}; m,s,\,-\lambda, ..\rangle $. Similar
considerations apply for charge conjugation and  time inversion. For example, 
it is straightforward to construct \cite{pre1}  the free {\it LF Dirac spinor}  
$\chi(p)= 
[\sqrt{2}p^{+}\Lambda^{+}+(m-\gamma^{a}p^{a})\,\Lambda^{-}]\tilde \chi/
{ {\sqrt{\sqrt {2}p^{+}m}}}$ which is also an eigenstate of 
${\cal J}_{3}$ with eigenvalues 
$\pm 1/2$. Here $\Lambda^{\pm}= \gamma^{0}\gamma^{\pm}/{\sqrt 2}=
\gamma^{\mp}\gamma^{\pm}/2=({\Lambda^{\pm}})^{\dagger}$, 
$ (\Lambda^{\pm})^{2}=\Lambda^{\pm}$,  
and $\chi(\tilde p)\equiv \tilde \chi\,$ with  $\gamma^{0}
\tilde \chi= \tilde \chi$. The conventional (equal-time) 
spinor can also be constructed by the  procedure analogous to that 
followed for the LF spinor and it has  the well known form 
$ \chi_{con}(p)=  (m+\gamma.p)\tilde \chi/
{\sqrt{2m(p^{0}+m)}}$. 
Under the conventional parity operation  ${\cal P}: 
\chi'(p')=c \gamma^{0} \chi(p)$ (since we must require 
$\gamma^{\mu}={L^{\mu}}_{\nu}\,S(L)\gamma^{\nu}{S^{-1}}(L)$, etc.). We find 
$\chi'(p)=c 
[\sqrt{2}p^{-}\Lambda^{-}+(m-\gamma^{a}p^{a})\,\Lambda^{+}]\,\tilde \chi
/{\sqrt{\sqrt {2}p^{-}m}}$. For $p\neq\tilde p$ 
it is not proportional to $\chi(p)$ in contrast to the result in 
the case of the usual spinor where 
$\gamma^{0}\chi_{con}(p^{0},-\vec p)=\chi_{con}(p)$ for  $E>0$ (and  
$\gamma^{0}\eta_{con}(p^{0},-\vec p)=-\eta_{con}(p)$ for  $E<0$). 
However, applying parity operator twice we do show 
$\chi''(p)=c^{2}\chi(p)$ hence leading to the usual result 
$c^{2}=\pm 1$. The LF parity operator over spin $1/2$ Dirac spinor is 
${\cal P}^{lf}= c \,(2J_{1})\,\gamma^{0}$ and the corresponding transform 
of $\chi$ is shown to be an  eigenstate of ${\cal J}_{3}$. 

\vspace{0.5cm}
\nl {\bf Appendix C: \quad BRS-BFT Quantization on the LF of the CSM}

\setcounter{equation}{0}
\renewcommand{\theequation}{C. \arabic{equation}}

\vspace{0.5cm}

We apply here the recently proposed BFT procedure \cite{bata} 
which is elegant and avoids the 
computation of Dirac brackets. It would thus get tested \cite{pre7} 
on the LF as well and it
also allows us to construct (new) effective Lagrangian theories.

We convert the two second class constraints of the bosonized
CSM with $a>1$ into first class constraints 
according to the BFT formalism. 
We obtain then the  first class Hamiltonian
from the canonical Hamiltonian 
and recover the DB using Poisson brackets in the extended phase space. 
The corresponding first class Lagrangian is then found by 
performing the momentum integrations in the generating functional.

\begin{center}
{\bf (a) Conversion to First Class
Constrained Dynamical System }
\end{center}

The bosonized CSM model (for  $a>1$) is described 
by the action 

\begin{equation}
    S_{CSM} ~=~ \int d^2x~\left[
                          -\frac{1}{4}F_{\mu \nu}F^{\mu \nu}
                          +\frac{1}{2}\partial_{\mu}\phi\partial^{\mu}\phi
                          +eA_{\nu}(\eta^{\mu \nu}
                          -\epsilon^{\mu \nu})\partial_{\mu}\phi
              +\frac{1}{2}ae^{2}A_{\mu}A^{\mu}~\right],
\end{equation}
where 
$a$ is a regularization ambiguity which enters when we calculate 
the fermionic determinant in the fermionic CSM. 
The action in the LF coordinates takes the form 

\begin{equation}
S_{CSM}~=~\int d^2x^{-}\;\left[{1\over 2} (\partial_{+}A_{-}-\partial_{-}A_{+})^{2}
+\partial_{-}\phi\,\partial_{+}\phi +2 e  A_{-}{\partial_{+} \phi}
+a e^{2} A_{+}A_{-}\right], 
\end{equation}

We now make  the {\it separation},   
in  the scalar field (a generalized function) 
:\quad  $\phi(\tau,x^{-})= \omega (\tau)+\varphi(
\tau,x^{-})$. The Lagrangian density then becomes
\begin{equation}
{\cal L}={1\over 2} (\partial_{+}A_{-}-\partial_{-}A_{+})^{2}
+\partial_{-}\varphi\,\partial_{+}\varphi 
+a e^{2} [A_{+}+{2\over
{ae}}(\partial_{+}\varphi+\partial_{+}\omega)] A_{-}, 
\end{equation} 
We note that the dynamical fields are $A_{-}$ and 
$\varphi$ while $A_{+}$ has no kinetic term. On making a 
redefinition of the (auxiliary) field $A_{+}$ we can recast 
the action on the LF  in the  following form 
\begin{equation}
S_{CSM}~=~ \int dx^{-}\;\left[{\dot\varphi}\varphi'
+{1\over 2}({\dot A}_{-}-{A_{+}}')^{2}
- 2e {\dot A_{-}}\varphi
+a e^{2} A_{+}A_{-}\right],
\end{equation}

The canonical momenta are given by
\begin{eqnarray}
    \pi^{+}~&=&~0, \nonumber \\
    \pi^{-}~&=&~ \dot{A}_{-}~-~ A_{+}'-2e\varphi, \nonumber \\
    \pi_{\varphi}~&=&~ {\varphi}'.
\end{eqnarray}
We follow now the Dirac's standard procedure in order to build an
Hamiltonian framework on the LF. The definition of the canonical momenta
leads to two primary constraints 
\begin{eqnarray}
    \pi^{+} \approx 0, \\
 \Omega_1 \equiv (\pi_{\varphi}-\varphi')\approx 0
\end{eqnarray}
and we derive one secondary constraint 
\begin{equation}
    \Omega_2 \equiv \partial_{-} \pi^{-} + +2e \varphi'+
a e^{2} A_{-} \approx 0.
\end{equation}
This one  follows when we require the $\tau$ independence 
(e.g., the persistency) of 
the primary constraint $\pi^{+}$  with respect to the preliminary  Hamiltonian

\begin{equation}
    H' = {H_c}^{l.f.} + \int dx~u_{+}\pi^{+}+\int dx~u_{1}\Omega_1 ,
\end{equation}
where $H_c$ is the canonical Hamiltonian  
\begin{eqnarray}
    {H_c}^{l.f.} &=& \int\!dx~\left[~
            \frac{1}{2}(\pi^{-}+2e\varphi)^2 + (\pi^{-}+2e\varphi) A_{+}'
-ae^{2} A_{+}A_{-}
               \right],
\end{eqnarray}
and we employ the standard equal-$\tau$ Poisson brackets. 
The  $u_{+}$ and $u_{1}$ denote  the Lagrange multiplier fields. The 
persistency requirement for $\Omega_{1} $ give conditions to 
determine  $u_{1}$. The 
Hamiltonian is next extended to include also the secondary constraint
\begin{equation}
    {H_e}^{l.f.} = {H_c}^{l.f.} + \int dx~u_{+}\pi^{+}+\int dx~u_{1}\Omega_1 
+\int dx~u_{2}\Omega_{2} 
\end{equation}
and the procedure is now repeated with respect to the extended Hamiltonian. 
For the case $a>1$, 
no more secondary constraints are seen 
to arise and we are left only with the persistency conditions which
determine the multipliers $u_{1}$ and $u_{2}$ while $u_{+}$ is left
undetermined. We also  
find\footnote{We make the convention that the first variable in an equal-
$\tau$ bracket refers to the longitudinal coordinate $x^{-}\equiv x$
while the second one to $y^{-}\equiv y$} 
 $\{\Omega_{i},\Omega_{j}\}~=~D_{ij}~$ $ (-2
\partial_{x} \delta(x-y))$ where $i,j =1,2$ and $~D_{11}=1,~D_{22}=a
e^2, ~D_{12}=~D_{21}= -e$ and  $\pi^{+}$ is shown to have 
vanishing brackets with $\Omega_{1,2}$. 
The $\pi^{+}\approx 0 $ constitutes a  first class constraint on the phase
space; it generates local transformations of $A_{+}$ which leave the 
$H_{e}$ invariant, $\{\pi^{+}, H_{e}\}= \Omega_{2}\approx 0$. 
The  $\Omega_1,\Omega_2 $ constitute a set of 
second class constraints and do not involve $A_{+}$ or $\pi^{+}$.  
It is very convenient, though not necessary, to add to the set of
constraints on the phase space the (accessible) gauge fixing constraint
$A_{+}\approx 0$. It is evident  from  
that such a gauge freedom is {\it  not} 
available at the Lagrangian level. 
We will also implement  (e.g., turn  into
strong equalities)  the (trivial) 
pair of weak constraints $A_{+}\approx 0,\; \pi^{+}\approx 0$ by defining 
the Dirac brackets with respect to them.  It is  easy to see that 
for the other remaining dynamical variables the corresponding  
Dirac brackets  coincide with the standard Poisson brackets. The
variables $A_{+}, \pi^{+}$ are thus 
removed from the discussion,   leaving 
behind a constrained dynamical system with the two 
second class  constraints $\Omega_{1}, 
\;\Omega_{2}$ and the  light-front Hamiltonian  
\begin{equation}
    H^{l.f.} = \frac{1}{2} \int\!dx~
            (\pi^{-}+2e\varphi)^2 + 
                  \int dx~u_{1}\Omega_1 +\int dx~u_{2}\Omega_{2} 
\end{equation}
which will be now handled by the BFT procedure. 

We introduce the following linear combinations $\top_{i}$, $i=1,2$, 
of the above constraints  

\begin{eqnarray}
    \top_{1}=c_1 (\Omega_1+\frac{1}{M}\Omega_2)\nonumber \\
\top_{2}=c_2 (\Omega_1-\frac{1}{M}\Omega_2)
\end{eqnarray}
where $c_1={1}/{\sqrt{2(1-e/M)}}$,  $c_2={1}/{\sqrt{2(1+e/M)}}$, 
$M^2=a e^2$, and $a>1$. They satisfy 
\begin{equation}
\{\top_i,\top_j\}=\delta_{ij} (-2\partial_{x}\delta(x-y))
\end{equation}
and thus diagonalize the constraint algebra. 

We now introduce new auxiliary fields $\Phi^{i}$
in order to convert the second class constraint $\top_{i}$ into
first class ones in the extended phase space. 
Following BFT \cite{bata} we require these fields to satisfy
\begin{eqnarray}
   \{A^{\mu}(\mbox{or}~ \pi_{\mu}), \Phi^{i} \} &=& 0,~~~
   \{\varphi(\mbox{or}~ \pi_{\varphi}), \Phi^{i} \} = 0, \\ \nonumber
   \{ \Phi^i(x), \Phi^j(y) \} &=& \omega^{ij}(x,y) =
                      -\omega^{ji}(y,x),
\end{eqnarray}
where $\omega^{ij}$ is a constant and antisymmetric matrix. 
The strongly involutive modified constraints 
$\widetilde{\top}_{i}$ satisfying the abelian algebra 
\begin{eqnarray}
\{\widetilde{\top}_{i}, \widetilde{\top}_{j} \}=0
\end{eqnarray}
as well as the boundary conditions,
$\widetilde{\top}_i \mid_{\Phi^i = 0} = \top_i$
are then postulated to take the form of the following expansion 
\begin{equation}
  \widetilde{\top}_i( A^\mu, \pi_\mu, \varphi, \pi_{\varphi}; \Phi^j)
         =  \top_i + \sum_{n=1}^{\infty} \widetilde{\top}_i^{(n)}, 
                       ~~~~~~\top_i^{(n)} \sim (\Phi^j)^n.
\end{equation}
The first order correction terms in this  infinite series  are
written  as 
\begin{equation}
  \widetilde{\top}_i^{(1)}(x) = \int dy X_{ij}(x,y)\Phi^j(y).
\end{equation}
The  first class  
constraint algebra  of $\widetilde{\top}_i$ then leads to 
the following condition: 
\begin{equation}
   \{\top_i,\top_j\} + \{ \widetilde{\top}_i^{(1)},
 \widetilde{\top}_i^{(1)}\}=0  
\end{equation}
or 
\begin{equation}
(-2\partial_x\delta(x-y)) \delta_{ij} +
   \int dw~ dz~
        X_{ik}(x,w) \omega^{kl}(w,z) X_{jl}(y,z)
         = 0.
\end{equation}
There is clearly some arbritrariness  
in the appropriate choice of $\omega^{ij}$ and $X_{ij}$ 
which corresponds to the canonical transformation
in the extended phase space. 
We can take without any loss of generality the simple solutions, 
\begin{eqnarray}
  \omega^{ij}(x,y)
         &=& - \delta^{ij} \epsilon(x-y) \nonumber  \\
  X_{ij}(x,y)
         &=&   \delta_{ij} \partial_{x}\delta ( x- y),
\end{eqnarray}
Their inverses are easily shown to be 
\begin{eqnarray}
  {\omega^{-1}}_{ij}(x,y)
         &=& -\frac{1}{2} \delta_{ij} \partial_x\delta(x-y) \nonumber  \\
  ({X^{-1}})^{ij}(x,y)
         &=&  \frac{1}{2} \delta^{ij} \epsilon( x- y),
\end {eqnarray}

With the above choice,  we find up to the first order
\begin{eqnarray}
\widetilde{ \top}_{i}&=&\top_{i}+\widetilde\top^{(1)}_{i} \\ \nonumber
&=&\top_{i} +\partial \Phi^{i},
\end{eqnarray}
and  a strongly first class  constraint algebra
\begin{equation}
  \{\top_{i}+ \widetilde{\top}^{(1)}_{i},
            \top_{j}+ \widetilde{\top}^{(1)}_{j} \} = 0.
\end{equation}
The higher order correction terms (suppressing the
integration operation ) 
\begin{eqnarray}
\widetilde{\top}^{(n+1)}_{i} =- \frac{1}{n+2} \Phi^{l} 
                  {\omega^{-1}}_{lk} ({X^{-1}})^{kj} 
B_{ji}^{(n)}~~~~~~~(n \geq 1)
\end{eqnarray}
with
\begin{eqnarray}
B^{(n)}_{ji} \equiv \sum^{n}_{m=0} 
       \{ \widetilde{\top}^{(n-m)}_{j}, 
          \widetilde{\top}^{(m)}_{i} \}_{(A, \pi, \varphi, \pi_{\varphi} )}+
         \sum^{n-2}_{m=0} 
          \{ \widetilde{\top}^{(n-m)}_{j}, 
              \widetilde{\top}^{(m+2)}_{i} \}_{(\Phi)}
\end{eqnarray}
automatically vanish as a consequence of the proper choice of $\omega^{ij}$ 
made above. 
The Poisson brackets are to be computed here using the standard canonical 
definition for $A_\mu$ and $\varphi$ as postulated above. 
We have now only  the first class constraints 
in the extended phase space and in view of the proper choice  
only $\widetilde\top_{i}^{(1)}$ contributes in the infinite series above. 
\bigskip

\begin{center}
{\bf (b)-  First Class Hamiltonian  and Dirac Brackets}
\end{center}
\medskip

We next introduce modified ("gauge invariant") dynamical variables
$\widetilde{F} \equiv (\widetilde{A}_{\mu},\widetilde{\pi}^{\mu},
\widetilde{\varphi}, \widetilde{\pi}_{\varphi} )$ corresponding to 
$F \equiv (A_{\mu},\pi^{\mu},\varphi, \pi_{\varphi} )$ over the phase space 
by requiring the the following strong involution condition for 
$\widetilde{F}$ with the first class
constraints in our extended phase space, viz, 
\begin{eqnarray}
\{ \widetilde{\top}_{i}, \widetilde{F} \} =0
\end{eqnarray}
with 
\begin{eqnarray}
\widetilde{F}(A_{\mu}, \pi^{\mu}, \varphi, \pi_{\varphi}; \Phi^{j} ) &=&
            F + \sum^{\infty}_{n=1} \widetilde{F}^{ (n)},
            ~~~~~~~ \widetilde{F}^{(n)} \sim (\Phi^{j})^{n}
\end{eqnarray}
and which satisfy the boundary conditions,
$\widetilde{F}\mid_{\Phi^i = 0} = F$.

The first order correction terms are easily shown to be given by 
\begin{equation}
\widetilde{F}^{(1)}(x) =  -\int~ du\, dv\, dz~ \Phi^{j}(u) {{\omega}^{-1}}_{jk}
(u,v) {X^{-1}}^{kl}(v,z)~
 \{ \top_{l}(z), F(x) \}_{(A, \pi, \varphi, \pi_{\varphi})}. 
\end{equation}
We find 
\begin{eqnarray}
\widetilde{A}_{-}^{(1)}~&=&~   
     \frac{1}{2M} \partial(c_1 \Phi^{1}-c_2 \Phi^{2}) \nonumber  \\
\widetilde{\pi}^{-{(1)}}~&=&~ \frac {M}{2} (c_1\Phi^{1}-c_2 \Phi^{2}) 
                                                         \nonumber \\
\widetilde{\varphi}^{(1)} ~&=&~ 
                       - \frac{1}{2} (c_1 \Phi^{1}+c_2 \Phi^{2}),\nonumber \\
\widetilde{\pi}_{\varphi}^{(1)} ~&=&~   \frac{1}{2} \partial
\left[c_1 (1-\frac{2e}{M})\Phi^{1}+c_2  (1+\frac{2e}{M}) \Phi^{2}\right]
\end{eqnarray}
where only the combinations $(c_1 \Phi^{1}\pm c_2 \Phi^{2})$ of the 
auxiliary fields are seen to occur. 
Furthermore, since the  modified variables $\widetilde F= F+\widetilde
 F^{(1)}+...$,   
up to the first order corrections,
are found to be strongly involutive
as a consequence of the proper choice made above, 
the higher order correction terms
\begin{eqnarray}
\widetilde{F}^{(n+1)} &=&
                              -\frac{1}{n+1}
                              \Phi^{j}\omega_{jk} X^{kl} G^{(n)}_{l},
\end{eqnarray}
with
\begin{eqnarray}
G^{(n)}_{l} &=& \sum^{n}_{m=0}
                \{ \top_{i}^{(n-m)}, 
                     \widetilde{F}^{(m)}\}_{(A, \pi, \phi, \pi_{\phi} )}
                +   \sum^{n-2}_{m=0}
                \{ \top_{i}^{(n-m)}, \widetilde{F}^{(m+2)}\}_{(\Phi)}
           +  \{ \top_{i}^{(n+1)}, \widetilde{F}^{(1)} \}_{(\Phi)} 
                  \nonumber \\
\end{eqnarray}
again vanish. In principle we may follow similar procedure for 
any functional of the phase space variables; it may get, however, 
involved.

We make a side remark on the Dirac formulation for dealing with the
systems with second class constraints by using  the Dirac
bracket (DB), rather than extending the phase space. In fact, the
Poisson brackets of the modified (gauge invariant) 
variables $\widetilde F$ in the 
BFT formalism are related \cite {bata} to  the DB,   
 which implement  the constraints $\top_i\approx 0$ in the problem
under discussion,  by the relation
$\{f,g\}_{D}= \{\widetilde f,\widetilde g\}\mid_{\Phi^{i}=0}$. In
view of only the linear first order correction in CSM 
the computation of the right hand side is quite simple.  We list some 
of the Dirac brackets 

\begin{eqnarray}
 \{\pi^{-},\pi^{-}\}_{D}&=&\{\widetilde{\pi^{-}}, 
\widetilde{\pi^{-}}\} |_{\Phi=0}  \nonumber \\
   & = & \{\widetilde{\pi^{-}}^{(1)}, \widetilde{\pi^{-}}^{(1)} \}
         =  \frac{a^2 e^2}{(a-1)} (-\frac{1}{4} \epsilon(x-y)),  
      \nonumber  \\
 \{\varphi,\varphi\}_{D} & = &
\{\widetilde{\varphi}, \widetilde{\varphi}\} |_{\Phi=0} \nonumber \\
  &  = & \{\widetilde{\varphi}^{(1)}, \widetilde{\varphi}^{(1)} \} 
    = \frac{a}{(a-1)} (-\frac{1}{4} \epsilon(x-y))
      \nonumber  \\
 \{\varphi,\pi^{-}\}_{D}&=& \{\widetilde{\varphi}^{(1)}, \widetilde{\pi^{-}}^
{(1)}\} = \frac{ae}{(a-1) } (-\frac{1}{4} \epsilon(x-y)).
\end{eqnarray}
The other ones follow on using the now strong relations 
$\Omega_1=\Omega_2=0$ with respect to $\{,\}_{D}$ and from $H^{l.f}$  
it follows that the LF Hamiltonian 
 reduces effectively to 
\begin{equation}
   H_D^{l.f.}= \frac{1}{2}\int dx  \, (\pi^{-}+2e\varphi)^{2}. 
\end{equation}
The first class LF Hamiltonian $\widetilde H$ which satisfies the
boundary condition $\widetilde H\mid_{\Phi^{i}=0}=H_D^{l.f.}$ and is in
strong involution with the constraints  
$\widetilde \top_{i}\;$,  e.g., 
 $\{\widetilde \top_{i},\widetilde H\}=0 $, may be constructed
following the BT procedure  or simply guessed for the CSM. It is given by 
\begin{equation}
\widetilde H= \frac{1}{2}\int dx(\widetilde \pi^{-}+2e\widetilde\varphi)^2
\end{equation}
which is just the expression in of  $H_D^{l.f.}$ 
with field variables $F$ replaced by the
$\widetilde F$ variables, which already commute with the constraints 
$\widetilde T_{i}$. 
We do also check that $\{\widetilde H,\widetilde H\}=0$ and we may
identify $\widetilde H$ with the BRS Hamiltonian. 
This completes the operatorial 
conversion of the original second class system with the Hamiltonian
$H_{c}$ and constraints $\Omega_{i}$ into the first class one with
the Hamiltonian $\widetilde H$ and (abelian) constraints 
$\widetilde T_{i}$.
\bigskip
\begin{center}
{\bf (c)- First Class Lagrangian}
\end{center}
\medskip

We consider now the partition function of the model in order to
construct the Lagrangian corresonding to $\widetilde H$ in the
canonical Hamiltonian formulation discussed above. 

We start by representing  each of the auxiliary field $\Phi^{i}$ by a pair
of fields $\pi^{i}, \theta^{i},\; i=1,2\; $ defined by 
\begin{equation}
\Phi^{i}=\frac{1}{2}\pi^{i}-\int du\;\; \epsilon(x-u)\;\theta^{i}(u)
\end{equation}
such that $\pi^{i}, \theta^{i}$ satisfy 
\begin{equation}
\{\pi^{i},\theta^{j}\}=-\delta^{ij}\delta(x-y)\ \  \  etc.,
\end{equation}
e.g., the  (standard Heisenberg type) canonical Poisson brackets. 

Then, The  Phase Space Partition Function Is Given By the
Faddeev formulae 

\begin{equation}
Z=  \int  {\cal D} A_{-}
          {\cal D} \pi^{-}
          {\cal D} \varphi
          {\cal D} \pi_{\varphi}
          {\cal D} \theta^{1}
          {\cal D} \pi^{1}
          {\cal D} \theta^{2}
          {\cal D} \pi^{2}
               \prod_{i,j = 1}^{2} \delta(\widetilde{\top}_i)
                           \delta(\Gamma_j)
                \mbox{det} \mid \{ \widetilde{\top}_i, \Gamma_j \} \mid
                e^{iS},
\end{equation}
where
\begin{equation}
S  =  \int d^2x \left(
               \pi^{-} {\dot A_{-}} +\pi_{\varphi} {\dot \varphi} 
+ \pi^{1} {\dot \theta^{1}} +\pi^{2} {\dot \theta^{2}} - \widetilde {\cal H}
                \right)\equiv \int d^{2}x \; {\cal L},
\end{equation}
with the Hamiltonian density $\widetilde {\cal H}$ corresponding to the 
Hamiltonian 
$\widetilde H$ which is now expressed in terms
of $(\theta^{i}, \pi_{i})$ rather than in terms of  $\Phi^i$.
The gauge-fixing conditions $\Gamma_i$ are chosen
such  that the determinants occurring in
the functional measure are nonvanishing.
Moreover, $\Gamma_i$ may be taken  to be independent of the momenta
so that they correspond to the  Faddeev-Popov type gauge conditions.

We will now verify in the  {\it unitary gauge}, defined by 
the original  second class constraints:\quad    $\Gamma_i\equiv
\Omega_i=0$, i=1,2  being employed in the partition function, 
 do in fact lead to the original 
Lagrangian. We check that the determinants in the functional
measure are non-vanishing and field independent while  the  product of 
delta functionals reduces to 
\begin{equation}
\delta(\pi_{\varphi}-\varphi')\delta({\pi^{-}}'+2e\varphi'+M^2 A_{-})
\delta({\pi^{1}}'-4\theta^{1})\delta({\pi^{2}}'-4\theta^{2})
\end{equation}
Since $\pi_{\varphi}$ is absent from $\widetilde H$ 
we can perform functional integration over it using 
the first delta functional. 
The second delta functional is exponentiated as usual and we name the
integration variable as $A_{+}$ for convenience. The 
functional integral over $\theta^{1}$ and $\theta^{2}$ are 
easily performed due to the presence of the delta functionals and it 
also reduces $\widetilde {\cal H}$ to  
$(\pi^{-}+2e\varphi)^{2}/2$. The  functional integrations over the
then decoupled variables $\pi^{1}$ and $\pi^{2}$ give rise to constant factors
which are absorbed in the normalization. The partition function in
the unitary gauge thus becomes 
\begin{equation}
Z=  \int  {\cal D} A_{-}
          {\cal D} \pi^{-}
          {\cal D} \varphi
          {\cal D} A_{+}
                          e^{iS},
\end{equation}
with 
\begin{equation}
S  =  \int d^2x \left[
               \pi^{-} {\dot A_{-}} +{\varphi}' {\dot \varphi} 
+({\pi^{-}}'+2e\varphi'+M^{2}A_{-}) A_{+} -\frac{1}{2}(\pi^{-}+2e\varphi)^{2}
                \right],
\end{equation}
Performing the shift $\pi^{-}\rightarrow \pi^{-}-2e\varphi$ and
doing subsequently a Gaussian integral over $\pi^{-}$ we obtain the
original bosonized  Lagrangian with $\omega$ eliminated by the field 
redefinition of $A_{+}$. It is interesting to recall  that 
while constructing the LF Hamiltonian framework we eliminated
the variable $A_{+}$  making use of the gauge freedom on the LF
phase space and  it  gave rise to appreciable simplification. 
However, on going
over to the first class Lagrangian formalism using 
 the partition functional  this  variable reappears as it should, since 
the initial bosonized action is not  gauge invariant due to the 
presence of the mass term for
the gauge field.  Making other acceptable choices for 
gauge-functions  we can arrive at different effective Lagrangians for
the system under consideration. It is interesting to recall that in the 
fermionic Lagrangian the right-handed component of the fermionic field 
describes a free field and only the left-handed one is gauged.  
It is also clear from our discussion that $\widetilde H$ proposed above 
 is not unique and we could modify it so that it still leads to 
the original Lagrangian  in the unitary gauge. 
The corresponding first class Lagrangian
would produce still other gauge-fixed effective Lagrangians. 



\begin{thebibliography}{99}
\bibitem{kurs}We recall the discovery of Kruskal-Szekers coordinates which 
threw a new light on the problem of the Schwarzshild singularity. \\ The 
$\pm$ components of a tensor, for example, $A^{\mu}$ are defined by 
$A^{\pm}=A_{\mp}= (A^{0}\pm A^{3})/{\sqrt 2}$ and the metric may be read from 
$A\cdot B= A^{+}B^{-}+A^{-}B^{+}-A^{1}B^{1}- A^{2}B^{2}$.
\bibitem{fno1}
The LF components of the four-momentum are
 $k^{\mu}=(k^{-},k^{+},k^{\perp})$ where 
$k^{\pm}=(k^{0}{\pm}k^{3})/{\sqrt 2}$.  Here 
$k^{-}$ is the LF energy while $k^{\perp}$ and  $k^{+}$  indicate 
the transverse and the {\it longitudinal} components of the  momentum respectively.  
For a free massive particle on the mass shell we have  
the dispersion relation: $2 k^{-}k^{+}=
({k^{\perp}}^{2}+m^{2}) > 0$ so that  
  $k^{\pm}$ are both positive  when $ k^{0}>0$ or both negative when 
   $ k^{0}<0$. It {\it has  no square root} as found in $k^{0}={\pm}\sqrt{ 
   {\vec k}^{2}+m^{2}}$.  
The conservation of the total 
longitudinal momentum does not permit the excitation of massive 
quanta by the LF vacuum which has vanishing  longitudinal momentum. 
It should, however,
be noted that when dealing with the momentum space 
 loop integrals, a significant contribution may arise 
from such   configuration in the integrand; the reason
being that we have to deal with the products of several
distributions. The components  $(k^{1},k^{2},k^{3})$ in the {\it instant 
form} theory on the other hand may take positive
or negative values and  the conventional theory 
vacuum state may contain  
 an arbitrary number of particles 
(and antiparticles) which may mix with the vacuum state, with no 
particles, to form the ground state.
\bibitem{fno2} For  massless particles  the correlation   ceases to exist
  at the point  $p^{\perp}\to 0$ since  $2 p^{+}p^{-}=p^{\perp}p^{\perp}\to 0$.
\bibitem{fno4} Here it is understood that 
an unsymmetrical expression like $ \bar{\psi} \gamma^{\mu}
\partial_{\mu} \psi $ is to be replaced  by its symmetrized form  
$\left[\bar{\psi} \gamma^{\mu}
\partial_{\mu} \psi- \partial_{\mu}\bar{\psi}\gamma^{\mu}\psi \right]/2 $. 
It is convenient to work on the LF in terms of the projected four spinors  
$\psi_{\pm}$.
\end{thebibliography}
\end{document}